\documentclass[aps,prd,preprint,eqsecnum,nofootinbib]{revtex4-1}
\usepackage[english]{babel}
\usepackage[utf8]{inputenc}
\usepackage{verbatim}
\usepackage{graphicx}
\usepackage{mathrsfs}
\usepackage{amssymb}
\usepackage{amsthm}
\usepackage{amsmath}
\usepackage{wasysym}
\usepackage{setspace}
\usepackage{placeins}

\allowdisplaybreaks[1]
\graphicspath{{./figs/}}

\newcommand{\C}{\mathbb{C}}

\newcommand{\nn}{\nonumber}
\newcommand{\mc}[1]{\mathcal{#1}}

\newcommand{\pd}[2]{\frac{\partial #1}{\partial #2}}

\newcommand{\spaa}[2]{\langle#1\,#2\rangle}
\newcommand{\spbb}[2]{[#1\,#2]}

\newcommand{\spvec}[3]{\left\langle #1\:\!|\:\!#2\:\!|\:\!#3\right]}
\newcommand{\Res}{\mathop{\rm Res}}

\newcommand{\Tr}{\operatorname{Tr}}

\newcommand{\HDB}{I_{\text{HDB}}}
\newcommand{\VDB}{I_{\text{VDB}}}

\newcommand{\bc}{\begin{center}}
\newcommand{\ec}{\end{center}}

\def\eqn#1{eq.~(\ref{#1})}

\def\eqns#1#2{eqs.~(\ref{#1}) and~(\ref{#2})}

\def\sect#1{Sec.~{\ref{#1}}}

\def\spa#1.#2{\left\langle#1\,#2\right\rangle}
\def\spb#1.#2{\left[#1\,#2\right]}
\def\nsand#1.#2.#3{%
\left\langle\smash{#1}{\vphantom1}\right|{#2}%
\left|\smash{#3}{\vphantom1}\right]}

\def\eps{\epsilon}
\def\Ord{{\cal O}}
\def\CP{{\mathbb C\mathbb P}}

\def\fig#1{fig.~{\ref{#1}}}

\def\figs#1#2{figs.~{\ref{#1}} and {\ref{#2}}}

\def\OneLoop{I}
\def\Ps{P^{*}}
\def\Pss{P^{**}}

\def\cbox#1{\hbox to\linewidth{\hfill#1\hfill}}

\begin{document}
\date{\today}
\noindent CERN-PH-TH-2014-262\hskip 15mm\hfill NORDITA-2014-141\hfill\hskip 15mm UUITP-20/14\\
CALT-TH--2015--014\hfill \hskip 15mm\hfill Saclay IPhT--T14/239\hfill \hskip 15mm
 NIKHEF/2014-051\\[-5mm]
\title{Cross-Order Integral Relations from Maximal Cuts}

\author{Henrik Johansson}
\affiliation{
\begin{spacing}{1.1}
\cbox{Theory Group, Physics Department, CERN,
CH--1211 Geneva 23, Switzerland;} \\
\cbox{NORDITA
KTH Royal Institute of Technology} \cbox{and Stockholm University,
Roslagstullsbacken 23, SE-10691 Stockholm, Sweden;}\\
{\rm and}\\
\cbox{Department of Physics and Astronomy, Uppsala University,
SE-75108 Uppsala, Sweden}\\
{\sf Henrik.Johansson@physics.uu.se}
\end{spacing}\vspace*{-8mm}}
\author{David A. Kosower}
\affiliation{
\begin{spacing}{1.05}
Walter Burke Institute for Theoretical Physics,
\cbox{California Institute of Technology, Pasadena, CA 91125, USA}\\
{\rm and}\\
\cbox{Institut de Physique Th\'eorique, CEA-Saclay,
F--91191 Gif-sur-Yvette cedex, France}\\
{\sf David.Kosower@cea.fr}
\end{spacing}\vspace*{-8mm}}
\author{Kasper J. Larsen}
\affiliation{
\begin{spacing}{1.1}
\cbox{Nikhef, Theory Group, Science Park 105,
NL--1098 XG Amsterdam, The Netherlands} \\
{\rm and}\\
\cbox{Institute for Theoretical Physics, ETH Z{\"u}rich,
8093 Z{\"u}rich, Switzerland}\\
{\sf Kasper.Larsen@phys.ethz.ch}
\end{spacing}\vspace*{-8mm}}
\author{Mads S{\o}gaard}
\affiliation{
\begin{spacing}{1.1}
\cbox{Niels Bohr International Academy and Discovery Center,
Niels Bohr Institute,} \\ 
\cbox{University of Copenhagen,
Blegdamsvej 17, DK-2100 Copenhagen, Denmark}\\
{\sf Mads.Sogaard@nbi.ku.dk}
\end{spacing}\vspace*{-7mm}}
\medskip

\begin{abstract}
\begin{spacing}{1.2}
We study the ABDK relation using maximal cuts of one- and two-loop integrals with up to five external legs.
We show how to find a special combination of integrals that allows the relation to exist,
and how to reconstruct the terms with one-loop integrals squared.
The reconstruction relies on the observation that integrals across
different
loop orders can have support on the same generalized unitarity cuts and can
share global poles. We discuss the appearance of
nonhomologous integration contours
in multivariate residues. Their origin can be understood in simple terms, and their existence enables
us to distinguish contributions from different integrals. Our analysis suggests
that maximal and near-maximal cuts can be used to infer the existence of
integral identities more generally.
\end{spacing}
\end{abstract}

\maketitle

\section{Introduction}

The development of on-shell methods~\cite{UnitarityFormalism,GeneralizedUnitarityZqqgg,%
GeneralizedUnitarityBCF,GeneralizedUnitarityBMST,BCFW,LoopRecursion,OnShellReview} for
computing scattering amplitudes in quantum field theory has led to
rapid progress in numerous directions in recent years, including
higher-loop computations in the maximally supersymmetric ($\mc N=4$) Yang--Mills 
theory (MSYM)~\cite{ABDK,TwoLoopFivePointEven,ABDK5,HigherLoopN=4base,%
Bern:2007ct,HigherLoopN=4results,RemainderFunction,%
HigherLoopN=4misc,HigherLoopN=4remainder,HigherLoopN=4cluster,HigherLoopN=4extra}, the understanding of dual conformal~\cite{DualConformal} and
Yangian~\cite{Yangian} symmetries, the development of alternate viewpoints on
amplitudes such as twistor strings~\cite{Twistors} and 
Grassmannians~\cite{GrassmanniansI,MergeAndSplit,GrassmanniansII}, as well as the
development of numerical one-loop libraries~\cite{BlackHatI,CutTools,GoSam,OpenLoops,NGluon} applied to next-to-leading
order (NLO) calculations for phenomenology at CERN's Large Hadron
Collider.  Related developments include computations at strong Yang--Mills coupling~\cite{StrongCoupling},
at all values of the coupling~\cite{BridgingI,BridgingII,BridgingIII}, 
computations of the infrared structure of amplitudes~\cite{IRstructure} 
and nonabelian exponentiation~\cite{Webs}, and advances in the computation
of integrals out of which amplitudes are built~\cite{Integrals}.

\def\Mampl#1#2{M_{#1}^{(#2)}}
Several years ago, Bern, Dixon, and Smirnov (BDS) wrote down a
remarkable conjecture~\cite{BDS}, namely that the planar part of all 
maximally helicity-violating
(MHV) amplitudes in $\mc N=4$ supersymmetric Yang--Mills theory can be written in
a certain sense as exponentials of the one-loop amplitude,
\begin{equation}
1+\sum_{L=1}^\infty a^L \Mampl{n}{L}(\{s_{ij}\};\eps) =
\exp\biggl[\sum_{l=1}^\infty a^l
 \Bigl(f^{(l)}(\eps) \Mampl{n}{1}(\{s_{ij}\};l\eps) +C^{(l)}+E_n^{(l)}(\eps)\Bigr)\biggr]\,,
\label{BDSconjecture}
\end{equation}
where $M_n^{(L)}$ is the $n$-point $L$-loop MHV leading-color ordered
amplitude after removing a factor of the tree color-ordered amplitude,
and $a_s$ is a rescaled version of the Yang--Mills coupling squared.
The additional functions $f^{(l)}$, $C^{(l)}$, and $E_n^{(l)}$ are
independent of the external kinematics, and the first two are
independent of the number of legs $n$.  The conjecture is true for the
four- and five-point functions~\cite{DualConformal,WilsonAmplitudeDuality}, but fails for six
or more external legs.  Its failure has been a stimulus to striking
advances~\cite{HigherLoopN=4remainder} in understanding the left-over, `remainder'
terms~\cite{RemainderFunction}.

The BDS conjecture was in turn based on an earlier calculation of the two-loop four-point amplitude~\cite{ABDK}
by Anastasiou, Bern, Dixon, and one of the present authors (ABDK).   These authors found by direct calculation
that,
\begin{equation}
\Mampl42(s,t;\eps) = \frac12 \Bigl[\Mampl41(s,t;\eps)\Bigr]^2 + f^{(2)}(\eps) \Mampl41(s,t;2\eps) + C^{(2)}
+\Ord(\eps)\,,
\label{ABDK4}
\end{equation}
where $f^{(2)}(\eps) = -(\zeta_2+\zeta_3\eps+\zeta_4\eps^2+\cdots)$ and $C^{(2)} = -\zeta_2^2/2$.

Our aim in this paper is to examine this relation within the context of two-loop
maximal generalized unitarity, another development of recent years in
the domain of scattering amplitudes.  The goal of the two-loop unitarity program
is to enable theorists to go beyond NLO calculations in order to meet
the challenge of future precision measurements at the LHC.  Here we will instead
examine an application in the context of maximally supersymmetric Yang--Mills theory.

The unitarity and generalized unitarity
methods~\cite{UnitarityFormalism,GeneralizedUnitarityZqqgg,%
GeneralizedUnitarityBCF,GeneralizedUnitarityBMST,UnitaritySupplemental,%
OtherUnitarity,%
BCFCutConstructible,OPP,Forde,Badger,DdimensionalII,BFMassive,BergerFordeReview,%
Bern:2010qa}
at one loop have made many previously-inaccessible calculations feasible.
Of particular note are processes with many partons in the final state.
In its modern form of generalized unitarity, it can be
applied either analytically or purely
numerically~\cite{EGK,BlackHatI,CutTools,MOPP,Rocket,BlackHatII,CutToolsHelac,%
Samurai,WPlus4,NGluon,MadLoop,GoSam}.
In its numerical form, the formalism underlies recent software libraries and
programs used for LHC phenomenology.  In this
approach, the one-loop amplitude in a quantum field theory is written as a sum over a set
of basis integrals, with coefficients that are rational in external
spinor variables,
\begin{equation}
{\rm Amplitude} = \sum_{j\in {\rm Basis}}
  {\rm coefficient}_j \times {\rm Integral}_j +
{\rm Rational}\,.
\label{MasterEquation}
\end{equation}
The integral basis for one-loop amplitudes with massless internal lines
contains box, triangle, and bubble integrals 
(dropping all terms of $\mathcal{O}(\epsilon)$ in the
dimensional regulator).  The coefficients are calculated from products
of tree amplitudes, typically by performing contour integrals (numerically, via
discrete Fourier projection).  In the Ossola--Papadopoulos--Pittau (OPP)
approach~\cite{OPP}, this
decomposition is carried out at the integrand level rather than at
the level of integrated expressions.

Higher-loop amplitudes can also be written in a form similar to that given in
\eqn{MasterEquation}.  As at one loop, one can carry out such a
decomposition at the level of the integrand.  This generalization of the
OPP approach has been pursued by Mastrolia and
Ossola~\cite{MastroliaOssola} and collaborators, and also by Badger,
Frellesvig, and~Zhang~\cite{BadgerFZ}.  The reader should consult
refs.~\cite{Zhang:2012ce,Mastrolia:2012an,Kleiss:2012yv,Mastrolia:2012wf,Mastrolia:2012du,%
Huang:2013kh,Mastrolia:2013kca}
for further developments within this approach.  
Arkani-Hamed, Bourjaily, Cachazo, Caron-Huot, 
 and Trnka have developed an integrand-level
approach~\cite{AllLoopIntegrand1,AllLoopIntegrand2} specialized to planar contributions to the $\mc N=4$ supersymmetric
theory, but to all loop orders.
In ref.~\cite{AllLoopIntegrand2}, these authors used global
residues to study the cancellation of the $1/\eps^4$ poles in the argument
of the exponential in \eqn{BDSconjecture}.  The present paper can be thought of
as extending this study to some of
the less-singular and finite terms.

Within the unitarity method applied at the level of integrated expressions,
one can distinguish two basic approaches.  In a `minimal' application
of generalized unitarity, used in a number of prior applications~\cite{Bern:1997nh,ABDK,
BDS,ABDK5,HigherLoopN=4base,RemainderFunction,Kosower:2010yk,Bern:2011rj}
and currently pursued by Feng and Huang~\cite{Feng:2012bm}, one
cuts just enough propagators to break apart a higher-loop amplitude
into a product of disconnected tree amplitudes.  Maximal cuts without complete localization
of integrands have also been used in recent multi-loop calculations
in maximally supersymmetric gauge and gravity 
theories~\cite{Bern:2007ct, Bern:2008pv,LeadingSingularity,Bern:2010tq,%
Carrasco:2011hw,Carrasco:2011mn,Bern:2012uc}.

\def\GenDisc{\mathop{\rm GenDisc}\nolimits}
We will work within a maximal unitarity approach, cutting all propagators
in a given integral, and further seeking to localize
integrands onto \emph{global poles}
to the extent possible.
In principle, this allows one to isolate individual integrals
on the right-hand side of the higher-loop analog of eq.~(\ref{MasterEquation}).
The coefficients are ultimately given in terms of linear
combinations of multivariate residues,
by so-called {\it generalized discontinuity operators\/} (GDOs).
In previous papers~\cite{TwoLoopMaximalUnitarityI,ExternalMasses,FourMass,
NonPlanarDoubleBox,MassiveNonPlanarDoubleBoxes,DoubledPropagators,ThreeLoopLadder},
the present authors and other collaborators have
shown how to use
multidimensional contours
around global poles to extract
the coefficients of both planar and nonplanar double-box master integrals,
and those of three-loop ladder integrals.  The same
approach, with the addition of integration over non-trivial cycles,
also allows the extraction of coefficients of a two-loop
double box with internal masses~\cite{MaximalElliptic}.

We devote several sections to background material.
In the next section, we review the notion of multivariate residues.  We emphasize
the differences from residues in a single complex variable, and provide both
a geometric and algebraic picture of the most important difference,
the contour dependence of such residues.  In \sect{TwoLoopIntegralSection}, we review the class 
of two-loop planar integrals whose residues we will study later on.
In \sect{DoubleBoxGlobalPoles}, we review the global poles of the double-box
integral.  We discuss the existence of global poles shared between different double-box
integrals in \sect{SharedGlobalPolesSection}, and distinguish between different ways
this can happen in \sect{CombiningContributionsSection}.  
We then analyze the four-point ABDK relation in \sect{FourPointABDKSection}, and
the five-point relation in \sect{FivePointABDKSection}.  We summarize in \sect{Conclusions}.

\section{Multivariate Residues}
\label{MultivariateResiduesSection}
The theory of multivariate complex residues is an important mathematical
tool in the higher-loop generalized unitarity program.   It does not always
generalize na\"\i{}vely from ordinary residues in a single complex variable.
For the benefit of those readers who may not be familiar with the multivariate
case, we give a bit of background and also discuss some of the subtleties
that arise.  The reader may find a more complete and mathematically rigorous presentation
in the classic book of Griffiths and Harris~\cite{GriffithsHarris}, as well as in
books of Tsikh~\cite{Tsikh} and Shabat~\cite{Shabat}.  
Cattani and Dickenstein~\cite{CattaniDickenstein} discuss the evaluation of
multivariate residues from a practical point of view, making use of powerful tools from modern
commutative algebra.  In \sect{AlgebraicEvaluation}, 
we show how to use one of the techniques they describe.

\subsection{General Aspects}
\label{GeneralAspects}

The Feynman rules for a quantum field theory tell us that the integrand at any loop order
is a rational function of the loop momenta.  Accordingly we can restrict attention
to rational functions, in this case rational functions of several
complex variables.  We consider separately a numerator polynomial $h$ and a multi-factor denominator
polynomial $f$, which we treat as a vector of polynomials.  In more mathematical language,
we take $f$ to be a holomorphic map from $\C^n\to\C^n$, and $h$ from $\C^n\to\C$.
We are interested in global poles $\xi$, where $f$ has an isolated
zero --- that is, $f_1(\xi) = \cdots = f_n(\xi) = 0$ and $f^{-1}(0)\cap U = \{\xi\}$ for
a sufficiently small neighborhood $U$ of $\xi$.
The object whose residue we want to compute at the global pole $z=\xi$ is
the meromorphic $n$-form,
\begin{align}
\omega = \frac{h(z)dz_1\wedge\cdots\wedge dz_n}{f_1(z)\cdots f_n(z)}\;.
\end{align}
The multivariate residue is defined by a multidimensional generalization of a contour integral: an integral
taken over a product of $n$ circles, that is an $n$-torus,
\begin{align}
\Res{}_{\{f_1,\dots,f_n\},\xi}(\omega) =
\frac{1}{(2\pi i)^n}
\oint_{\Gamma_\epsilon}
\frac{h(z)dz_1\wedge\cdots\wedge dz_n}{f_1(z)\cdots f_n(z)}\;,
\label{eq:residue_def}
\end{align}
where $\Gamma_\delta = \{z\in\C^n:|f_i(z)| = \delta_i\}$ and the $\delta_i$
have infinitesimal real values.   The definition
of $\Gamma_\delta$ is the first difference from single-variable contour integration,
as the integration cycle is defined not directly in terms of the variables $z$ but rather in terms
of the denominator factors $f_i(z)$.

\def\boint{\oint}

 The simplest case is the
factorizable one: if each component of $f$ depends only on a single
variable, that is $f_i(z) = f_i(z_i)$, the residue factorizes
completely into a product of one-dimensional contour integrals,
\begin{align}
\Res{}_{\{f_1,\dots,f_n\},\xi}(\omega) =
\frac1{(2\pi i)^n}\boint_{|f_1(z_1)| = \delta_1}\frac{dz_1}{f_1(z_1)}
\cdots
\boint_{|f_n(z_n)| = \delta_n}\frac{dz_n}{f_n(z_n)}h(z)\;.
\end{align}
In general, however, each $f_i$ will depend on several variables.
There are two types of multivariate residues we should consider:
nondegenerate and degenerate. In this case,
to compute the residue we must first evaluate the
Jacobian determinant,
\begin{align}
J(\xi)\equiv \det_{i,j}\left(\pd{f_i}{z_j}\right)\bigg|_{z = \xi}\;.
\end{align}
So long as this Jacobian does not vanish, the residue is said to be
nondegenerate.  For a nondegenerate residue, we can apply a coordinate
transformation to eq.~\eqref{eq:residue_def} in order to factorize the
denominator in a small neighborhood of the global pole.    We can do so, for example,
by making use of the transformation law presented and proved in Sec.~5.1 of ref.~\cite{GriffithsHarris}:

\begin{quotation}
Let $I = \langle f_1 (z), \ldots, f_n (z) \rangle$
be a zero-dimensional ideal%
\footnote{The ideal $I$ is said to be \emph{zero-dimensional}
if and only if the solution to the equation system $f_1 (z) = \cdots = f_n (z) = 0$
consists of a finite number of points $z \in \mathbb{CP}^n$.} generated
by a finite set of meromorphic functions $f_i (z) : \mathbb{CP}^n \to \mathbb{C}$
with $f_i (\xi) = 0$. Furthermore, let $J = \langle g_1 (z), \ldots, g_n (z) \rangle$
be a zero-dimensional ideal such that $J \subseteq I$; that is, whose generators
are related to those of $I$ by $g_i (z) = \sum_{i=1}^n a_{ij} (z) f_j(z)$
with the $a_{ij} (z)$ being polynomials. Letting $A(z) = (a_{ij} (z))_{i,j=1,\ldots,n}$
denote the conversion matrix, the residue at $\xi$ satisfies,
\end{quotation}
\begin{equation}
\mathop{\mathrm{Res}}_{\{f_1, \ldots, f_n\}, \hspace{0.3mm} \xi} \hspace{-0.2mm}\left(
\frac{h(z) \hspace{0.4mm} dz_1 \wedge \cdots \wedge dz_n}{f_1 (z) \cdots f_n (z)} \right)
\hspace{1mm}=\hspace{1mm} \mathop{\mathrm{Res}}_{\{g_1, \ldots, g_n\}, \hspace{0.3mm} \xi}
 \hspace{-0.2mm}\left(
\frac{h(z) \hspace{0.1mm} \det A(z) \hspace{0.4mm} dz_1 \wedge \cdots \wedge dz_n}
{g_1 (z) \cdots g_n (z)} \right) \,. \label{eq:transformation_law}
\end{equation}

After the transformation, we
obtain,
\begin{align}
\Res{}_{\{f_1,\dots,f_n\},\xi}(\omega) = \frac1{(2\pi i)^n}\frac{h(\xi)}{J(\xi)} \;.
\end{align}
 for the nondegenerate residue.
On the other hand, if the Jacobian vanishes, the residue is termed degenerate. 
In this case, the transformation law (\ref{eq:transformation_law}) remains valid~\cite{GriffithsHarris}
and may be used to compute the residue. To find a useful transformation
of the set of ideal generators, we follow the approach explained in Sec.~1.5.4 of
ref.~\cite{CattaniDickenstein} (see also applications by one of the present authors and 
Zhang~\cite{ThreeLoopLadder,DoubledPropagators,MassiveNonPlanarDoubleBoxes}).
 The idea is to choose the $g_i$ to be \emph{univariate};
that is, $g_i (z_1, \ldots, z_n) = g_i(z_i)$ so that the residue can be evaluated as a product
of univariate residues. A set of univariate polynomials $g_i$ can be obtained by generating
a Gr{\"o}bner basis~\cite{Buchberger}
 of $\{ f_1(z), \ldots, f_n(z) \}$ in lexicographic monomial order.
(The reader may consult the books in ref.~\cite{ComputationalAlgebraicGeometry}
for background material on multivariate polynomials and Gr\"obner bases.)
Specifying the variable ordering $z_{i+1} \succ z_{i+2} \succ \cdots \succ z_n \succ z_1
\succ z_2 \cdots \succ z_i$ will produce a Gr{\"o}bner basis containing a polynomial
which depends only on $z_i$. We define $g_i (z_i)$ as this polynomial.
By considering all $n$ cyclic permutations of the variable ordering
$z_1 \succ z_2 \succ \cdots \succ z_n$ we thus generate a set of $n$
univariate polynomials $\{ g_1(z_1), \ldots, g_n(z_n) \}$.

If the number of denominator factors of the form $\omega$
is greater than the number of variables $n$, we partition the denominator of $\omega$
into $n$ factors. For a given pole $\xi$, any partitioning $\{ f_1,\ldots,f_n\}$
which generates a zero-dimensional ideal produces an {\it a priori\/} distinct
residue. We will see an example of this in the next subsection.

Degenerate residues will play an important role in the present paper.  In the next
subsection, we consider a simple example of a degenerate residue, give a geometric picture,
and show how to evaluate it both geometrically and algebraically.

\subsection{Geometry of Degenerate Residues}
\label{DEGENERATERESIDUES}
Let us consider the following two-form\footnote{If one adds a boundary at infinity as needed to apply
global residue theorems, we can define it on $\CP^2$ rather than $\C^2$.},
\begin{align}
\label{DEGENERATEEXAMPLE}
\omega = \frac{z_1dz_1\wedge dz_2}{z_2(a_1z_1+a_2z_2)(b_1z_1+b_2z_2)}\;.
\end{align}
For generic values of the $a_i$ and $b_i$, there is a single global pole at finite values of $z_1$ and $z_2$:
requiring
any two of the denominator factors to vanish yields the solution
$z_1 = z_2 = 0$.  We immediately see that all three factors vanish at the global pole, and that the
two-dimensional residue at the global pole is degenerate according to the
definition given in the previous subsection.
In this subsection we focus on providing a more geometric
picture for this example.  As we shall see, the global pole admits two {\it distinct\/} integration
contours, which yield distinct residues.  This is very much unlike contour integration in one complex
variable, where a contour either encloses a pole or doesn't, and there is a unique nonzero value for
a residue.

We can split the two-form into two terms by making the following change of
variables in eq.~\eqref{DEGENERATEEXAMPLE},
\begin{align}
z_1' = a_1 z_1 + a_2 z_2, \quad
z_2' = z_2\;;
\end{align}
the form then becomes (dropping the primes on $z_i'$),
\begin{align}
\label{DEGENERATESPLIT}
\omega = \frac{1}{a_1}\left(
\frac{1}{z_2(c_1z_1+c_2z_2)}-\frac{a_2}{z_1(c_1z_1+c_2z_2)}\right)
dz_1\wedge dz_2\;,
\end{align}
where $c_1 \equiv b_1$ and $c_2 \equiv a_1 b_2 - a_2 b_1$. (This separation is
a partial fractioning followed by a change of variables.) Let us start by examining
the first term. The canonical integration contour is a product of two circles,
\begin{align}
|z_2| = \delta_2\;, \quad
|c_1z_1+c_2z_2| = \delta_c\;,
\end{align}
where $\delta_2,\delta_c > 0$. The residue of this term is,
\begin{equation}
\frac1{a_1 b_1 (2\pi i)^2}\,,
\end{equation} independent of the
precise values of the radii of the circles.  Going into a little bit more detail, we
can parametrize the integration cycle as
\begin{align}
\label{TORUSPARAM}
z_2 = \delta_2e^{i\theta_2}\;, \quad
c_1z_1+c_2z_2 = \delta_ce^{i\theta_c}\;,
\end{align}
so that, as $\theta_2,\theta_c$ run over the interval $[0,2\pi]$ the cycle is
traced out. (Indeed, the contour integrals become ordinary integrals over
$\theta_2,\theta_c$.)

What about the second term of eq.~\eqref{DEGENERATESPLIT}? Care must be taken
to ensure that the integrand is not singular on the contour; that would be an
\emph{illegitimate} contour. The second denominator factor is of course
nonvanishing on the cycle~(\ref{TORUSPARAM}). We can use the same pair of equations
to write
\begin{align}
z_1 = (\delta_c e^{i\theta_c} - c_2 \delta_2 e^{i\theta_2})/c_1\;.
\end{align}
It follows that $z_1$ (the first denominator factor) will not vanish so long
as $\delta_c\neq |c_2|\delta_2$. On the other hand, if
$\delta_c = |c_2|\delta_2$, $z_1$ is guaranteed to vanish for some values of
the angles. The illegitimate choice $\delta_c = |c_2|\delta_2$ divides the
moduli space $(\delta_c,\delta_2)$ into two regions,
\begin{align}
\mathrm{(1)} \hspace{2.5mm} \delta_c > |c_2|\delta_2 \hspace{7mm} \mathrm{and}
\hspace{7mm} \mathrm{(2)} \hspace{2.5mm} \delta_c < |c_2|\delta_2\;,
\label{eq:regions_of_torus_moduli_space}
\end{align}
which we consider in turn.

At a first glance, the global contour \eqref{TORUSPARAM} winds around $z_1 = 0$
in both regions. However, in the first region, it is the $\theta_c$-parametrized
circle which winds around this point; the $\theta_2$-parametrized circle does not enclose
$z_1 = 0$. But $\theta_c$ is the same variable which winds around the zero of
the second denominator factor; that is, it is not linearly independent. This
means that the torus fails to have the global pole inside it; the situation is
more like a tube with the global pole sitting at the center of the symmetry
plane of the tube, but not inside the tube.  We conclude that in region (1), the
second term in eq.~\eqref{DEGENERATESPLIT} integrated over the cycle
\eqref{TORUSPARAM} produces a vanishing residue.  In contrast, in region (2), the
$\theta_2$-parametrized circle does wind around $z_1 = 0$, so that the contour
\eqref{TORUSPARAM} will enclose the global pole of the second term in
eq.~\eqref{DEGENERATESPLIT} as well as the first. Thus, in this region, both
terms produce a nonvanishing residue. In particular, we observe that the residue
of eq.~\eqref{DEGENERATESPLIT} differs in the two regions
\eqref{eq:regions_of_torus_moduli_space}, and thus depends on the relative radii
$\delta_2,\delta_c$ of the integration cycle.

More generally, let us consider a generic torus,
\begin{align}
z_1 = \delta_{1,1}e^{i\theta_1}+\delta_{1,2}e^{i\theta_2}\;, \quad
z_2 = \delta_{2,1}e^{i\theta_1}+\delta_{2,2}e^{i\theta_2}\;,
\label{eq:generic_torus}
\end{align}
where $\delta_{i,j}$ are real positive constants which fix the shape of the
contour. For the $2$-form at hand, we could rescale all $\delta$s uniformly
without loss of generality, so we really have only three independent real
parameters.

The contour is legitimate for the first term in eq.~\eqref{DEGENERATESPLIT} if
and only if $\delta_{2,1}\neq\delta_{2,2}$ and $r_1\neq r_2$, where
\begin{align}
r_1 = |c_1\delta_{1,1}+c_2\delta_{2,1}| \hspace{6mm} \mathrm{and} \hspace{6mm}
r_2 = |c_1\delta_{1,2}+c_2\delta_{2,2}|\;.
\end{align}
The contour is legitimate for the second term if and only if
$\delta_{1,1}\neq\delta_{1,2}$ and $r_1 \neq r_2$. This gives us eight
regions to consider, corresponding to choosing the upper or lower
inequality in each of the three relations,
\begin{align}
\delta_{2,1}\gtrless\delta_{2,2}\;, \quad
\delta_{1,1}\gtrless\delta_{1,2}\;, \quad
r_1\gtrless r_2\;.
\end{align}
Let us denote the upper choice by `$+$', and the lower choice by `$-$'; each
region is then labeled by a string of signs. We can see that in $R^{+++}$,
corresponding to $\delta_{2,1}>\delta_{2,2}$, $\delta_{1,1}>\delta_{1,2}$ and
$r_1>r_2$, $\theta_1$ is the wrapping variable for $z_1$ and $z_2$ --- but also
for $c_1z_1+c_2z_2$, so that the torus fails to enclose the pole in either term
in eq.~\eqref{DEGENERATESPLIT}. In $R^{++-}$, the torus will enclose both
terms, and the residue will be the sum of the two terms' residues. In
$R^{+-+}$, the torus encloses only the second term, and in $R^{+--}$, the torus
encloses only the first term. The remaining four regions are related to these
four by flipping all inequalities, which leaves the results invariant (up to a
sign).

The above analysis shows that a degenerate residue is not fully characterized
by the location of the pole. The value of the residue depends on the shape of
the torus wrapping around the global pole. Therefore, to correctly specify a
residue, we should rather think of the integration cycles. In the present
example we deduced that the \emph{moduli space} of tori is divided into several
regions. These regions correspond to distinct homology classes of the $(z_1,z_2)$ space
with the zeros of the individual denominator factors in \eqref{DEGENERATEEXAMPLE} removed. 
(In the mathematics literature, the hypersurfaces where these factors vanish are called \emph{divisors}.) 
That is, tori \eqref{eq:generic_torus} with moduli $\delta_{i,j}$ taken from distinct regions
$R^{+++}, R^{++-}$, etc. are \emph{non-homologous}.

\subsection{Algebraic Evaluation of Degenerate Residues}
\label{AlgebraicEvaluation}

Let us now turn to the evaluation of the residues of $\omega$ at the pole
at $\xi = (0,0)$ by use of the approach explained at the end of
\sect{GeneralAspects}. This calculation serves the dual purpose of providing
a concrete example of the evaluation algorithm, and 
of displaying a one-to-one map between the distinct denominator partitionings
 and the distinct regions
$R^{+++}, R^{++-}$, etc. of the torus moduli space discussed 
at the end of the previous subsection.
This map provides a dictionary between the algebraic and geometric
pictures of distinct residues for a form at a given global pole.

Let us denote the denominator factors of eq.~(\ref{DEGENERATEEXAMPLE}) as follows,
\begin{align}
f_1(z_1,z_2)  &=  z_2 \,, \\
f_2(z_1,z_2)  &=  a_1 z_1 + a_2 z_2 \,, \\
f_3(z_1,z_2)  &=  b_1 z_1 + b_2 z_2 \,.
\end{align}
As we are performing a two-dimensional contour integral, we seek to partition
the denominator~(\ref{DEGENERATEEXAMPLE}) into two factors.  This can be
done in three distinct ways, namely $\{ f_1, f_2 f_3 \}$, $\{ f_2, f_3 f_1 \}$ and
$\{ f_3, f_1 f_2 \}$.  Let us evaluate the residue
for the denominator partitioning $\{ f_1, f_2 f_3 \}$, using the method explained
at the end of \sect{GeneralAspects}.
 The lexicographically ordered
Gr{\"o}bner basis of $\{ f_1, f_2 f_3 \}$ in the variable ordering
$z_2 \succ z_1$ is $\{ a_1 b_1 z_1^2, z_2 \}$; in the variable
ordering $z_1 \succ z_2$ it is $\{ z_2, a_1 b_1 z_1^2 \}$. Choosing the first
element of each Gr{\"o}bner basis we have,
\begin{align}
g_1(z_1,z_2)  &=  a_1 b_1 z_1^2 \, , \\
g_2(z_1,z_2)  &=  z_2 \, .
\end{align}
We can obtain the conversion matrix as a by-product of finding the Gr\"obner
basis (or using the approach implemented in ref.~\cite{Lichtblau}).  In the simple case
considered here, ordinary multivariate polynomial division yields the same
result,
\begin{equation}
A \hspace{0.7mm}=\hspace{0.7mm} \begin{pmatrix} -(a_1 b_2 + a_2 b_1) z_1 - a_2 b_2 z_2 & \hspace{1.5mm} 1 \\
1 & \hspace{1.5mm} 0 \end{pmatrix}\,,
\end{equation}
that relates the two sets of ideal generators,
\begin{equation}
A \cdot \hspace{0.4mm} \begin{pmatrix} f_1(z_1, z_2) \\ f_2(z_1, z_2) f_3(z_1, z_2) \end{pmatrix}
= \begin{pmatrix} g_1(z_1, z_2) \\ g_2(z_1, z_2) \end{pmatrix} \,.
\end{equation}
From the transformation law (\ref{eq:transformation_law}) we then find that the residue of $\omega$
at $\xi = (0,0)$ with respect to the ideal generators $\{ f_1, f_2 f_3 \}$ is
\begin{equation}
\mathop{\mathrm{Res}}_{\{ f_1, f_2 f_3 \}, \hspace{0.5mm} \xi} \omega
\hspace{1mm}=\hspace{1mm} \mathop{\mathrm{Res}}_{\xi} \frac{z_1 \hspace{0.2mm}
\det A \hspace{0.7mm} dz_1 \wedge dz_2}{g_1(z_1, z_2) \hspace{0.5mm} g_2 (z_1, z_2)}
\hspace{1mm}=\hspace{1mm} - \mathop{\mathrm{Res}}_{\xi} \frac{dz_1 \wedge dz_2}
{a_1 b_1 z_1 z_2} \,. \label{eq:residue_omega_{f1_f2f3}}
\end{equation}
In practice, it is important to keep in mind that the residue is antisymmetric
under interchanges of the denominator factors of the form $\omega$. We observe
that the denominator on the right-hand side of eq.~(\ref{eq:residue_omega_{f1_f2f3}})
is a product of univariate polynomials, as desired. The residue can therefore
be computed as a product of univariate residues and yields,
\begin{align}
\rho_1 \equiv \mathop{\mathrm{Res}}_{\{ f_1, f_2 f_3 \}, \hspace{0.5mm} \xi} \omega
\hspace{1mm}&=\hspace{1mm} -\frac{1}{a_1 b_1 (2\pi i)^2} \,, \label{eq:rho_1_def} \\
\rho_2 \equiv \mathop{\mathrm{Res}}_{\{ f_2, f_3 f_1 \}, \hspace{0.5mm} \xi} \omega
\hspace{1mm}&=\hspace{1mm} -\frac{a_2}{a_1 (a_1 b_2 - a_2 b_1) (2\pi i)^2} \,, \label{eq:rho_2_def} \\
\rho_3 \equiv \mathop{\mathrm{Res}}_{\{ f_3, f_1 f_2 \}, \hspace{0.5mm} \xi} \omega
\hspace{1mm}&=\hspace{1mm} \frac{b_2}{b_1 (a_1 b_2 - a_2 b_1) (2\pi i)^2} \,, \label{eq:rho_3_def}
\end{align}                                              
where the residues for the two other denominator partitionings $\{ f_2, f_3 f_1\}$
and $\{ f_3, f_1 f_2\}$ are computed in a similar fashion.

Likewise, we can apply the residue evaluation algorithm to each of the two terms
in eq.~(\ref{DEGENERATESPLIT}) separately, yielding  $\rho_1$ and $\rho_2$ for the first and second terms
respectively. Combining this with the observations
made in the discussion following eq.~(\ref{eq:regions_of_torus_moduli_space}), we see
that in the region $R^{+++}$ of the torus moduli space, the residue evaluates to
$0$; in $R^{+--}$ to $\rho_1$; in $R^{+-+}$ to $\rho_2$; and in $R^{++-}$ to
$\rho_1 + \rho_2 = -\rho_3$. These observations allow us to conclude that we have
the following one-to-one map between the partitionings of the denominator of
$\omega$ and the regions of the torus moduli space,
\begin{align}
\{ f_1, f_2 f_3 \}  \hspace{2mm}&\longleftrightarrow\hspace{2mm}  R^{+--} \\
\{ f_2, f_3 f_1 \}  \hspace{2mm}&\longleftrightarrow\hspace{2mm}  R^{+-+} \\
\{ f_3, f_1 f_2 \}  \hspace{2mm}&\longleftrightarrow\hspace{2mm}  R^{++-} \,.
\end{align}
This map provides a dictionary between the algebraic and geometric
pictures of the distinct residues defined at the given global pole.

Only two out of the three residues $\rho_1, \rho_2, \rho_3$
in eqs.~(\ref{eq:rho_1_def})--(\ref{eq:rho_3_def}) are independent,
as the residues satisfy the identity,
\begin{equation}
\rho_1 + \rho_2 + \rho_3 = 0 \,.
\end{equation}
In the geometric picture, only two of the regions $R^{+--}, R^{+-+},\ldots$ define
linearly independent integration cycles.

\section{Two-Loop Integrals}
\label{TwoLoopIntegralSection}

In this section, we introduce the principal actors in our study,
planar two-loop integrals.  Let us first define our notation for
one-loop integrals,
\begin{equation}
\OneLoop_n(K_1,\ldots,K_n) \equiv -i\int \frac{d^D\ell}{(2\pi)^D}\;
   \frac1{\ell^2 (\ell-K_1)^2 (\ell-K_{12})^2\cdots (\ell-K_{1\cdots(n-1)})^2}\,.
\end{equation}
We use the notation $K_{j\cdots l} =
K_{j}+\cdots+K_l$.

We will make use of the massless box integral,
$I_\square = I_4$, and the massless pentagon, $I_{\pentagon} = I_5$.

\begin{figure}[ht]
\begin{minipage}[b]{0.4\linewidth}
\centering
\includegraphics[scale=0.5]{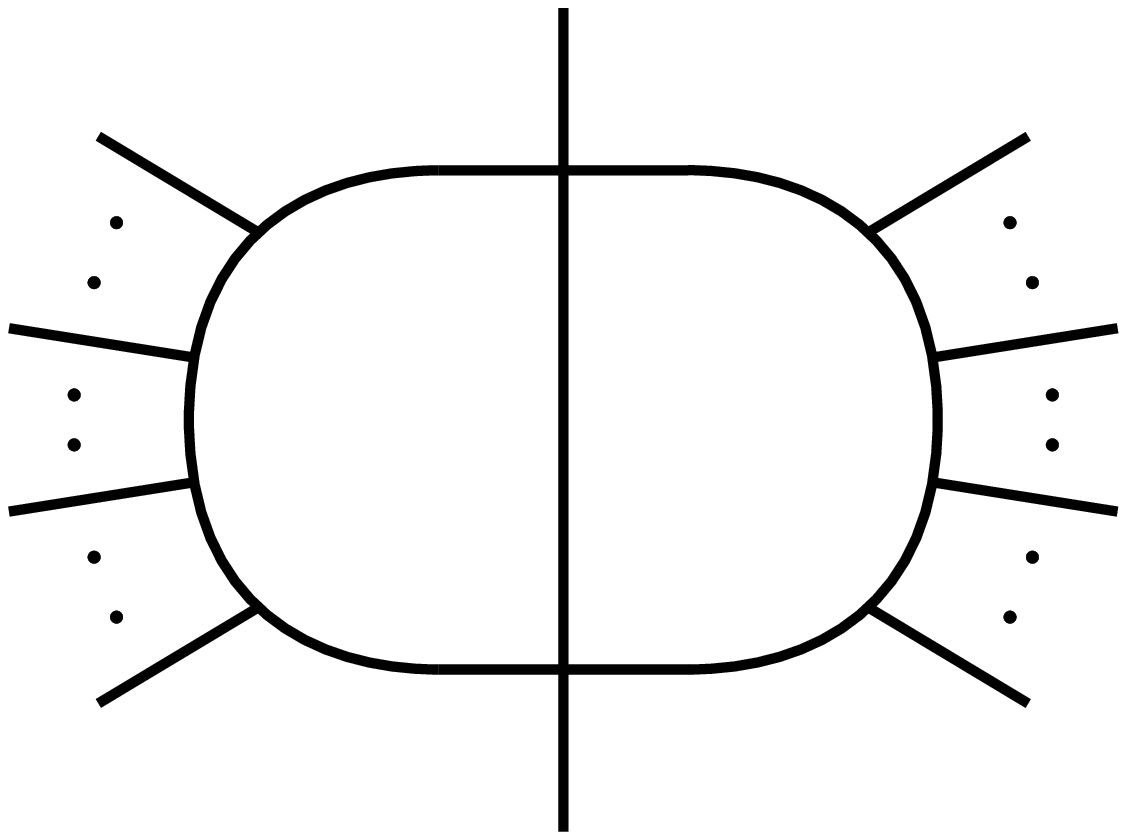}
\center{(a)}
\end{minipage}
\hspace{0.5cm}
\begin{minipage}[b]{0.4\linewidth}
\centering
\includegraphics[scale=0.5]{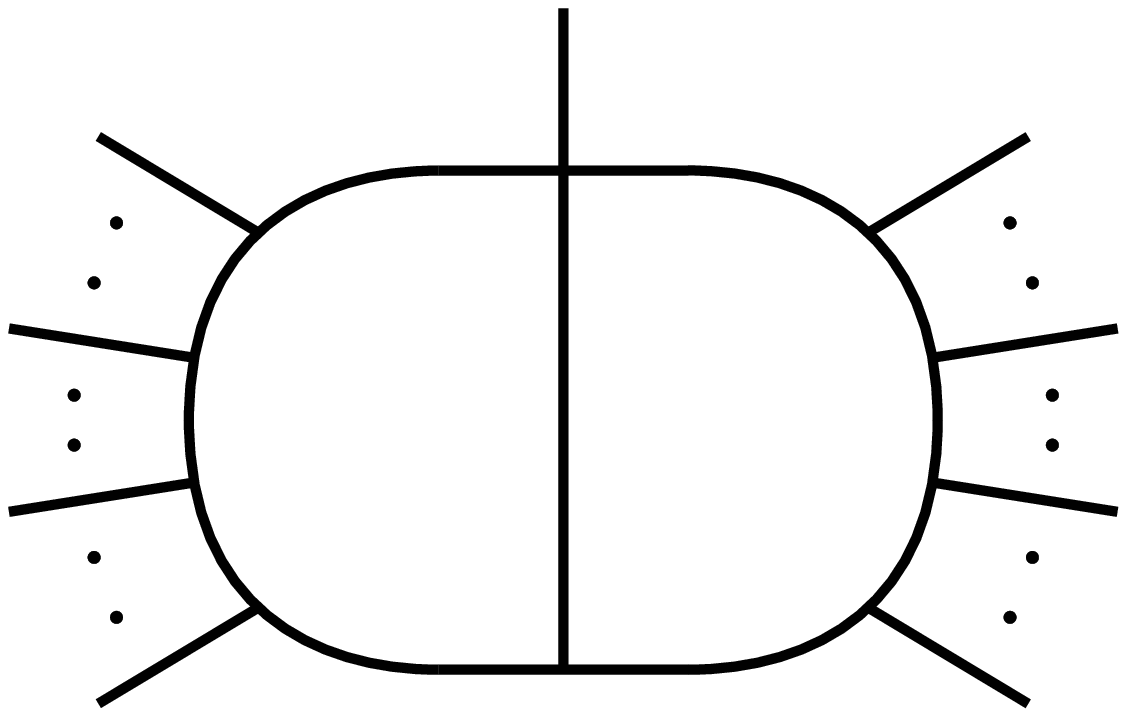}
\vspace{7mm}
\center{(b)}
\end{minipage}
\begin{minipage}[b]{0.4\linewidth}
\centering
\includegraphics[scale=0.5]{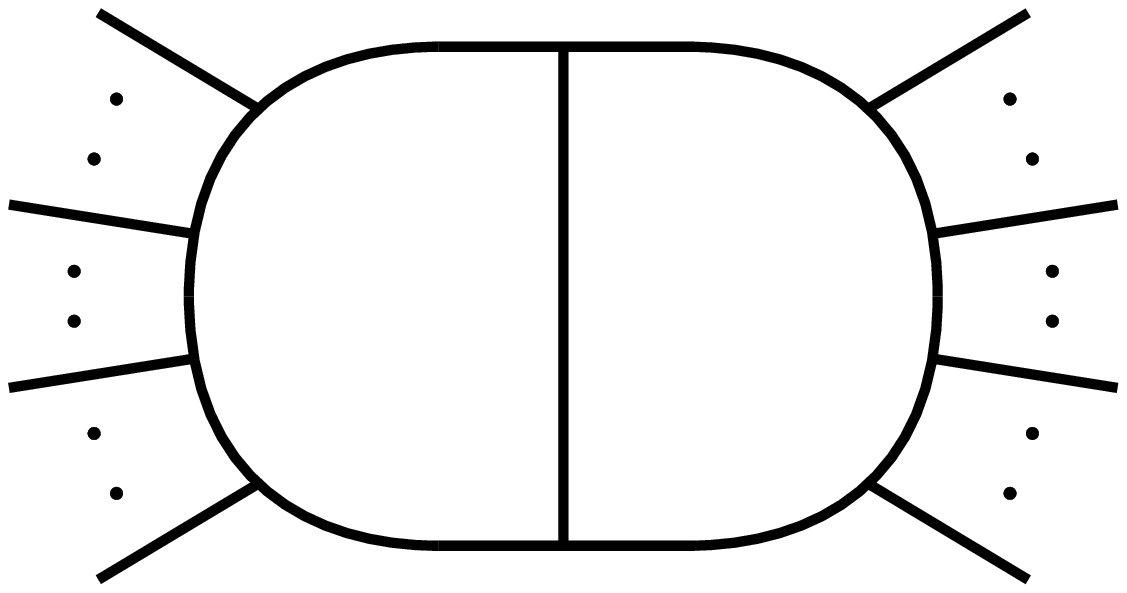}
\vspace{7mm}
\center{(c)}
\end{minipage}
 \caption{The three basic types of two-loop planar integrals,
labeled by the number
of legs attached to each internal line of the vacuum diagram: (a) $P_{n_1,n_2}$, (b)
$P^*_{n_1,n_2}$, (c) $\Pss_{n_1,n_2}$.}
\label{TwoLoopPlanarIntegralsFigure}
\end{figure}

Two-loop integrals can be organized into two broad classes: those that factor into a product of
one-loop integrals when cutting certain internal lines; and those that are irreducibly two-loop, which
remain connected upon cutting any internal line.
We can organize irreducibly two-loop integrals, constructed by attaching
external legs to the non-factorizable two-loop vacuum diagram, into three
classes~\cite{TwoLoopBasis}.  These have
external legs attached to one or two of the internal lines, and
possibly to its vertices.  Attaching
external legs to the
third internal line as well (the middle line) would yield non-planar
integrals, which we will not consider in the present article.
  We label the integrals according to the number of            
external legs attached to each of the vacuum diagram's internal lines.
The absence of lines attached to vertices is denoted by a superscripted
star.  The three
types of integrals are,
\begin{align}
P_{n_1,n_2}[{N}(\ell_1,\ell_2)]
 &= (-i)^2\int \frac{d^D\ell_1}{(2\pi)^D} \frac{d^D\ell_2}{(2\pi)^D}
\;\frac{{N}(\ell_1,\ell_2)}
{\ell_1^2 (\ell_1-K_1)^2\cdots (\ell_1-K_{1\cdots n_1})^2(\ell_1+\ell_2+K_{n_1+n_2+2})^2}
\nn\\
&\hphantom{ \int \frac{d^D\ell_1}{(2\pi)^D} \frac{d^D\ell_2}{(2\pi)^D} }
  \times \frac{1}
{\ell_2^2 (\ell_2-K_{n_1+n_2+1})^2\cdots (\ell_2-K_{(n_1+2)\cdots(n_1+n_2+1)})^2}
\,,\nn\\
P^*_{n_1,n_2}[{N}(\ell_1,\ell_2)]
 &= (-i)^2\int \frac{d^D\ell_1}{(2\pi)^D} \frac{d^D\ell_2}{(2\pi)^D}
\;\frac{{N}(\ell_1,\ell_2)}
{\ell_1^2 (\ell_1-K_1)^2\cdots (\ell_1-K_{1\cdots n_1})^2(\ell_1+\ell_2)^2}
\nn\\&\hphantom{ \int \frac{d^D\ell_1}{(2\pi)^D} \frac{d^D\ell_2}{(2\pi)^D} }
  \times \frac1{\ell_2^2 (\ell_2-K_{n_1+n_2+1})^2\cdots (\ell_2-K_{(n_1+2)\cdots(n_1+n_2+1)})^2}
\,,\label{TwoLoopPlanarIntegrals}\\
\Pss_{n_1,n_2}[{N}(\ell_1,\ell_2)]
 &= (-i)^2\int \frac{d^D\ell_1}{(2\pi)^D} \frac{d^D\ell_2}{(2\pi)^D}
\;\frac{N(\ell_1,\ell_2)}{\ell_1^2 (\ell_1-K_1)^2\cdots (\ell_1-K_{1\cdots n_1})^2(\ell_1+\ell_2)^2}
\nn\\&\hphantom{ \int \frac{d^D\ell_1}{(2\pi)^D} \frac{d^D\ell_2}{(2\pi)^D} }
  \times \frac1{\ell_2^2 (\ell_2-K_{n_1+n_2})^2\cdots (\ell_2-K_{(n_1+1)\cdots(n_1+n_2)})^2}
\,.\nn\end{align}
The numerator polynomial $N(\ell_1,\ell_2)$ is a function of
the loop momenta as well as of external momenta.
For the reader's convenience, these integrals are shown in
\fig{TwoLoopPlanarIntegralsFigure}.

We will examine the scalar `horizontal' ($s$-channel)
and `vertical' ($t$-channel) double-box integrals,
\begin{align}
\HDB\equiv \Pss_{2,2}(k_1,k_2,k_3,k_4)\,,\\
\VDB\equiv \Pss_{2,2}(k_4,k_1,k_2,k_3)\,.
\end{align}
The labeling of the loop momenta in later sections will {\it not\/}
 always follow \eqn{TwoLoopPlanarIntegrals}, but will be indicated in figures
throughout the text.

We will also consider the dual-conformal pentabox integral,
$\Pss_{3,2}[(\ell_1+k_5)^2]$; scalar and irreducible-numerator
one-mass double-box integrals, $\Pss_{2,2}[1](K_{12},k_3,k_4,k_5)$
and $\Pss_{2,2}[(\ell_1+k_5)^2](K_{12},k_3,k_4,k_5)$;
and scalar and irreducible-numerator turtle-box integrals,
$\Ps_{2,2}[1]$ and $\Ps_{2,2}[(\ell_1+k_4)^2]$.

\section{Global Poles of the Double-Box Integral}
\label{DoubleBoxGlobalPoles}

In this section, we review the global poles of the massless
double-box integral.
In order to find the global poles, we first impose the maximal cut, cutting
all seven propagators.  We then examine the resulting integrand to further
localize the one remaining degree of freedom.

Formally, we impose the maximal cut by performing a contour integral
around a sum of seven-tori encircling the solution surfaces.  In practice,
we do this simply by solving the on-shell equations for the seven
different propagator momenta.  It is easiest to do this by using the same
linear parametrization as in ref.~\cite{TwoLoopMaximalUnitarityI},
\begin{align}
\ell_1^\mu &= \alpha_1 k_1^\mu + \alpha_2 k_2^\mu + \frac{s_{12}
\alpha_3}{2\nsand1.4.2} \nsand1.{\sigma^\mu}.2
+ \frac{s_{12} \alpha_4}{2\nsand2.4.1}\nsand2.{\sigma^\mu}.1
\,,\nn\\
\ell_2^\mu &= \beta_1 k_3^\mu + \beta_2 k_4^\mu + \frac{s_{12}
\beta_3}{2\nsand3.1.4} \nsand3.{\sigma^\mu}.4
 + \frac{s_{12} \beta_4}{2\nsand4.1.3}
\nsand4.{\sigma^\mu}.3\,.
\label{TwoLoopParametrization}
\end{align}
In the original loop integral, taken along the real slice of
complexified loop momenta, $\alpha_{1,2}$ and $\beta_{1,2}$
are real, while $\alpha_{3,4}$ and $\beta_{3,4}$ lie along rays in
the complex plane.  We will be considering general contour
integrals in $\CP^4$, for which all $\alpha_i,\beta_i\in\C$.

\def\Sol{{\cal S}}
\def\Global{{\cal G}}
Imposing the seven on-shell conditions leads to six distinct solutions~\cite{TwoLoopMaximalUnitarityI}.
In all of them,
\begin{equation}
\alpha_1 = 1\,,\qquad
\alpha_2 = 0\,,\qquad
\beta_1 = 0\,,\qquad
\beta_2 = 1\,,
\end{equation}
while the other parameters take on different values,
\begin{equation}
\begin{alignedat}{8}
\Sol_1:& \qquad \alpha_3 = -\chi\,,\qquad
&\alpha_4 &= 0\,,\qquad
&\beta_3 &= z\,,\qquad
&\beta_4 &= 0\,;\\
\Sol_2:& \qquad \alpha_3 = z\,,\qquad
&\alpha_4 &= 0\,,\qquad
&\beta_3 &= -\chi\,,\qquad
&\beta_4 &= 0\,;\\
\Sol_3:& \qquad \alpha_3 = 0\,,\qquad
&\alpha_4 &= -\chi\,,\qquad
&\beta_3 &= 0\,,\qquad
&\beta_4 &= z\,;\\
\Sol_4:& \qquad \alpha_3 = 0\,,\qquad
&\alpha_4 &= z\,,\qquad
&\beta_3 &= 0\,,\qquad
&\beta_4 &= -\chi\,;\\
\Sol_5:& \qquad \alpha_3 = 0\,,\qquad
&\alpha_4 &= z\,,\qquad
&\beta_3 &= -\frac{(\chi+1)(z+\chi)}{z+\chi+1}\,,\qquad
&\beta_4 &= 0\,;\\
\Sol_6:& \qquad \alpha_3 = z\,,\qquad
&\alpha_4 &= 0\,,\qquad
&\beta_3 &= 0\,,\qquad
&\beta_4 &= -\frac{(\chi+1)(z+\chi)}{z+\chi+1}\,.
\end{alignedat}
\end{equation}
We have defined $\chi = s_{14}/s_{12}$, and have labeled the remaining
degree of freedom uniformly by $z$.

Performing the contour integral
over the seven-torus leads to the appearance of an inverse Jacobian
in the integrand for $z$,
\begin{equation}
J^{-1}(z) = -\frac1{16 s_{12}^3\,z (z+\chi)}\,.
\end{equation}
This integrand has two poles, at $z=0$ and $z= -\chi$.  In addition, integrals
containing powers of the loop momenta will also have poles in solutions
$\Sol_{5,6}$ at $z=-\chi-1$.  Such integrals will also have poles at
$z=\infty$.
We can fully localize the integrand by integrating $z$ along
a contour surrounding one of these poles (or a linear combination thereof).  These poles
are global poles of the original double-box integrand; we could have
equivalently
performed a multivariate contour integral of the original
integrand around an appropriately-chosen
eight-torus.

At first glance,
the six different solutions can be thought of as six independent complex
planes; or, adding the point at infinity to each, as six
independent copies of $\CP^1\simeq S^2$.
A simple count suggests that we have twenty global poles: three each for
solutions $\Sol_{1,\ldots,4}$, and four each for solutions $\Sol_{5,6}$.
This count is too hasty, because the six independent solutions do meet
at global poles~\cite{TwoloopContourUniqueness}: the point $z= -\chi$ in solution $\Sol_1$ is the same
point in the original loop-momentum variables as $z=-\chi$ in solution
$\Sol_2$.  Furthermore, we can make use of an independent
 Cauchy residue theorem for each of the solution spheres to rewrite contour
integrals around $z=\infty$ in terms of a sum around the other poles.
Removing these poles, and accounting for shared poles leaves us with
eight independent global poles.  Finding the appropriate contour for
isolating the coefficients of the two master integrals $\Pss_{2,2}[1]$
and $\Pss_{2,2}[\ell_1\cdot k_4]$ was the subject of
ref.~\cite{TwoLoopMaximalUnitarityI}.

In terms of the loop momenta, the two `exceptional' poles at $z=-\chi-1$
in solutions $\Sol_{5,6}$ correspond to $\ell_2$
diverging.
(The asymmetry between $\ell_1$ and $\ell_2$ is due to our choice of
eliminating the poles at $z=\infty$, where $\ell_1$ diverges.)
The scalar double-box integral (with no irreducible numerators inserted)
does not have these poles, and so they do not contribute to the
$\mc N=4$ amplitude.  We will not need to consider them further in
this paper.

The remaining six global poles are each shared between two solutions;
we can choose to parametrize them as
$z=-\chi$ in $\Sol_2$, which we denote $\Global_1$;
$z=-\chi$ in $\Sol_4$, denoted $\Global_2$;
$z=0$ in $\Sol_1$, denoted $\Global_3$;
$z=0$ in $\Sol_3$, denoted $\Global_4$;
$z=0$ in $\Sol_5$, denoted $\Global_5$;
and $z=0$ in $\Sol_6$, denoted $\Global_6$.

The first two of these poles will be of particular interest
to us.  In the first ($\Global_1$),
\begin{equation}
\begin{aligned}
\ell_1^\mu &= k_1^\mu - \frac{s_{14}}{2 \nsand1.4.2}\nsand1.{\sigma^\mu}.2
= -\frac{\spb1.2}{2\spb2.4}\nsand1.{\sigma^\mu}.4\,,\\
\ell_2^\mu &= k_4^\mu - \frac{s_{14}}{2 \nsand3.1.4}\nsand3.{\sigma^\mu}.4
= -\frac{\spa3.4}{2\spa1.3}\nsand1.{\sigma^\mu}.4\,.
\end{aligned}
\label{G1value}
\end{equation}
We thus find,
\begin{equation}
\ell_1^\mu+\ell_2^\mu =
-\frac{\spa1.3\spb1.2+\spa3.4\spb2.4}{2\spa1.3\spb2.4}
\nsand1.{\sigma^\mu}.4 = 0\,.
\end{equation}
The pole corresponds to the middle rung of the double box becoming soft~\cite{TwoloopContourUniqueness}.

The situation is similar in the second pole in the above list ($\Global_2$),
which is just the spinor (or parity) conjugate of the first,
\begin{equation}
\begin{aligned}
\ell_1^\mu &= -\frac{\spa1.2}{2\spa2.4}\nsand4.{\sigma^\mu}.1\,,\\
\ell_2^\mu &= -\frac{\spb3.4}{2\spb1.3}\nsand4.{\sigma^\mu}.1\,;
\end{aligned}
\end{equation}
again $\ell_1^\mu+\ell_2^\mu = 0$.

In the third pole in the above list ($\Global_3$),
\begin{equation}
\begin{aligned}
\ell_1^\mu &= -\frac{\spb1.2}{2\spb2.4}\nsand1.{\sigma^\mu}.4\,,\\
\ell_2^\mu &= k_4^\mu\,.
\end{aligned}
\label{G3value}
\end{equation}
In this case, $(\ell_2-k_4)^\mu = 0$, so it is the rung between legs 3 and~4
that becomes soft.  This is also the case for the fourth pole ($\Global_4$),
which is the parity conjugate of this one.

In the fifth pole in the list ($\Global_5$),
\begin{equation}
\begin{aligned}
\ell_1^\mu &= k_1^\mu\,,\\
\ell_2^\mu &= -\frac{\spa3.4}{2\spa1.3}\nsand1.{\sigma^\mu}.4\,.
\end{aligned}
\label{G5value}
\end{equation}
Here, $(\ell_1-k_1)^\mu = 0$, thus the rung between legs 1 and~2 becomes
soft.  This is also true for the sixth and last pole in the list
($\Global_6$), which is the parity conjugate of the fifth.

While we will not analyze the outer-rung poles $\Global_{3,\ldots,6}$ in detail,
they also play a role in an analysis of the ABDK relation.

\section{Shared Global Poles}
\label{SharedGlobalPolesSection}

In this section we investigate a curious phenomenon: global poles of two-loop
integrals can be shared between two or more integrals.
In some cases this turns out to have interesting and nontrivial consequences.

\begin{figure}[!h]
\bc
\includegraphics[scale=0.65]{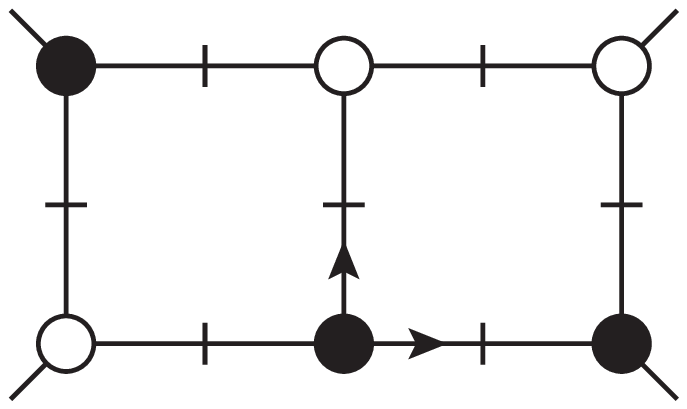}
\put(-137,20){$k_1$}
\put(-138,105){$k_2$}
\put(-5,104){$k_3$}
\put(-5,19){$k_4$}
\put(-43,24){$\ell_1$}
\put(-83,62){$\ell_2$}
\hspace*{.1cm}
\includegraphics[scale=0.65]{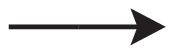}
\put(-36,50){$\ell_2\to0$}
\hspace*{.1cm}
\includegraphics[scale=0.65]{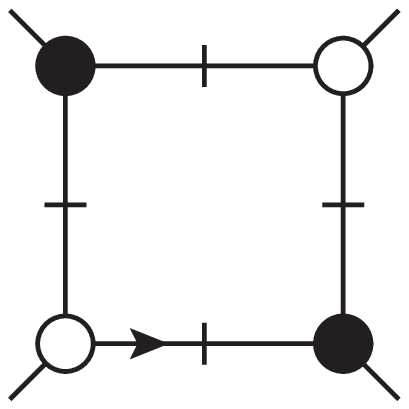}
\put(-85,20){$k_1$}
\put(-86,105){$k_2$}
\put(-5,104){$k_3$}
\put(-5,19){$k_4$}
\put(-45,24){$\ell_1$}
\hspace*{.1cm}
\includegraphics[scale=0.65]{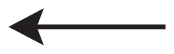}
\put(-36,50){$\ell_2\to0$}
\hspace*{.1cm}
\includegraphics[scale=0.65]{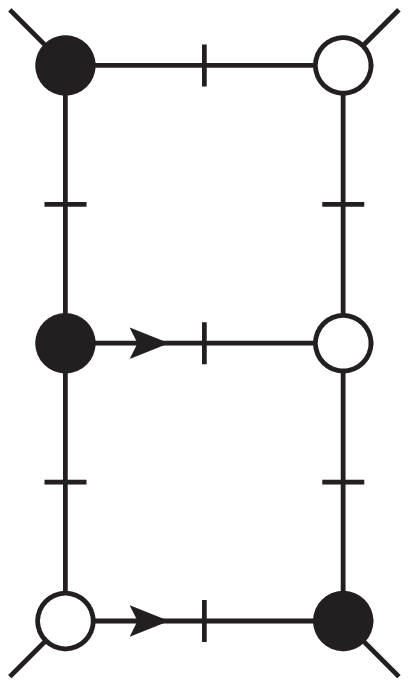}
\put(-85,-5){$k_1$}
\put(-85,133){$k_2$}
\put(-5,133){$k_3$}
\put(-5,-5){$k_4$}
\put(-45,74){$\ell_2$}
\put(-45,0){$\ell_1$}
\caption{
\label{DBOXSOFTOVERLAP1}
An example of a global pole shared between the horizontal and vertical
double-box integrals. The loop momentum labeling is chosen cunningly. 
At the global pole, the nonvanishing heptacut propagator momentum in each double box
corresponds to that of a quadruply cut one-loop box.
The white and black blobs indicate chiral (MHV) and antichiral ($\overline{\text{MHV}}$) vertices, respectively, as in the conventions (for example) of ref.~\cite{ExternalMasses}.}
\ec
\end{figure}

\subsection{Horizontal and Vertical Double Boxes}
\label{DoubleBoxSharing}

We start by re-examining the equations for the global poles in the
horizontal double box,
\begin{align}
\mc G_{\text{HDB}}:\;\, {} &
\ell_1^2 = \ell_2^2 = (\ell_1+\ell_2+K_{12})^2 = (\ell_1-k_4)^2 =
(\ell_1+\ell_2+k_1)^2 = 0\;, \nn \\ &
(\ell_1-K_{34})^2 = (\ell_1+\ell_2)^2 = 0\;;
\label{HDBequations}
\end{align}
here the labeling is {\it not\/} the one used earlier, but rather
the one shown in \fig{DBOXSOFTOVERLAP1}.
The first five equations are identical to those for the vertical
double box (again with the momentum labeling as given in
\fig{DBOXSOFTOVERLAP1}),
\begin{align}
\mc G_{\text{VDB}}:\;\, {} &
\ell_1^2 = \ell_2^2 = (\ell_1+\ell_2+K_{12})^2 = (\ell_1-k_4)^2 =
(\ell_1+\ell_2+k_1)^2 = 0\;, \nn \\ &
(\ell_1+\ell_2-k_4)^2 = (\ell_1+k_1)^2 = 0\;.
\label{VDBequations}
\end{align}
The remaining two equations in each case, on the second lines of
\eqns{HDBequations}{VDBequations}, appear at first glance to be
different.  However, if we focus on the first two global poles
($\Global_{1,2}$) discussed in the previous section, we find
a remarkable overlap.  When the momentum of the middle rung, which
in the labeling here is given simply by $\ell_2$, becomes soft, all the
second-line equations reduce to first-line equations.

Indeed, the full set of equations simplifies to a set of four equations
for $\ell_1$,
\begin{align}
\mc G_{\text{HDB}}\Longleftrightarrow\mc G_{\text{VDB}}:\;
\ell_1^2 = (\ell_1+k_1)^2 =
(\ell_1-k_4)^2 = (\ell_1+K_{12})^2 = 0\,.
\end{align}
These are precisely the quadruple-cut equations for a one-loop box
with loop momentum $\ell_1$, labeled as in \fig{DBOXSOFTOVERLAP1}.
As is well known~\cite{GeneralizedUnitarityBCF}, these equations have two distinct
solutions, related by spinor or equivalently parity conjugation.  (One
is illustrated in fig.~\ref{DBOXSOFTOVERLAP1}.)

At first sight, the appearance of the same global pole in different
integrals is alarming.  There was no hint of the second integral lurking in
the previous section's discussion; its presence casts doubt on our ability
to isolate the coefficient of either of the two double boxes by performing
a multivariate contour integral.  To understand the problem more fully,
consider that both will typically contribute to a given
amplitude.  We can write the combined contribution together,
\begin{align}
\int d^4\ell_1d^4\ell_2\left(
\mc I_{\text{HDB}} N_{\text{HDB}} +\mc I_{\text{VDB}} N_{\text{VDB}}
+\cdots\right)\,,
\end{align}
where $\mc I_{\text{HDB}}$ and $\mc I_{\text{VDB}}$ are the integrands of the
two double-box integrals, parametrized as in fig.~\ref{DBOXSOFTOVERLAP1},
and $N_{\text{HDB}}$
and $N_{\text{VDB}}$ are the corresponding numerators for the
given amplitude.
Each of the horizontal and vertical double boxes has two master integrals
(one scalar and one with an irreducible numerator); following
ref.~\cite{TwoLoopMaximalUnitarityI}, we would use a linear combination
of contour integrals around the global poles to extract the
corresponding coefficients in \eqn{MasterEquation}.  Each of the two
horizontal master integrals, for example, has a unique contour,
with the coefficient then schematically of the form,
\begin{align}
\oint_{T^8(\mc G_{\text{HDB}})}\!d^4\ell_1d^4\ell_2\left(
\mc I_{\text{HDB}} N_{\text{HDB}} +\mc I_{\text{VDB}} N_{\text{VDB}}
+\cdots\right)\,.
\label{JointCoefficient}
\end{align}
The presence of a second pair of master integrals, the vertical double-box
ones, risks contaminating the values of the coefficients for the
horizontal double boxes.  It may seem as though we cannot separate the
two, because of the shared global poles.

\subsection{Nonhomologous Contours}

Before conceding to the alarm raised by the overlap of global poles,
we should however ask whether the {\it contours\/} implicit in
\eqn{JointCoefficient} are the same.  As we have seen in
\sect{MultivariateResiduesSection}, in the multivariate case,
global poles can admit more than one inequivalent contour of
integration surrounding them.  
(The inequivalent contours are termed \emph{nonhomologous}.)
As we shall see, this is precisely
what happens in the case of the double boxes we are considering.
Furthermore, performing the contour integrals in a certain order ---
a heptacut, followed by the remaining contour integration --- selects one
of the nonhomologous contours, and isolates the coefficient of either
the horizontal or vertical double box, removing any possible
contamination.

\begin{figure}[!th]
\bc
\includegraphics[scale=0.75]{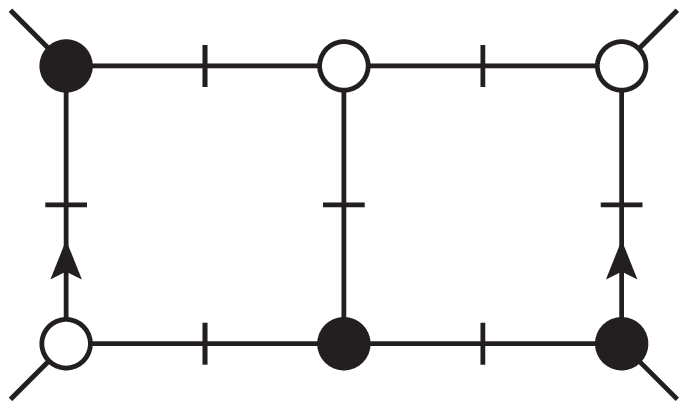}
\put(-157,22){$k_1$}
\put(-158,120){$k_2$}
\put(-6,120){$k_3$}
\put(-6,22){$k_4$}
\put(-153,73){$\ell_1$}
\put(-9,73){$\ell_2$}
\hspace*{1.5cm}
\includegraphics[scale=0.75]{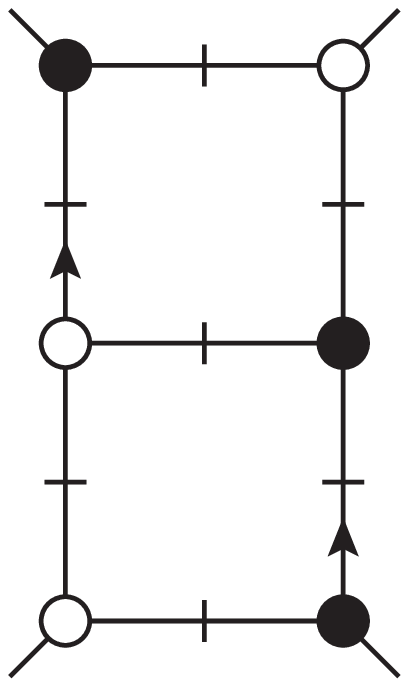}
\put(-95,-6){$k_1$}
\put(-96,152){$k_2$}
\put(-5,152){$k_3$}
\put(-5,-6){$k_4$}
\put(-93,103){$\ell_1$}
\put(-10,42){$\ell_2$}
\caption{\label{HDBVDB}
The horizontal and vertical double boxes. All internal lines are on-shell.}
\ec
\end{figure}

In order to visualize the multidimensional tori in question, we make
use of the same parametrization as in \eqn{TwoLoopParametrization},
but now applied to the labeling of \fig{HDBVDB}.  This labeling
allows us to align five of the seven internal lines.
It also allows us to take the heptacut solutions directly from
ref.~\cite{TwoLoopMaximalUnitarityI}.
     
As we saw in \sect{DoubleBoxGlobalPoles}, there are six distinct
heptacut solutions.  Two of eight global poles are shared
between the horizontal and vertical double boxes.  The fully-localized
integrand has a non-vanishing residue that is equal for both types
of scalar double boxes, up to a sign.  We now examine the possible
eight-fold contours more carefully.
                              
\def\alphap{\bar\alpha}
\def\betap{\bar\beta}
We first make a change of variables,
\begin{equation}         
\begin{aligned}          
\alpha_1 &= \alphap_1 - (\chi+1)^{-1} \alphap_4\,,\\
\alpha_2 &= \alphap_2 + (\chi+1)^{-1} \alphap_4\,,\\
\alpha_3 &= \alphap_3 -\chi \alphap_\chi+1\alphap_2
            +\chi(\chi+1)^{-1}\alphap_4
            -(\chi+1)^{-1}\betap_4\,,\\
\alpha_4 &= \alphap_4\,,\\
\beta_1 &= \betap_1 + (\chi+1)^{-1} \betap_4\,,\\
\beta_2 &= \betap_2 - (\chi+1)^{-1} \betap_4\,,\\
\beta_3 &= \betap_3 +\chi \betap_1-\chi\betap_2
            +\chi(\chi+1)^{-1}\betap_4
            -(\chi+1)^{-1}\alphap_4\,,\\
\beta_4 &= \betap_4\,,\\
\end{aligned}
\end{equation}
which simplifies the structure of the seven propagators.  We can
expand each denominator factor around the pole $\Global_1$,
retaining only the leading term in deviations $\delta\alphap_i$ and
$\delta\betap_i$.  This expansion yields the following expression
for the integrand of the horizontal double box,
\begin{equation}
\frac{C}{\delta\alphap_1\,\delta\alphap_2\,\delta\alphap_4\,
          \delta\betap_1\,\delta\betap_2\,\delta\betap_4\,
          Q(\delta\alphap_3,\delta\betap_3,\delta\alphap_1,\delta\alphap_2,
            \delta\betap_1,\delta\betap_2)}
\end{equation}
where $C$ is a function of the external spinors and invariants alone,
and can be treated as a constant for the purpose of analyzing
contours of integration, and where $Q$ is,
\begin{equation}
          Q(\delta\alphap_3,\delta\betap_3,\delta\alphap_1,\delta\alphap_2,
            \delta\betap_1,\delta\betap_2) =
\delta\alphap_3 \delta\betap_3 +\chi (\delta\alphap_1-\delta\betap_2)
(\delta\alphap_2-\delta\betap_1)\,.
\label{Qdef}
\end{equation}

In principle, we should choose a cycle for each factor, but the quadratic
nature of the last factor makes this less straightforward.
In the region of contour moduli space where
$\delta\alphap_3,\delta\betap_3 \gg \delta\alphap_1, \delta\alphap_2,
\delta\betap_1, \delta\betap_2$, the quadratic factor simplifies
into the product of two linear factors ($\delta\alphap_3\,\delta\betap_3$), and
the structure of the eight-tori encircling the global pole
becomes clearer.

In this region, we can parametrize the canonical eight-torus as follows,
\begin{align}
\alphap_1 = {} & \delta_{\alphap_1}e^{i\theta_{\alphap_1}}\,, \!\!&\!\!
\alphap_2 = {} & \delta_{\alphap_2}e^{i\theta_{\alphap_2}}\,, \!\!&\!\!
\alphap_3 = {} & -\chi+\delta_{\alphap_3}e^{i\theta_{\alphap_3}}\,, \!\!&\!\!
\alphap_4 = {} & \delta_{\alphap_4}e^{i\theta_{\alphap_4}}\,, \nn \\
\betap_1 = {} & \delta_{\betap_1}e^{i\theta_{\betap_1}}\,, \!\!&\!\!
\betap_2 = {} & \delta_{\betap_2}e^{i\theta_{\betap_2}}\,, \!\!&\!\!
\betap_3 = {} & -\chi+\delta_{\betap_3}e^{i\theta_{\betap_3}}\,, \!\!&\!\!
\betap_4 = {} & \delta_{\betap_4}e^{i\theta_{\betap_4}}\,.
\end{align}
The $\delta$s are positive
real numbers, and the angles $\theta$ run over $[0,2\pi]$ in order to
cover the surface of integration.
As discussed above, we take
$\delta_{\alphap_1},\delta_{\alphap_2},
\delta_{\betap_1},\delta_{\betap_2}\ll
\delta_{\alphap_3},\delta_{\betap_3}$.  Taking a horizontal
double-box heptacut followed
by a contour integration over the remaining degree of freedom $z$
corresponds to an integration over an eight-torus within this region.

Expanding each denominator factor
around the same global pole for the vertical double box
labeled as in \fig{HDBVDB}, we find
\begin{equation}
\begin{aligned}                      
&-\frac{C}{\delta\alphap_1\,
\big(\delta\alphap_3-\delta\betap_4/(\chi+1)\big)\,
\delta\alphap_4\,
          \delta\betap_1              
\big(\delta\betap_3 -\delta\alphap_4/(\chi+1)\big)\,
\delta\betap_4\,
          Q(\delta\alphap_3,\delta\betap_3,\delta\alphap_1,\delta\alphap_2,
            \delta\betap_1,\delta\betap_2)}
\end{aligned}
\end{equation}
for the integrand, where $Q$ is the same function given
in \eqn{Qdef}.  We first notice that if
$|\chi+1|\delta_{\alphap_3} = \delta_{\betap_4}$ or
$|\chi+1|\delta_{\betap_3} = \delta_{\alphap_4}$, the contour
is illegitimate because the integrand is singular on it; furthermore,
if $|\chi+1|\delta_{\alphap_3} < \delta_{\betap_4}$
or $|\chi+1|\delta_{\betap_3} < \delta_{\alphap_4}$, the contour
fails to enclose the global pole.  Thus to obtain a non-vanishing
residue for the vertical double box, we must take
if $|\chi+1|\delta_{\alphap_3} > \delta_{\betap_4}$
and $|\chi+1|\delta_{\betap_3} > \delta_{\alphap_4}$.

This does not suffice, however, because we also need the $Q$
factor to yield poles in $\alphap_2$ and $\betap_2$.  This will not
happen in the region where
$\delta_{\alphap_1},\delta_{\alphap_2},
\delta_{\betap_1},\delta_{\betap_2}\ll
\delta_{\alphap_3},\delta_{\betap_3}$; instead, we select the region
where
$\delta_{\alphap_1},\delta_{\alphap_3},
\delta_{\betap_1},\delta_{\betap_3}\ll
\delta_{\alphap_2},\delta_{\betap_2}$.  The vertical double-box
heptacut is contained within this region, which will yield a non-zero
residue for the vertical double box.  Although the two integrals
share the same global pole, just as in the case of the simple
example considered in \sect{MultivariateResiduesSection}, different
contours surrounding the global pole are required to obtain nonvanishing
residues for the two integrals.

\subsection{Other Configurations of Shared Poles}

\begin{figure}[!h]
\bc
\includegraphics[scale=0.65]{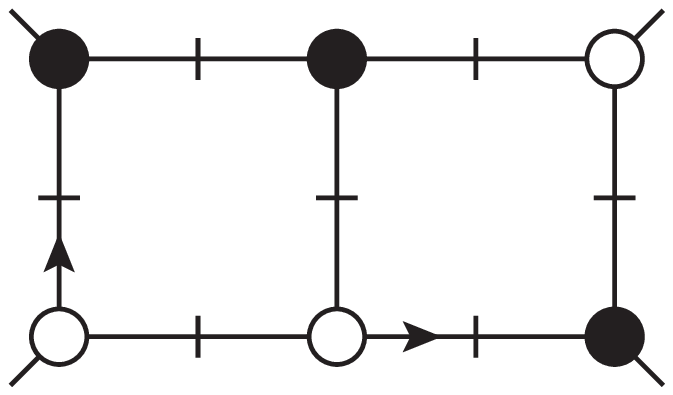}
\put(-137,20){$k_1$}
\put(-138,105){$k_2$}
\put(-5,104){$k_3$}
\put(-5,19){$k_4$}
\put(-43,24){$\ell_1$}
\put(-134,62){$\ell_2$}
\hspace*{.1cm}
\includegraphics[scale=0.65]{arrow_right}
\put(-36,50){$\ell_2\to0$}
\hspace*{.1cm}
\includegraphics[scale=0.65]{quadruple_cut_ex}
\put(-85,20){$k_1$}
\put(-86,105){$k_2$}
\put(-5,104){$k_3$}
\put(-5,19){$k_4$}
\put(-45,24){$\ell_1$}
\hspace*{.1cm}
\includegraphics[scale=0.65]{arrow_left}
\put(-36,50){$\ell_2\to0$}                        
\hspace*{.1cm}
\includegraphics[scale=0.65]{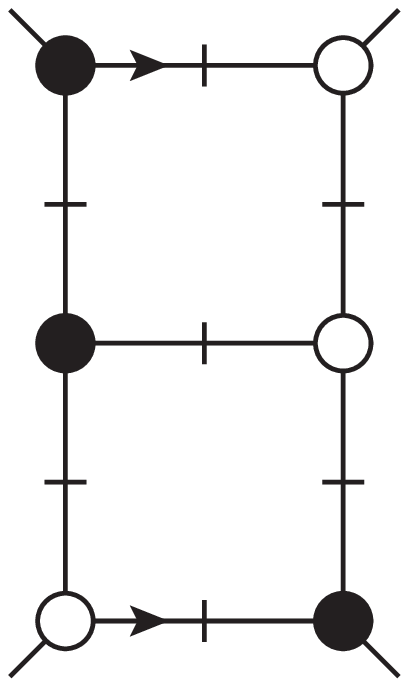}
\put(-85,-5){$k_1$}
\put(-85,133){$k_2$}
\put(-5,133){$k_3$}
\put(-5,-5){$k_4$}
\put(-45,0){$\ell_1$}
\put(-45,126){$\ell_2$}
\caption{
\label{DBOXSOFTOVERLAP2}
A second example of how a global pole could be shared between the horizontal
and vertical double-box integrals.}
\ec
\end{figure}

The momentum labeling in \fig{DBOXSOFTOVERLAP1} is not the only
one that gives rise to overlapping solutions of the on-shell equations.
A second example of overlapping kinematical configurations is shown in
fig.~\ref{DBOXSOFTOVERLAP2}.  Here, the shared global poles correspond
an outer edge (again labeled $\ell_2$) becoming soft,
$\ell_2^\mu\rightarrow 0$, in both the horizontal and vertical double
boxes.  One again obtains a kinematic solution for the other momentum
that is identical to that of a quadruply cut one-loop box.

\begin{figure}[!h]
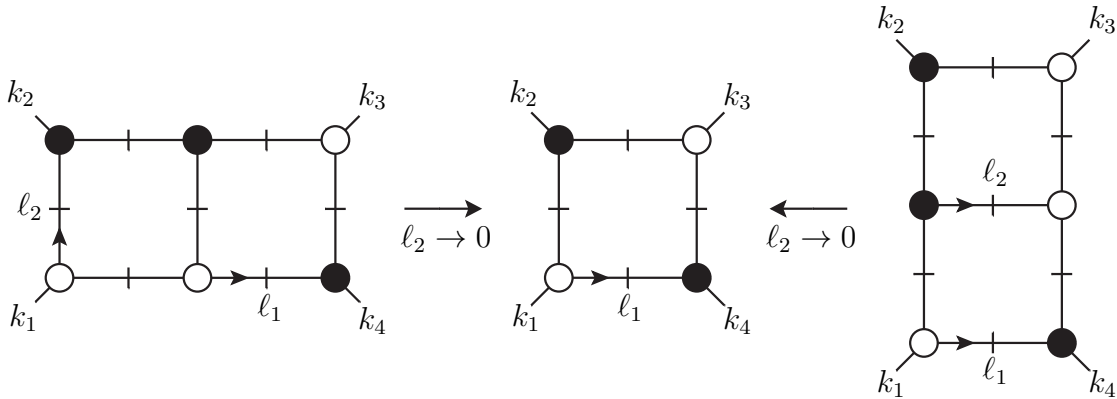

\bc
\includegraphics[scale=0.65]{dbox_s_hepta_cut_ex_2}
\put(-137,20){$k_1$}
\put(-138,105){$k_2$}
\put(-5,104){$k_3$}
\put(-5,19){$k_4$}
\put(-43,24){$\ell_1$}
\put(-134,62){$\ell_2$}
\hspace*{.1cm}
\includegraphics[scale=0.65]{arrow_right}
\put(-36,50){$\ell_2\to0$}
\hspace*{.1cm}
\includegraphics[scale=0.65]{quadruple_cut_ex}
\put(-85,20){$k_1$}
\put(-86,105){$k_2$}
\put(-5,104){$k_3$}
\put(-5,19){$k_4$}
\put(-45,24){$\ell_1$}
\hspace*{.1cm}
\includegraphics[scale=0.65]{arrow_left}
\put(-36,50){$\ell_2\to0$}
\hspace*{.1cm}
\includegraphics[scale=0.65]{dbox_t_hepta_cut_ex_1}
\put(-85,-5){$k_1$}
\put(-85,133){$k_2$}
\put(-5,133){$k_3$}
\put(-5,-5){$k_4$}
\put(-45,0){$\ell_1$}
\put(-45,75){$\ell_2$}
\caption{
\label{DBOXSOFTOVERLAP3}
A third example of how a global pole could be shared between the horizontal
and vertical double-box integrals.}
\ec
\end{figure}

We could also consider a labeling where one of the integrals, say the
horizontal double box, has a soft outer rung, while the vertical
double box has a soft middle rung.  This again gives rise to a
shared global pole, where the remaining loop momentum is that of
a quadruply cut one-loop box.  This configuration is illustrated
in \fig{DBOXSOFTOVERLAP3}.

As we shall show in \sect{CombiningContributionsSection}, the sharing 
described earlier
in \sect{DoubleBoxSharing} is reflected in the existence of a common
daughter integral, while the two different
overlaps described here do not admit a common two-loop daughter,
and hence are unnatural as far as the amplitude
is concerned.  In later
sections of this paper, we will rely only on the sharing
of poles described in \sect{DoubleBoxSharing}.

\subsection{Poles in the Cross-Section Integrand}

The sharing of global poles displayed in
\figs{DBOXSOFTOVERLAP2}{DBOXSOFTOVERLAP3} does not have a direct
application to the
amplitude.  It does, however, have a natural application
to the differential cross section.
\begin{figure}[!h]
\bc
\includegraphics[scale=0.75]{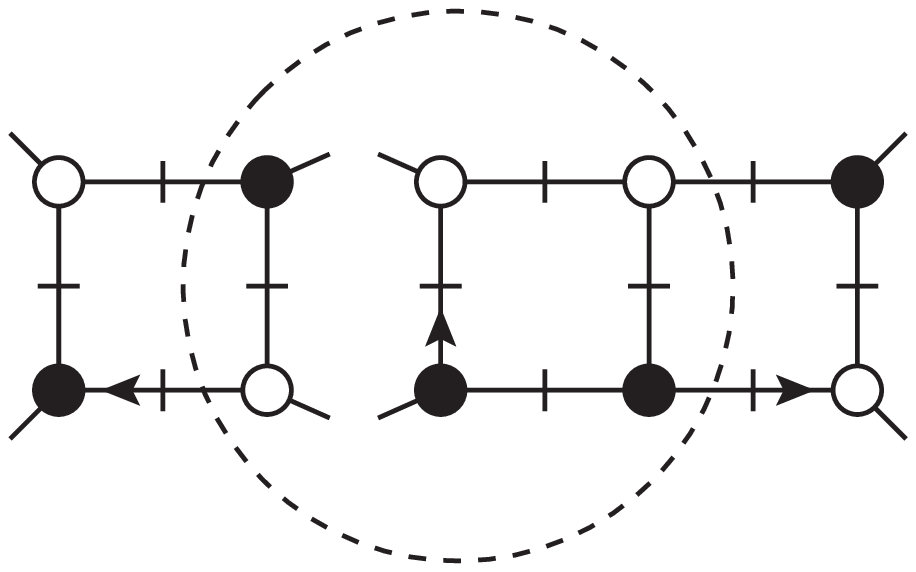}
\put(-127.5,30){$1$}
\put(-127.5,90){$2$}
\put(-205,98){$3$}
\put(-205,21){$4$}
\put(-5,98){$3$}
\put(-5,21){$4$}
\put(-43,27){$\ell_1$}
\put(-101,60){$\ell_2$}
\put(-170,27){$\ell_3$}
\put(-108,10){(a)}
\includegraphics[scale=0.75]{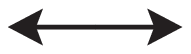}
\put(-40,72){$\ell_2\to0$}
\put(-40,48){$k_5\to0$}
\includegraphics[scale=0.75]{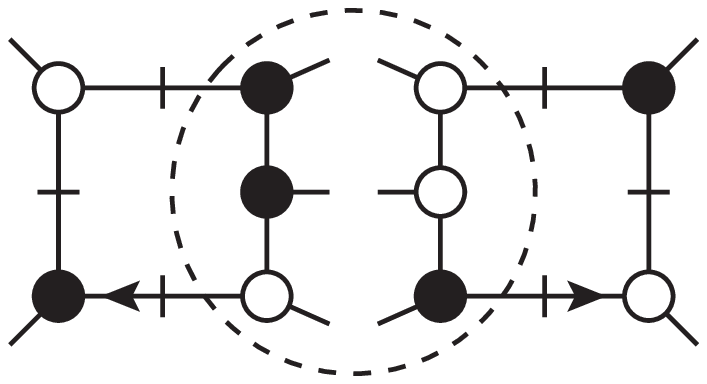}
\put(-82.5,30){$1$}
\put(-82.5,90){$2$}
\put(-160,98){$3$}
\put(-160,21){$4$}
\put(-82.5,60){$5$}
\put(-5,98){$3$}
\put(-5,21){$4$}
\put(-43,27){$\ell_1$}
\put(-124,27){$\ell_3$}
\put(-87,10){(b)}
\caption{
\label{CROSSSECTION}
Two maximal unitarity cuts for diagrams that enter the cross section at the
same order in the coupling: (a) a one-loop times two-loop contribution and
(b) a one-loop squared contribution. These cuts share the same kinematical
configuration in the indicated soft limits. The encircled subdiagrams correspond
precisely to the sharing of poles illustrated in
fig.~\ref{DBOXSOFTOVERLAP3}.}
\ec
\end{figure}

We consider two contributions to the differential cross section
for $2\rightarrow2$ scattering, the interference of a one-loop amplitude
with a two-loop amplitude, and the square of a five-point one-loop
amplitude.  Let us further consider generalized cuts of these objects.
In particular, we examine the maximal cut of the horizontal double box
shown in fig.~\ref{DBOXSOFTOVERLAP2},
and multiply by the quadruple cut of
the complex-conjugated four-point one-loop amplitude.  This contribution
is shown in fig.~\ref{CROSSSECTION}(a).  This can be thought of
as a global pole of the virtual contribution to the cross section.
Alternatively, by the optical theorem, we can also think of it
as a global pole of the four-loop amplitude for special external
kinematics.  From this latter point of view, the rung labeled by $\ell_2$ is no
longer an outer rung, but instead a middle rung of a two-loop subdiagram. This
subdiagram is enclosed by the dashed circle in fig.~\ref{CROSSSECTION}(a).
From
the analysis in the previous section we know that there is a natural candidate
to cancel the pole that arises when this leg goes soft.
We obtain this second contribution by
replacing the horizontal double-box subdiagram by the corresponding vertical
double-box subdiagram, as shown in fig.~\ref{CROSSSECTION}(b).

Returning to the interpretation of this cut as a contribution to the cross
section, we see that something remarkable has happened. The individual
amplitude contributions in fig.~\ref{CROSSSECTION}(b) are no longer four-point
diagrams, but five-point diagrams. The additional external leg, called $k_5$,
is soft, similar to $\ell_2$.  The global pole is associated with either
an internal line or a final-state line becoming soft, that is with infrared
singularities which must ultimately cancel by the KLN theorem~\cite{KLN}.
  This cancellation echoes 
the cancellation of global pole residues between the two different contributions
depicted
in the figure.
It confirms the close connection between nodal global poles and the infrared
singularities of the integrated amplitude, a connection previously
observed elsewhere~\cite{TwoloopContourUniqueness}.

\section{Combining Cut Contributions}
\label{CombiningContributionsSection}

In the previous section we showed that it is possible to find
kinematical configurations of loop momenta
that simultaneously localize two different integrals
to the same global pole.  We also showed that it is nonetheless possible
to find contours that distinguish the two.  Of course contours that
simply combine the two also exist.

In this discussion, it was important to
line up the loop momenta in each integral appropriately.
However,
there is considerable freedom in choosing the loop-momentum parametrization.
Indeed, we saw that there are different ways in which global poles can
be shared between two integrals.
One may wonder about the significance of any particular choice of
parametrization, or equivalently any particular choice of how poles
are shared.

In this section, we will argue that although the three examples
in \sect{SharedGlobalPolesSection}
are superficially similar, there is a clear distinction between
them.  The first example, shown in
fig.~\ref{DBOXSOFTOVERLAP1}, is a physically meaningful identification
of loop momenta for amplitudes, whereas the second and third examples, shown in
figs.~\ref{DBOXSOFTOVERLAP2} and \ref{DBOXSOFTOVERLAP3}, are not.  
(They nonetheless have other applications, which we
discussed in the previous section.)

\begin{figure}[!h]
\bc
\includegraphics[scale=0.75]{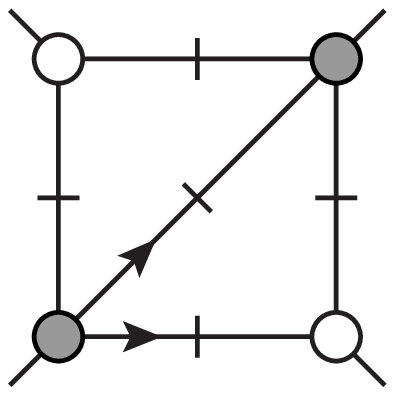}
\put(-93,24){$k_1$}
\put(-94,119){$k_2$}
\put(-6,119){$k_3$}
\put(-6,24){$k_4$}
\put(-50,30){$\ell_1$}
\put(-59,80){$\ell_2$}
\hspace*{.75cm}
\includegraphics[scale=0.75]{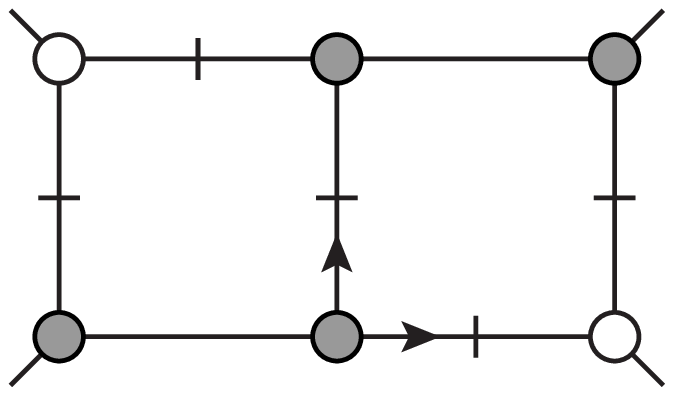}
\put(-155,24){$k_1$}
\put(-155,119){$k_2$}
\put(-6,119){$k_3$}
\put(-6,24){$k_4$}
\put(-48,29){$\ell_1$}
\put(-92,73){$\ell_2$}
\hspace*{.75cm}
\includegraphics[scale=0.75]{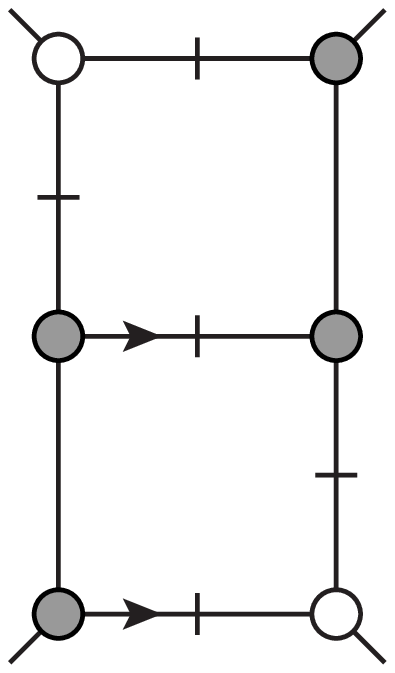}
\put(-93,-5){$k_1$}
\put(-94,150){$k_2$}
\put(-5,150){$k_3$}                           
\put(-5,-5){$k_4$}
\put(-49,0){$\ell_1$}
\put(-49,82){$\ell_2$}
\caption{
\label{SLASHEDBOXCUT}
The pentacut provides a natural prescription for aligning the loop-momentum
labels in different integrals. The global octacut poles of the
double-box integrals can be analyzed by performing a further three-dimensional contour
integral on the pentacut slashed-box integral at left.  
The gray blobs indicate vertices of indefinite chirality.}
\ec
\end{figure}

The examples are distinguished by the existence of daughter integrals,
that is integrals                                   
with fewer propagators,
which share common subsets of cuts.  Their existence will allow us to align loop
momenta of 
different integrals in a physically meaningful way, rather than in an
arbitrary way.
The pentacut slashed box in fig.~\ref{SLASHEDBOXCUT}
combines the horizontal and vertical double-box integrals naturally
 using a momentum
labeling that is identical to that in fig.~\ref{DBOXSOFTOVERLAP1}.
It is possible to open the four-point vertices in the slashed box diagram in
various ways with two additional propagators to obtain both the horizontal and
vertical double-box integrals with massless external legs, $\HDB$ and $\VDB$.
These integrals differ simply by a cyclic permutation of the
external legs.
                                                 
The situation is different for the other two labelings shown in
\figs{DBOXSOFTOVERLAP2}{DBOXSOFTOVERLAP3}.  Consider, for example,
\fig{DBOXSOFTOVERLAP2}.  Can we find an integral and unitarity cut
that contains both partly-cut horizontal and vertical double-box
integrals?  Loop
momentum $\ell_1$ is located somewhere between external legs 1 and 4
in both integrals, so finding a unitarity cut with that property should
be straightforward. However, momentum $\ell_2$ is located between
external legs 1 and 2 in the HDB integral, but between 2 and 3 in the
VDB integral. Such a behavior is difficult to reconcile in any
unitarity cut.  Indeed, for a planar unitarity cut it is
impossible. It might conceivably be possible for a non-planar
integral; but in the present paper we are in any case focused on the
planar case.  One could try to simultaneously reparametrize the
$\ell_2$ dependence in both integrals so as to fit into a single
unitarity cut, but there appears to be no such possibility. A similar
analysis can be performed on the momentum labeling in
\fig{DBOXSOFTOVERLAP3}, leading again to
the conclusion that there is no unitarity cut that is consistent
with this way of sharing global poles.

We can also see the distinction between the sharing described in 
\figs{DBOXSOFTOVERLAP2}{DBOXSOFTOVERLAP3} and that described in
\fig{SLASHEDBOXCUT} using dual coordinates~$x_i$~\cite{DualConformal}. 
The external momenta are given by differences of the coordinates,
$k_i = x_{i+1}-x_i$.  In terms of these coordinates, the horizontal and
vertical double boxes have the following expressions,
\begin{equation}
\begin{aligned}
\HDB = {} & \int d^D x_5\int d^D x_6
\frac{x_{13}^4 x_{24}^2}{x_{15}^2 x_{25}^2 x_{35}^2 x_{56}^2 x_{36}^2 x_{46}^2 x_{16}^2}\;,\\
\VDB = {} & \int d^D x_5\int d^D x_6
\frac{x_{24}^4 x_{13}^2}{x_{15}^2 x_{25}^2 x_{45}^2 x_{56}^2 x_{36}^2 x_{46}^2 x_{26}^2}\;.\\
\end{aligned}
\label{DoubleBoxesinDualCoordinates}
\end{equation} 
Five denominators --- $x_{15}^2$, $x_{25}^2$, $x_{56}^2$, $x_{36}^2$, and $x_{46}^2$ ---
are the same in both integrals.  The sharing described by \fig{SLASHEDBOXCUT} corresponds
to taking the limit $x_6\rightarrow x_5$; upon taking that limit, all remaining
denominators in one integral manifestly match denominators in the other 
($x_{35}^2\leftrightarrow x_{36}^2$, $x_{45}^2\leftrightarrow x_{46}^2$, 
$x_{16}^2\leftrightarrow x_{15}^2$, and $x_{26}^2\leftrightarrow x_{25}^2$).  
In contrast, in the sharing described by \fig{DBOXSOFTOVERLAP2}, the limit corresponds
to taking $x_5\rightarrow x_2$ in the horizonal double box, and
$x_6\rightarrow x_3$ in the vertical double box.  Under this limit,
only one additional denominator in each integral manifestly
matches a denominator in the other: $x_{35}^2 \leftrightarrow x_{26}^2$.  (The
remaining denominator matches only when solving the equations.) In the sharing
described by~\fig{DBOXSOFTOVERLAP3}, the limit again corresponds to taking
$x_5\rightarrow x_2$ in the horizonal double box, 
but in this case to taking $x_5\rightarrow x_6$ in the vertical
double box.  Here, while both additional denominators in the vertical
double box manifestly match denominators in the horizontal double box, only one
additional denominator in the horizontal double box matches a denominator
in the vertical double box: $x_{16}^2 \leftrightarrow x_{15}^2$.

\section{The Four-Point ABDK Relation}
\label{FourPointABDKSection}

We turn next to a maximal-cut analysis of the ABDK relation~(\ref{ABDK4}) for
the four-point amplitude in the $\mc N=4$ super-Yang--Mills theory.  
Let us study the various integrals that arise
in the two representations of the four-point amplitude on the two sides
of \eqn{ABDK4}.  In particular, we examine the integrals 
which admit an eight-fold localization,
the one-loop box squared and the two-loop double box.
They appear in the amplitude 
with the same power of the
Yang--Mills coupling.  
(These integrals are also the ones of leading
and equal polylogarithmic weight, whereas in the remaining terms, part 
of the polylogarithmic weight comes from constant prefactors.)
At the very least, if we apply a GDO like that of ref.~\cite{TwoLoopMaximalUnitarityI} 
that extracts either the coefficient
of the horizontal double box or of the vertical double box to the right-hand
side of eq.~(\ref{ABDK4}), we must obtain the same coefficient as on the left-hand
side.  For this to be possible, both sides must share global poles.
The one-loop box squared appearing on the right-hand side
is of course not part of
the standard basis for two-loop amplitudes, and so ordinarily one
would not consider it in computing two-loop amplitudes.  Furthermore,
any global poles it shares with two-loop integrals would not affect our
ability to extract coefficients of the latter in the standard basis.  This
is true even if the same contours enclose the global poles in both integrals.

As we shall see, the sharing of poles and residues is more extensive than required:
the two sides share all poles and residues present in any
double-box integral on the left-hand side, and on some non-maximal cut surfaces,
even share integrands away from the global poles.

\begin{figure}[!h]
\bc
\includegraphics[scale=0.75]{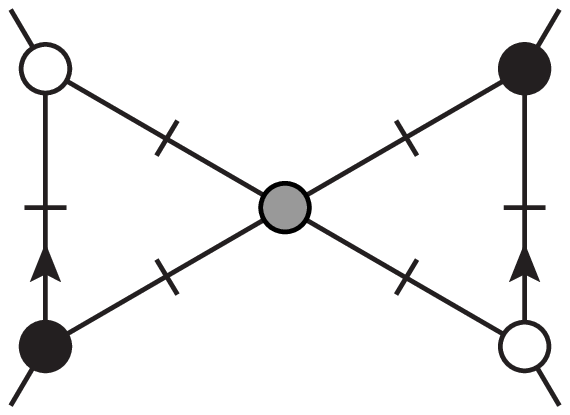}
\put(-129,-7){$k_1$}
\put(-129,93){$k_2$}
\put(-7,93){$k_3$}
\put(-7,-7){$k_4$}
\put(-132,44){$\ell_1$}
\put(-5,44){$\ell_2$}
\hspace*{1.5cm}
\includegraphics[scale=0.75]{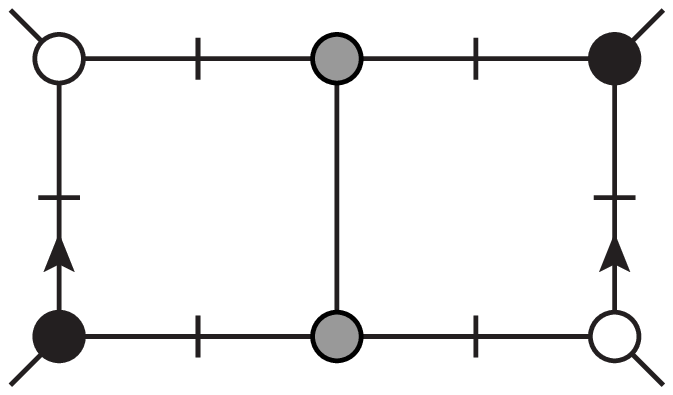}
\put(-155,-7){$k_1$}
\put(-155,93){$k_2$}
\put(-5,93){$k_3$}
\put(-5,-7){$k_4$}
\put(-151,44){$\ell_1$}
\put(-8,44){$\ell_2$}
\\[3mm]
\includegraphics[scale=0.75]{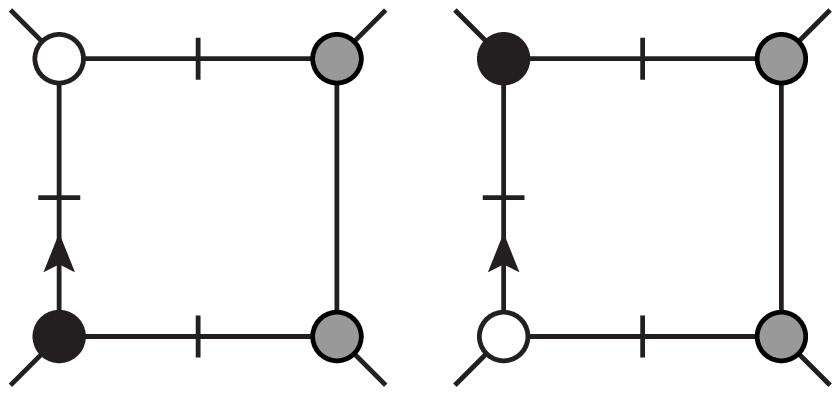}
\put(-191,-6){$k_1$}
\put(-191,92){$k_2$}
\put(-98,92){$k_3$}
\put(-98,-6){$k_4$}
\put(-7,92){$k_2$}
\put(-7,-6){$k_1$}
\put(-188,44){$\ell_1$}
\put(-68,44){$\ell_2$}
\caption{
\label{ABDKBOWTIE4}
The bow-tie hexacut detects contributions from both the planar double box and
squared one-loop boxes. This figure shows a particular example where vertices at
opposite corners have equal chiralities in the bow-tie diagram. 
}
\ec
\end{figure}

We parametrize the squared one-loop box and horizontal double-box integrals
as shown in \fig{ABDKBOWTIE4}.  We start by cutting all
propagators that are manifestly shared
between the two integrals.  There are six such propagators, which 
together form the integrand of a bow-tie integral.  
This leads us to consider the zero locus,
\begin{align}
\Sol\equiv
\big\{\,(\ell_1,\ell_2)\in \CP^4\times\CP^4\;|\; &
\ell_1^2 = 0\,,\;
(\ell_1-k_2)^2 = 0\,,\;
(\ell_1+k_1)^2 = 0\,,\; \nn \\ &
\ell_2^2 = 0\,,\;
(\ell_2-k_3)^2 = 0\,,\;
(\ell_2+k_4)^2 = 0\,\big\}\,.
\label{BOWTIECUTEQ}
\end{align}
The hexacut solutions for the bow-tie are two-dimensional, parametrized by a pair of
complex variables $(z_1,z_2)\in\CP^2$.   They are simply products of 
solutions for independent one-mass triangles, which makes it straightforward to 
write them down.
There are four distinct solutions to the
equations, $\Sol = S_1\cup\cdots\cup S_4$.  We can identify four distinct hexacut diagrams
which are in one-to-one correspondence with the four hexacut
solutions. The diagrams are characterized by the relative chiralities of the 
vertices at opposite corners.
By parity it suffices to work out one example of each kind: one where the chiralities at opposite corners
are identical, and one where they are opposite.
Performing the six-fold contour integral that imposes the hexacut conditions
implicitly picks a contour that isolates only one of the
horizontal or vertical double boxes, so we need not worry about sharing
of global
poles between these two integrals.

We can solve the on-shell equations \eqref{BOWTIECUTEQ} straightforwardly
using the
parametrization~\eqref{TwoLoopParametrization}; they are after all just two copies of
one-loop triangle cuts.  We are left with one free parameter from
each loop momentum.  For the
hexacut depicted in fig.~\ref{ABDKBOWTIE4}, which we label $\Sol_1$, we have
the very simple solution,
\begin{equation}
\begin{aligned}
\Sol_1:\,
\left\{
\begin{array}{llll}
\alpha_1 = 0\,, &\;\;
\alpha_2 = 0\,, &\;\;
\alpha_3 = z_1\,, &\;\;
\alpha_4 = 0\,,\\[2mm]
\beta_1 = 0\,, &\;\;
\beta_2 = 0\,, &\;\;
\beta_3 = z_2\,, &\;\;
\beta_4 = 0\,.\\[1mm]
\end{array}
\right.
\end{aligned}
\end{equation}
The bow-tie hexacut squared one-   
and two-loop integrals then take the form,
\begin{equation}
\begin{aligned}
\HDB\big|_{6\text{-cut}} = {} &
\oint d^2z
\frac{J^{-1}(z_1,z_2)}{(\ell_1+\ell_2-K_{23})^2}\Big|_{\Sol_1} =
-\frac{\chi}{16s_{12}^3}    
\oint \frac{d^2z}{z_1z_2(z_1+\chi)(z_2+\chi)}\;,
\\[1mm]
I_\square^2\big|_{6\text{-cut}} = {} &
\oint d^2z
\frac{J^{-1}(z_1,z_2)}{(\ell_1-K_{23})^2(\ell_2-K_{23})^2}\Big|_{\Sol_1} =
-\frac{1}{16s_{12}^4}
\oint \frac{d^2z}{z_1z_2(z_1+\chi)(z_2+\chi)}\;,
\end{aligned}
\label{4PTBOWTIE_EQUAL}
\end{equation}
where $J(z_1,z_2)$ is the Jacobian that arises upon evaluation of the residue
in the loop momentum parametrization; that is, the integrand of the scalar hexacut
bow-tie integral itself. The expression $J^{-1}(z_1,z_2)$ is the same for all hexacut
solutions,
\begin{align}
J^{-1}(z_1,z_2)\equiv -\frac{1}{16s_{12}^2z_1z_2}\;.
\end{align}
We observe that, remarkably, the integrands of the horizontal double box and
the squared one-loop box integrals coincide on the hexacut solution
$\Sol_1$, up to a constant. Demanding that the cuts should be equal fixes the
relative constant. The result takes the form,
\begin{align}
s_{12}^2s_{23}\HDB\big|_{6\text{-cut}} =
\big(s_{12}s_{23}I_\square\big)^2\big|_{6\text{-cut}}\;.
\end{align}
The coefficient on the left-hand side is exactly that which appears
in the four-point amplitude in the $\mc N = 4$ super Yang-Mills theory (after
removing overall normalization factors).
Because the integrands are identical, the global poles as well as the contours
surrounding them are now shared between the double-box integral and the
one-loop integrals squared.  What global poles are these?  We find four poles,
located at the following values of $(z_1,z_2)$,
\begin{equation}
(0,0)\,;\qquad (0,-\chi)\,;\qquad (-\chi,0)\,;\qquad (-\chi,-\chi)\,.
\label{HexacutPoles}
\end{equation}
The first is a `spurious' pole from the point of view of the double box, as it
does not correspond to a maximal cut, and is thus not required for construction
of the amplitude.  It arises purely from the Jacobian, and corresponds to a
soft limit; we will call such poles `soft' poles more generally.
The second pole in \eqn{HexacutPoles} corresponds to $\Global_5$ [\eqn{G5value}]; 
the third to $\Global_3$ [\eqn{G3value}];
and the last pole, to $\Global_1$ [\eqn{G1value}].  
From the point of view of the squared one-loop box,
only the last pole corresponds to a maximal cut, while the first three are `soft'.
All are nonetheless shared between the double box and the squared one-loop box.
Performing the additional contour integrals to localize all coordinates
of course gives us identical residues,
\begin{align}
s_{12}^2s_{23}\HDB\big|_{8\text{-cut}} =
\big(s_{12}s_{23}I_\square\big)^2\big|_{8\text{-cut}}\;.
\label{4PTOCTACUTIDENTITY_1}
\end{align}
As mentioned earlier, this sharing does not affect our ability to extract
coefficients in the two-loop version of the master equation~(\ref{MasterEquation}),
because the one-loop integrals squared are not part of the two-loop basis.

\begin{figure}[!ht]
\bc
\includegraphics[scale=0.75]{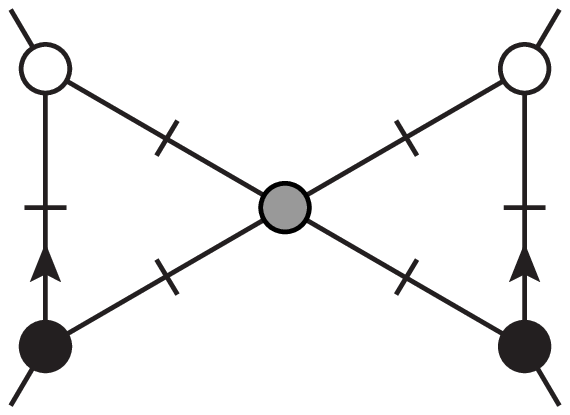}
\put(-129,-7){$k_1$}
\put(-129,93){$k_2$}
\put(-7,93){$k_3$}
\put(-7,-7){$k_4$}
\put(-132,44){$\ell_1$}
\put(-5,44){$\ell_2$}
\hspace*{1.5cm}
\includegraphics[scale=0.75]{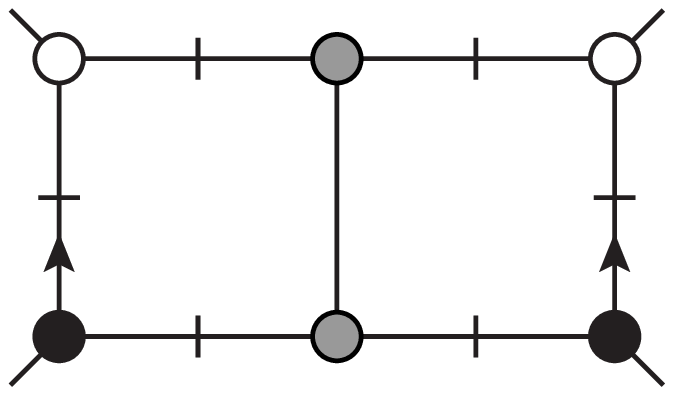}
\put(-155,-7){$k_1$}
\put(-155,93){$k_2$}
\put(-5,93){$k_3$}
\put(-5,-7){$k_4$}
\put(-151,44){$\ell_1$}
\put(-8,44){$\ell_2$}
\\[3mm]
\includegraphics[scale=0.75]{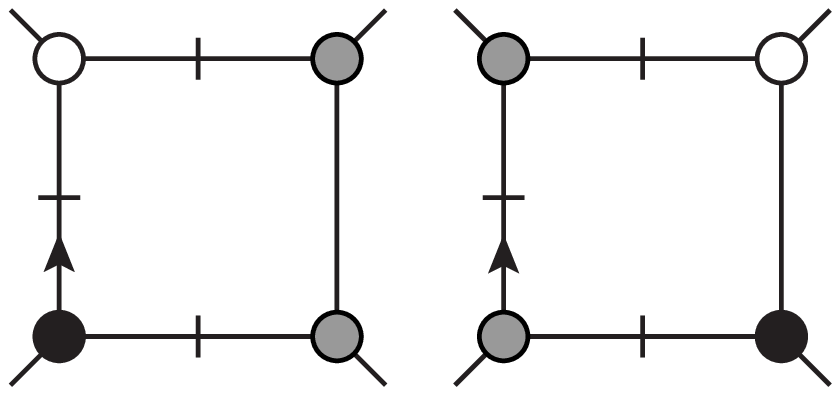}
\put(-191,-6){$k_1$}
\put(-191,92){$k_2$}
\put(-98,92){$k_3$}
\put(-98,-6){$k_4$}
\put(-7,92){$k_2$}
\put(-7,-6){$k_1$}
\put(-188,44){$\ell_1$}
\put(-68,44){$\ell_2$}
\put(-212,44){$+$}
\caption{
\label{ABDKBOWTIE4OPP}
The second inequivalent type of bow-tie hexacut considered
in the main text, with contributions from the double box and a squared box.}
\ec
\end{figure}
The parity conjugate solution to $\Sol_1$, which we label $\Sol_2$, is given by,
\begin{equation}
\begin{aligned}
\Sol_2:\,
\left\{
\begin{array}{llll}
\alpha_1 = 0\,, &\;\;
\alpha_2 = 0\,, &\;\;
\alpha_3 = 0\,, &\;\;
\alpha_4 = z_1\,,\\[2mm]
\beta_1 = 0\,, &\;\;
\beta_2 = 0\,, &\;\;
\beta_3 = 0\,, &\;\;
\beta_4 = z_2\,. \\[1mm]
\end{array}
\right.
\end{aligned}
\end{equation}
It contains the global poles $\Global_2$, $\Global_4$, and $\Global_6$, completing
the list of global poles present in the horizontal double box and the equality of
their residues to the squared one-loop box.
Alternatively, we could examine a hexacut solution with opposite chiralities at the
opposite corners, see \fig{ABDKBOWTIE4OPP}. The solution, which we label $\Sol_3$,
 takes the
form,
\begin{equation}
\begin{aligned}
\Sol_3:\,
\left\{
\begin{array}{llll}
\alpha_1 = 0\,, &\;\;
\alpha_2 = 0\,, &\;\;
\alpha_3 = z_1\,, &\;\;
\alpha_4 = 0\,,\\[2mm]
\beta_1 = 0\,, &\;\;
\beta_2 = 0\,, &\;\;
\beta_3 = 0\,, &\;\;
\beta_4 = z_2\,.\\[1mm]
\end{array}               
\right.
\end{aligned}
\end{equation}
For the hexacut integrals we have
\begin{align}                                                            
\HDB\big|_{6\text{-cut}} = {} &
-\frac{\chi+1}{16s_{12}^3}
\oint\frac{d^2z}{z_1z_2\big[ (z_1 + \chi)(z_2 + \chi) + z_1 + z_2 + \chi \big]}\,,
\nn \\[1mm]
I_\square^2\big|_{6\text{-cut}} = {} &
-\frac{1}{16s_{12}^4}
\oint\frac{d^2z}{z_1z_2(z_1+\chi)(z_2+\chi)}\,.
\label{4PTBOWTIE_OPPOSITE}
\end{align}
Here, the initial six-fold contour integrals leave us with different
expressions.
We can nonetheless proceed as before. We know from
ref.~\cite{TwoLoopMaximalUnitarityI} that the (horizontal) double box has two
nonzero octacut poles at $(z_1,z_2) = (-\chi,0)$, corresponding to $\Global_3$,
and $(z_1,z_2) = (0,-\chi)$, corresponding to $\Global_6$.  (These poles
lie on the intersection with the $\Sol_1$ and $\Sol_2$ hexacut solutions, respectively.)
As before, there is also a `soft' pole at $(z_1,z_2) = (0,0)$. 
(The soft pole lies on the intersection with all other hexacut solutions.)
These three poles are also present in the squared
one-loop integral, cf. eq.~\eqref{4PTBOWTIE_OPPOSITE}, where all are `soft'.
In contrast, in this hexacut solution
 the global pole of the squared one-loop integral, at $(z_1,z_2)=(-\chi,-\chi)$
 is not a global pole of the horizontal double box.
  Evaluating
the three residues from both integrals in eq.~\eqref{4PTBOWTIE_OPPOSITE} yields
the same answer, up to an overall constant. For example, for the
residue at $(0,0)$,
\begin{align}     
\HDB\big|_{8\text{-cut}} =
-\frac{1}{16\chi s_{12}^3} \;, \quad
I_\square^2\big|_{8\text{-cut}} =
-\frac{1}{16\chi^2 s_{12}^4}\;.
\label{4PTDOUBLESOFTRESIDUES}
\end{align}
The remaining octacut residues differ from those in
eq.~\eqref{4PTDOUBLESOFTRESIDUES} only by an overall sign, so we will not write them
down explicitly. We can summarize the results in a single equation,
\begin{align}         
s_{12}^2s_{23}\HDB\big|_{8\text{-cut}} =
\big(s_{12}s_{23}I_\square\big)^2\big|_{8\text{-cut}}\;.
\label{4PTOCTACUTIDENTITY_2}
\end{align}
To establish this identity, it would in fact suffice to take the residue at either
$z_1 = 0$ or $z_2 = 0$, as the remaining heptacut integrands would be equal for the
two types of integrals.  A similar analysis holds for
the vertical double box.

The identity of residues described above is precisely what is required for the ABDK
relation~(\ref{ABDK4}).  The right-hand side is an expression, 
given in terms of a physical amplitude, that
has a term proportional to $I_\square^2$.   The quantity $M_4^{(L)}$ appearing in 
that equation is defined in terms of the one- or two-loop partial amplitude
normalized by the tree-level amplitude,
$M_4^{(L)}(s,t;\epsilon)\equiv A_4^{(L)}(s,t;\epsilon)/A_4^{(0)}(s,t)$.
The planar partial amplitudes themselves are conventionally
 normalized according to,
\begin{align}
\mc A_n^{(L)} = g^{n-2}\bigg[
\frac{2e^{-\gamma\epsilon}g^2N_c}{(4\pi)^{2-\epsilon}}\bigg]^L
\sum_{\rho\in S_n/Z_n}\Tr(T^{a_{\rho(1)}}\cdots T^{a_{\rho(n)}})
A_n^{(L)}(\rho(1),\dots,\rho(n))\;.
\label{PARTIALAMPLITUDES}
\end{align}
The removal of color and normalization factors
yields $\Mampl41(s_{12},s_{23};\epsilon) = s_{12}s_{23}I_\square$.
We focus on the first term on the right-hand side of \eqn{ABDK4}, as we
cannot take an eight-fold cut of the other terms; we leave them for
future study.  (The one-loop integral $M_4^{(1)}(s_{12},s_{23};2\epsilon)$ has at
most a quadruple cut.) In the notation of this paper, the reduced two-loop
amplitude is given by
\begin{align}
\Mampl42(s_{12},s_{23};\epsilon) =
s_{12}^2s_{23}\HDB+s_{12}s_{23}^2\VDB\,.
\end{align}

In order to fix the numerical coefficients in front of the integrals in the
ABDK relation~(\ref{ABDK4}), we must consider
a somewhat subtle point.  The term on the right-hand side of \eqn{ABDK4} proportional 
to $I_\square^2$ is symmetric under the interchange of $\ell_1\leftrightarrow \ell_2$.  The 
same is not true of the double-box terms on the left-hand side.  These features can
be seen most easily using dual coordinates~\cite{DualConformal}.  The expressions for the
double boxes were given in \eqn{DoubleBoxesinDualCoordinates}; the squared one-loop
box has the following expression,
\begin{equation}
I_\square^2 = {}
\int d^D x_5
\frac{x_{13}^2x_{24}^2}{x_{15}^2x_{25}^2x_{35}^2x_{45}^2}
\int d^D x_6
\frac{x_{13}^2x_{24}^2}{x_{16}^2x_{26}^2x_{36}^2x_{46}^2}\,.
\end{equation}
If we antisymmetrize the right-hand side under $\ell_1\leftrightarrow\ell_2$
or equivalently $x_5\leftrightarrow x_6$, it vanishes.  In contrast,
the left-hand side integrand
will not vanish upon antisymmetrization; it only vanishes after integration.  In this
respect, it is similar to integrals with insertions of Levi-Civita tensors, whose
integrands (or even isolated residues) do not vanish, but which vanish after integration.
We thus need to form a projector, analogous to the treatment of such 
insertions~\cite{TwoLoopMaximalUnitarityI,ExternalMasses,FourMass,NonPlanarDoubleBox,%
MassiveNonPlanarDoubleBoxes,DoubledPropagators,ThreeLoopLadder,MaximalElliptic},
that will set the sum of residues to zero for the antisymmetric combination.  It
is easiest to do this by symmetrizing the left-hand side of \eqn{ABDK4}, and thus
of \eqn{4PTOCTACUTIDENTITY_2}.  This introduces a factor of $1/2$, because only
one of the resulting two terms will have a non-vanishing octacut residue for the
contour we are considering.  Both terms will contribute the same result after
integration along the standard contour, so the octacut relation~(\ref{4PTOCTACUTIDENTITY_2})
along with its partner for the vertical double box
imply the following integral relation,
\begin{equation}
s_{12} s_{23}\big( s_{12}\HDB + s_{23}\VDB\big) = \frac12\big(s_{12}s_{23}I_\square\big)^2
+\text{octacut-free}\,.
\end{equation}
While the double boxes do have a symmetry under interchange of $(1,2)\leftrightarrow (3,4)$,
which is equivalent to $\ell_1\leftrightarrow\ell_2$, the symmetrization and resulting
factor of $1/2$ are independent of the interchange symmetry, and would apply even in its
absence.

Thus far we have examined
only global poles present in one of the double-box integrals, and
found that there is always a corresponding global pole in the
one-loop box integral squared. 
However, one may wonder whether there are additional
poles present in the squared one-loop box terms.
Such poles are indeed present, at $(z_1,z_2)=(-\chi,-\chi)$ in the hexacut 
solutions $\Sol_3$ and $\Sol_4$.  
They occur in the product of two quadruply cut box
integrals of
opposite chirality. As we noted above, these poles are not global 
poles of the double-box integral, yet
their residues are nonvanishing for the squared one-loop integral. 
This seeming inconsistency could perhaps be cured by inserting a parity-odd
term into the integrand of the squared one-loop box in such a way as to cancel
these incompatible residues.  This addition would not modify the integrated expression,
and hence would leave the ABDK relation~(\ref{ABDK4}) unmodified.
We leave an investigation of this issue to future work.

\section{The Five-Point ABDK Relation}
\label{FivePointABDKSection}

The remarkable implications of shared global poles for the
four-point ABDK relation motivate a similar analysis for the five-point relation. 
To what extent can we reconstruct the identity
from maximal cuts?

The five-point relation has the same form as the four-point one~(\ref{ABDK4}),
\begin{equation}
\begin{aligned}
\Mampl52(s_{12},s_{23},s_{34},s_{45},s_{51};\eps) = &\frac12 \Bigl[\Mampl51(s_{12},s_{23},s_{34},s_{45},s_{51};\eps)\Bigr]^2 \\
&+ f^{(2)}(\eps) \Mampl51(s_{12},s_{23},s_{34},s_{45},s_{51};2\eps) + C^{(2)}
+\Ord(\eps)\,.
\label{ABDK5}
\end{aligned}
\end{equation}
In this equation, the normalized five-point one-loop MSYM 
amplitude~\cite{OneLoopFivePoint} is given by,
\begin{align}
\Mampl51(\{s_{ij}\};\epsilon) =
\frac{1}{4}\sum_{\rho\in\text{cyclic}}\rho(s_{34}s_{45})I_{\Square}(\rho)+
\frac{\epsilon}{2}\varepsilon_{1234}I_{\pentagon}^{D=6-2\epsilon}\;,
\label{5PT1LOOP}
\end{align}
where $\rho(k_i) = k_{\rho(i)}$,
$\rho(s_{ij}) = s_{\rho(i)\rho(j)}$,
$\varepsilon_{1234}\equiv
4i\varepsilon_{\mu\nu\rho\sigma}k_1^\mu k_2^\nu k_3^\rho k_4^\sigma$,
and,
\begin{equation}
I_{\square}(\rho) = I_{\square}(\rho(K_{12},k_3,k_4,k_5))\,.
\label{BoxPermutationDefinition}
\end{equation}
The normalized five-point two-loop MSYM amplitude~\cite{ABDK5} is,
\begin{equation}
\begin{alignedat}{10}
\Mampl52(\{s_{ij}\};&\epsilon) =
\frac18 \sum_{\rho\in\text{cyclic}} \\
&\biggl\{
  \rho(s_{12}^2 s_{23}) \Pss_{2,2}(\rho(k_{1},k_{2},k_{3},K_{45}))
  +\rho(s_{12}^2 s_{51}) \Pss_{2,2}(\rho(k_{1},k_{2},K_{34},k_{5}))\\
& \hspace*{2mm}+\rho(s_{12} s_{23}s_{45})
\Pss_{3,2}[(\ell_1+k_{\rho(5)})^2](\rho(
k_{1},k_{2},k_{3},k_{4},k_{5}))
\biggr\}
+\text{parity-odd}.
\end{alignedat}
\label{TwoLoop5}
\end{equation}
In the present paper, we will consider only the parity-even part of the amplitude.
On the left-hand side of \eqn{ABDK5}, each pentabox and one-mass double box appears
with a factor of $1/8$, whereas each square of a one-loop box appears with
a factor of $1/32$, and each product of different one-loop boxes with a factor of
$1/16$.  

We must first work out the relevant maximal cuts of the five-point two-loop integrals
that appear on the left-hand side of the five-point relation~(\ref{ABDK5}).  In this case,
we have eight propagators to cut, and so we examine the   
octacuts of various five-point integrals. 
We will again encounter
several nonhomologous octacut contours that encircle the same
global poles, but produce distinct residues. As in the four-point case,
the sharing of global poles between different integrals plays a key role. 
As we shall see, the possibility of opening the four-point vertex of the
turtle-box integral ($\Ps_{2,2}$) into either the left or the right loop gives rise to a highly
nontrivial sharing of global poles between different pentabox integrals ($\Pss_{3,2}$).
    
\begin{figure}[!ht]
	\bc
	\hspace*{-.5cm}
	\includegraphics[scale=0.5]{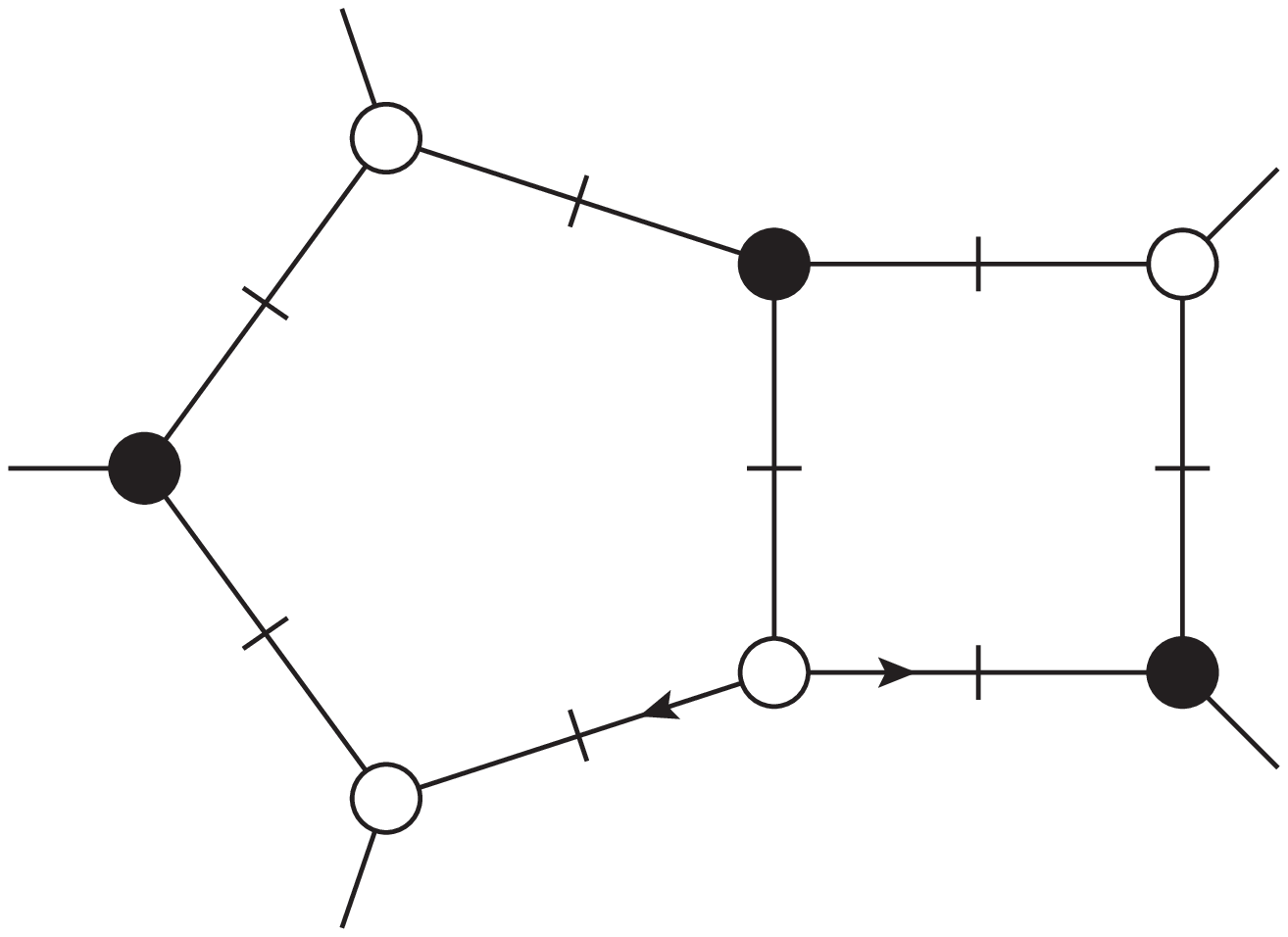}
	\put(-113,39){$\ell_1$}
	\put(-53,48){$\ell_2$}
	\put(-146,-7){$k_1$}
	\put(-204,70){$k_2$}
	\put(-146,143){$k_3$}
	\put(0,116){$k_4$}
	\put(0,19){$k_5$}
	\put(-102,10){Solution $\Global_1$}
	\hspace*{1cm}
	\includegraphics[scale=0.5]{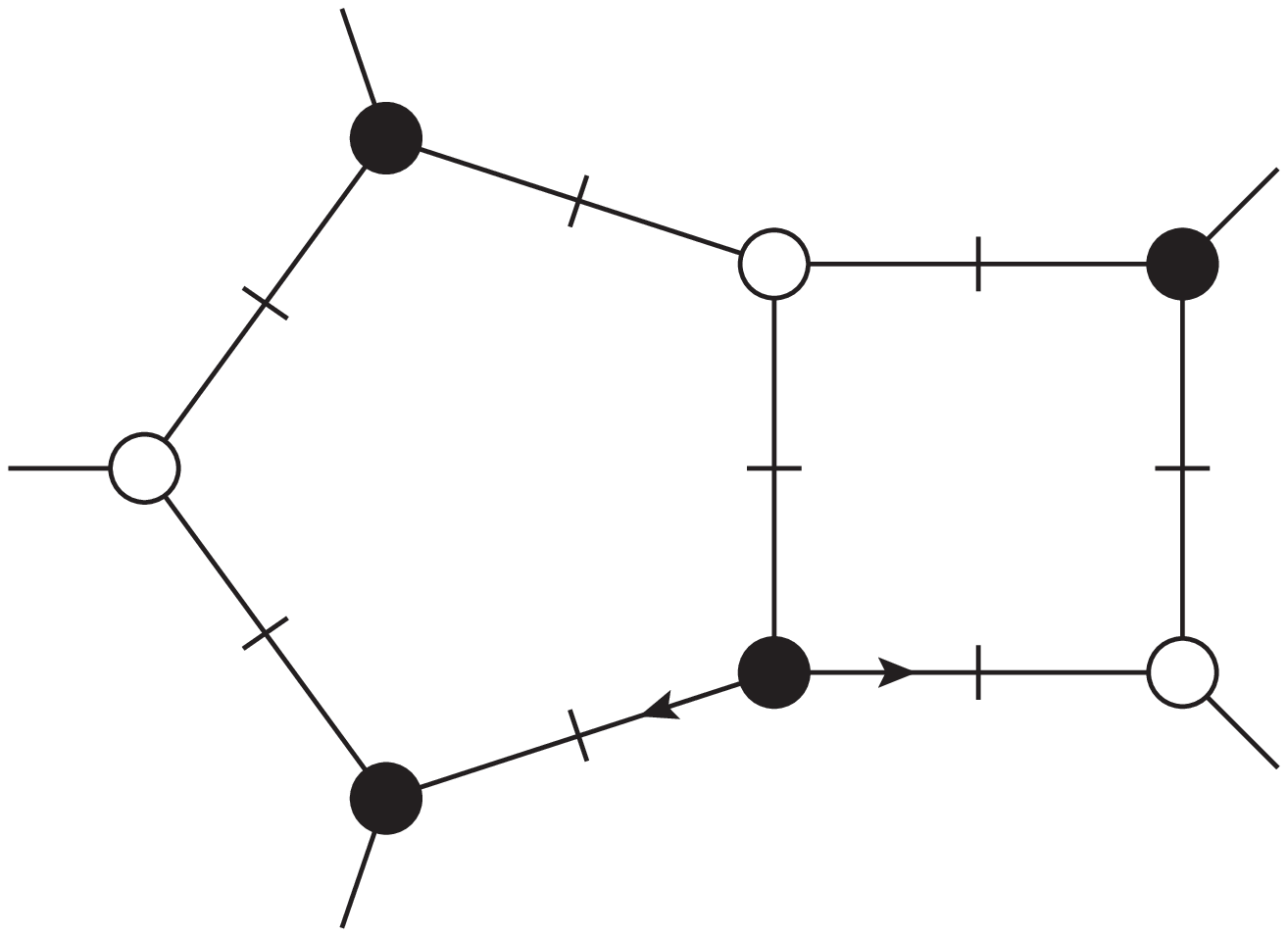}
	\put(-113,39){$\ell_1$}
	\put(-53,48){$\ell_2$}
	\put(-146,-7){$k_1$}
	\put(-204,70){$k_2$}
	\put(-146,143){$k_3$}
	\put(0,116){$k_4$}
	\put(0,19){$k_5$}
	\put(-102,10){Solution $\Global_2$}
	\\[.5cm]
	\hspace*{-.5cm}
	\includegraphics[scale=0.5]{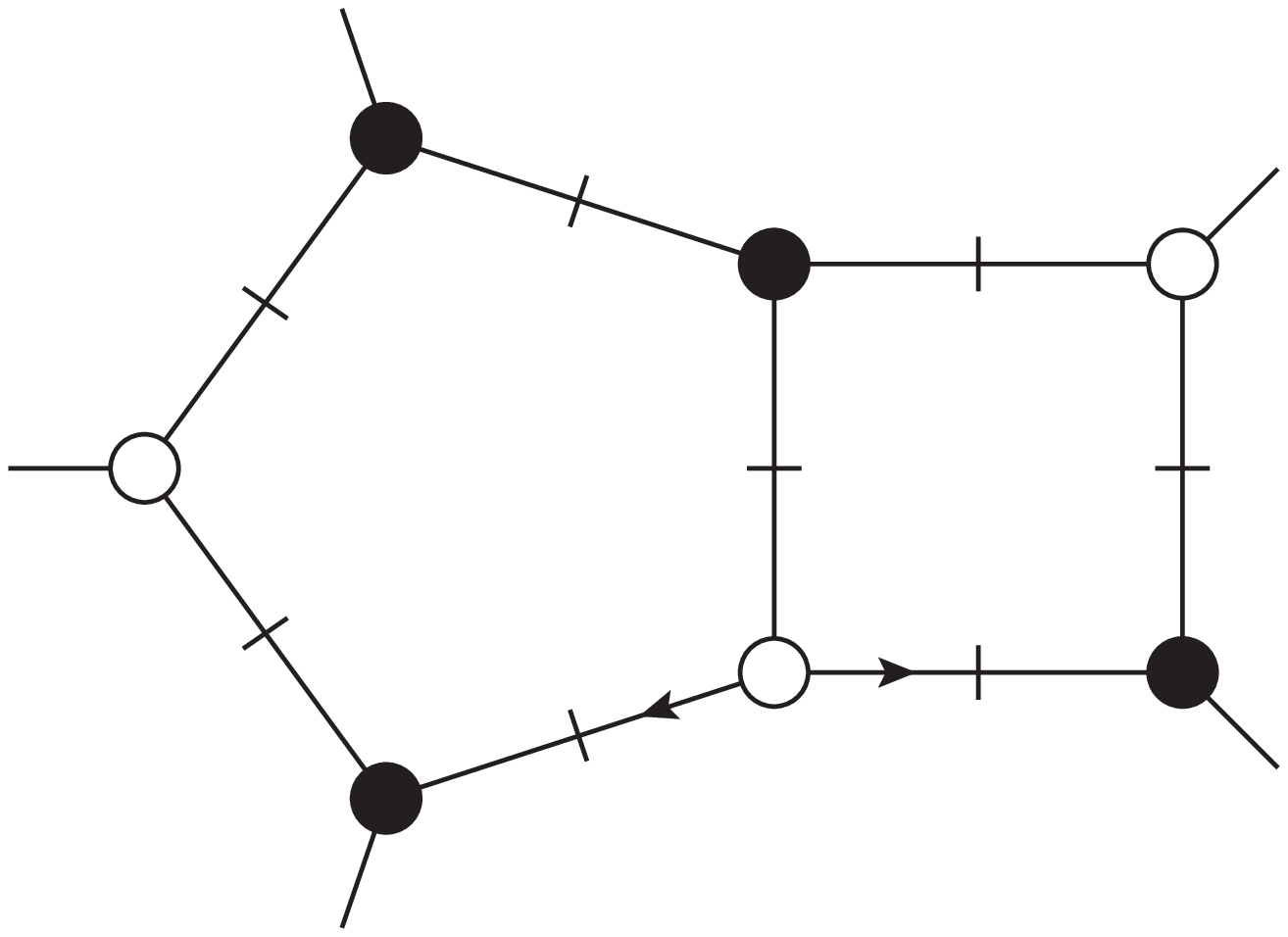}
	\put(-113,39){$\ell_1$}
	\put(-53,48){$\ell_2$}
	\put(-146,-7){$k_1$}
	\put(-204,70){$k_2$}
	\put(-146,143){$k_3$}
	\put(0,116){$k_4$}
	\put(0,19){$k_5$}
	\put(-102,10){Solution $\Global_3$}
	\hspace*{1cm}
	\includegraphics[scale=0.5]{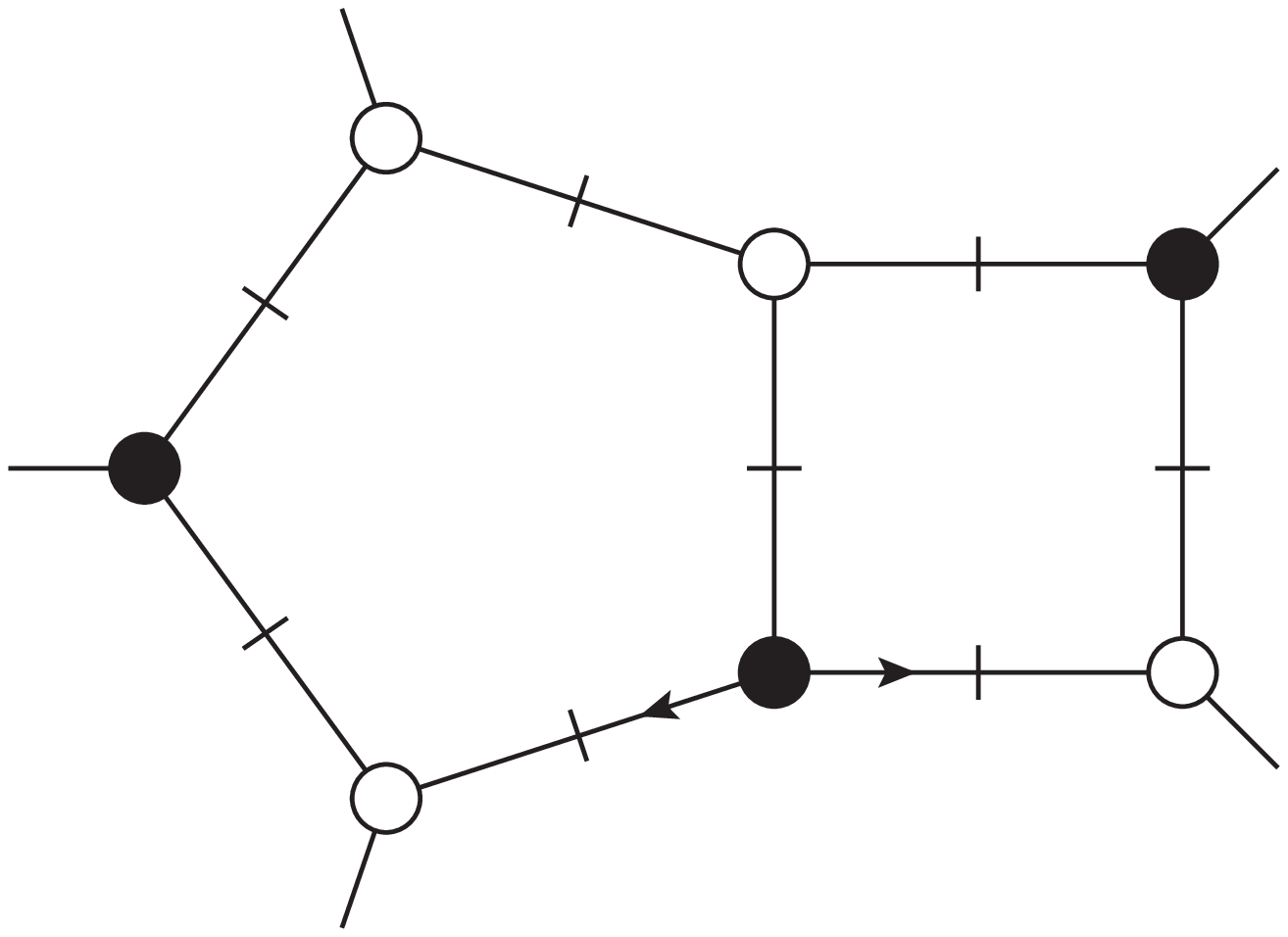}
	\put(-113,39){$\ell_1$}
	\put(-53,48){$\ell_2$}
	\put(-146,-7){$k_1$}
	\put(-204,70){$k_2$}
	\put(-146,143){$k_3$}
	\put(0,116){$k_4$}
	\put(0,19){$k_5$}
	\put(-102,10){Solution $\Global_4$}
	\ec
	\caption{
		\label{PBOXVERTICES}
		The four octacut solutions for the massless pentabox. In our notation, 
		as in prior figures, chiral
		and antichiral vertices are depicted by white and black blobs respectively,
		following for example the conventions of ref.~\cite{ExternalMasses}. Each of these four
		solutions is associated with a unique point in $\CP^4\times \CP^4$.}
\end{figure}

\subsection{Pentabox Global Poles}
We will begin our analysis by determining the global poles of the
massless pentabox integral. In four spacetime dimensions, the octacut solutions
are a discrete set of points, that is, they form a zero-dimensional algebraic variety.  
We can localize the entire
pentabox integral to discrete points in $\CP^4\times\CP^4$ by changing the real-slice
contour to a linear combination of eight-tori each encircling one of the
octacut global poles.

The global poles for the pentabox are given by the zero locus of the
polynomial ideal generated by the eight inverse propagators. In our notation,
this zero locus is
\def\sqz{\mskip-1mu}
\begin{equation}
\hspace*{-2mm}
\begin{aligned}
\Sol\equiv
\big\{(\ell_1,\ell_2)\in \CP^4\!\times\!\CP^4\;|\; &
\ell_1^2 = 0\,,\,
(\ell_1\sqz -\sqz k_1)^2 = 0\,,\,
(\ell_1\sqz -\sqz K_{12})^2 = 0\,,\,
(\ell_1\sqz -\sqz K_{123})^2 = 0\,, \\[2mm] &
\ell_2^2 = 0\,,\,
(\ell_2\sqz -\sqz k_5)^2 = 0\,,\,
(\ell_2\sqz -\sqz K_{45})^2 = 0\,,\,
(\ell_1\sqz +\sqz \ell_2)^2 = 0\,\big\}\,.
\label{PBOXOCTACUTEQ}
\end{aligned}
\end{equation}
\FloatBarrier
There are four inequivalent octacut solutions
$\Sol = \Global_1\cup\cdots\cup\Global_4$, which group into two pairs of parity
conjugates. The four solutions correspond to the four ways of distributing
chiral and antichiral three-point vertices in the cut pentabox diagram that are valid for
generic external momenta (see \fig{PBOXVERTICES}). In other words, two of
the six maximal cut solutions for the one-mass four-point double-box diagram
\cite{ExternalMasses} fail to accommodate an additional on-shell three-point vertex
for generic external kinematics.

We can solve the pentabox on-shell constraints \eqref{PBOXOCTACUTEQ} straightforwardly
using the loop-momentum parametrization,
\begin{equation}
\begin{aligned}
\ell_1^\mu = {} &
\alpha_1 k_1^\mu+\alpha_2 k_2^\mu+
\frac{\alpha_3}2\spvec{1}{\sigma^\mu}{2}+
\frac{\alpha_4}2\spvec{2}{\sigma^\mu}{1}\;,
\\[1mm]                           
\ell_2^\mu = {} &
\beta_1 k_4^\mu+\beta_2 k_5^\mu+
\frac{\beta_3}2\spvec{4}{\sigma^\mu}{5}+
\frac{\beta_4}2\spvec{5}{\sigma^\mu}{4}\;.
\end{aligned}
\label{5PTPARAM}
\end{equation}
For all four solutions,
\begin{align}
\Global_i:\, {} &
\left\{
\begin{array}{ll}
\alpha_1 = 1\,, &\;\;
\alpha_2 = 0\,, 
\\[2mm]
\beta_1 = 0\,, &\;\;
\beta_2 = 1\,.
\end{array}
\right.
\label{CommonParametersPB}
\end{align}
To express the remaining parameters and for later use, it is convenient to 
introduce a notation for certain complex values,
\def\spinorconj#1{#1^{\bullet}}
\begin{equation}
\begin{aligned}
P_1&\equiv  -\frac{\spaa{1}{5}}{\spaa{2}{5}}\;, &
P_2&\equiv  \frac{\spbb{2}{3}}{\spbb{1}{3}}\;, &
P_3&\equiv  -\frac{\spaa{1}{4}}{\spaa{2}{4}}\;, \\
Q_1&\equiv  -\frac{\spbb{1}{5}}{\spbb{1}{4}}\;, &
Q_2&\equiv  \frac{\spaa{3}{4}}{\spaa{3}{5}}\;,
\end{aligned}
\end{equation}
as well as for the spinor (or parity) conjugates,
\def\Pc{\spinorconj{P}}
\def\Qc{\spinorconj{Q}}
\begin{equation}
\begin{aligned}
\Pc_1&\equiv -\frac{\spbb{1}{5}}{\spbb{2}{5}}\;, &
\Pc_2&\equiv \frac{\spaa{2}{3}}{\spaa{1}{3}}\;, &
\Pc_3&\equiv -\frac{\spbb{1}{4}}{\spbb{2}{4}}\;, \\
\Qc_1&\equiv -\frac{\spaa{1}{5}}{\spaa{1}{4}}\;, &
\Qc_2&\equiv \frac{\spbb{3}{4}}{\spbb{3}{5}}\;.
\end{aligned}
\end{equation}
In terms of these values, the loop-momentum parameters take on the
following values at the octacut global poles of the pentabox,
\begin{equation}
\begin{aligned}
\Global_1:\, {} &
\left\{
\begin{array}{lrl}
\alpha_3 = 0\,,\hphantom{{}_2^{\bullet}\,} &\;\;
\alpha_4 &= P_2\,, 
\\[2mm]
\beta_3 = 0\,, \, &\;\;
\beta_4 &= Q_1\,,
\end{array}
\right. &
\Global_2:\, {} &
\left\{
\begin{array}{lrl}
\alpha_3 = \Pc_2\,, &\;\;
\alpha_4 &= 0\,, 
\\[2mm]
\beta_3 = \Qc_1\,, &\;\;
\beta_4 &= 0\,,
\end{array}
\right. \\[2mm]
\Global_3:\, {} &
\left\{
\begin{array}{lrl}
\alpha_3 = \Pc_2\,, &\;\;
\alpha_4 &= 0\,, \\[2mm]
\beta_3 = 0\,, &\;\;
\beta_4 &= Q_2\,,
\end{array}
\right. &
\Global_4:\, {} &
\left\{
\begin{array}{lrl}
\alpha_3 = 0\,, &\;\;
\alpha_4 &= P_2\,, \\[2mm]
\beta_3 = \Qc_2\,, \, &\;\;
\beta_4 &= 0\,.
\end{array}
\right.
\end{aligned}
\end{equation}
Four other pentabox integrals that arise from cyclicly permuting the external
legs can be treated in the same fashion; we omit the details.

\subsection{Overlapping Kinematical Configurations}
A naive counting based on the preceding discussion suggests that the five
cyclic permutations of the pentabox will contain a total of $5\times 4 = 20$
distinct global poles.   This turns out to be an overcount,
because the global poles are actually \emph{shared} between pentabox integrals with different
cyclic permutations of the external legs.  This is analogous to the sharing of global poles
between the horizontal and vertical double boxes at four points.  In light of
the discussion in \sect{SharedGlobalPolesSection}, we may ask:
 does the sharing of pentabox global
poles arise in a simple manner, from the point of view of a generalized unitarity cut?
\begin{figure}[!ht]
\bc
\includegraphics[scale=0.6]{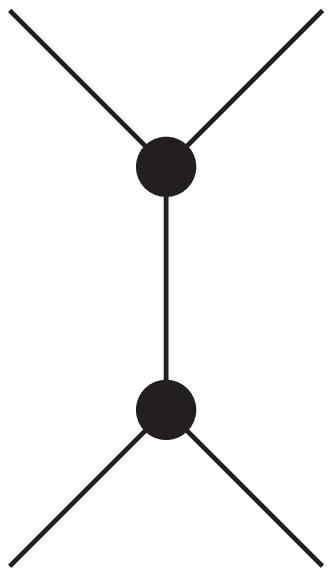}
\hspace{.2cm}
\includegraphics[scale=0.6]{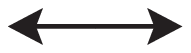}
\hspace{.4cm}
\includegraphics[scale=0.6]{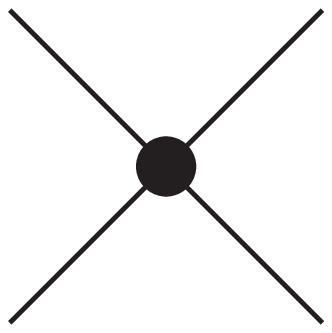}
\hspace{.3cm}
\includegraphics[scale=0.6]{arrow}
\hspace{.4cm}
\includegraphics[scale=0.6]{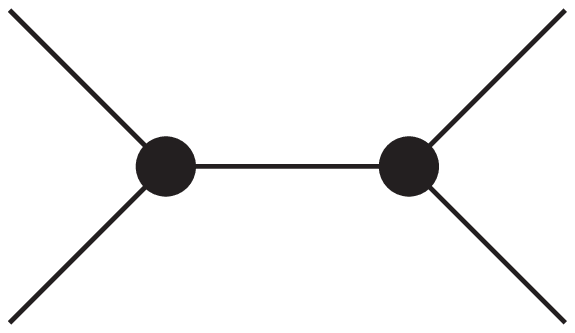}
\vspace*{-.5cm}
\ec                                                                
\caption{
\label{MERGEANDSPLIT}
The merge-and-split operation applied to two adjacent antichiral
vertices.}
\end{figure}

The coincidence of global poles is easy to understand from the octacut pentabox 
diagrams in fig.~\ref{PBOXVERTICES}. Each diagram is uniquely
characterized by the chiralities of the three-point
vertices in the pentagon loop. Overlapping kinematical configurations are
related by an elementary {\it merge-and-split} 
operation~\cite{SplitMove,MergeAndSplit} (see
fig.~\ref{MERGEANDSPLIT}), which manifestly preserves the locations of leading
singularities. The reason is the following. In a massless three-point vertex,
either chiral or antichiral spinors are collinear by momentum conservation.
Accordingly, for two adjacent like-chirality vertices, four spinors of the same
type must be aligned. This constraint is obviously invariant under the
merge-and-split operation, and therefore the cut solution is left unchanged.

\begin{figure}[!h]
\bc
\includegraphics[scale=0.5]{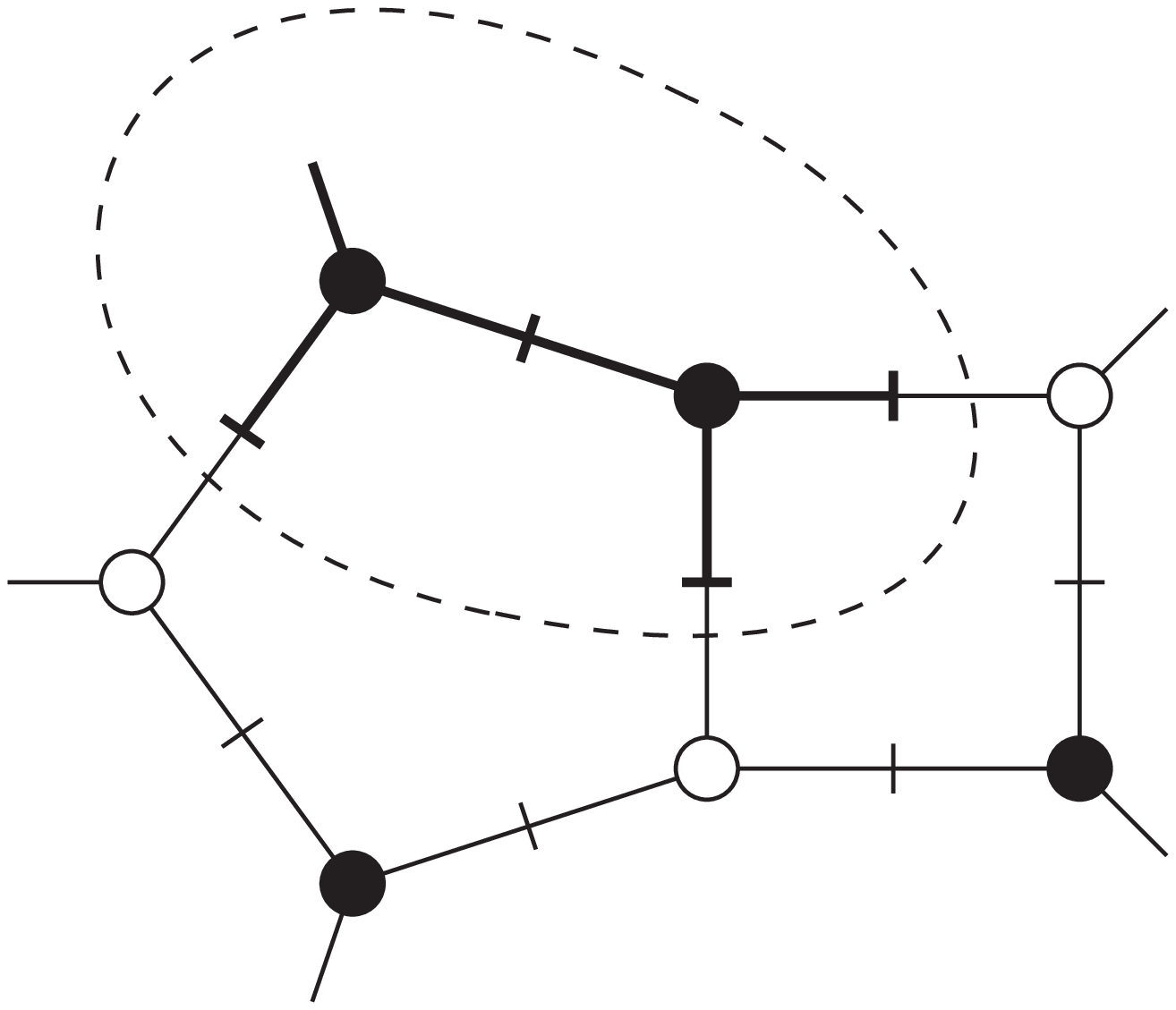}
\put(-148,-9){$k_1$}
\put(-204,68){$k_2$}
\put(-148,144){$k_3$}
\put(-3,119){$k_4$}
\put(-3,15){$k_5$}
\hspace*{.7cm}
\includegraphics[scale=0.5]{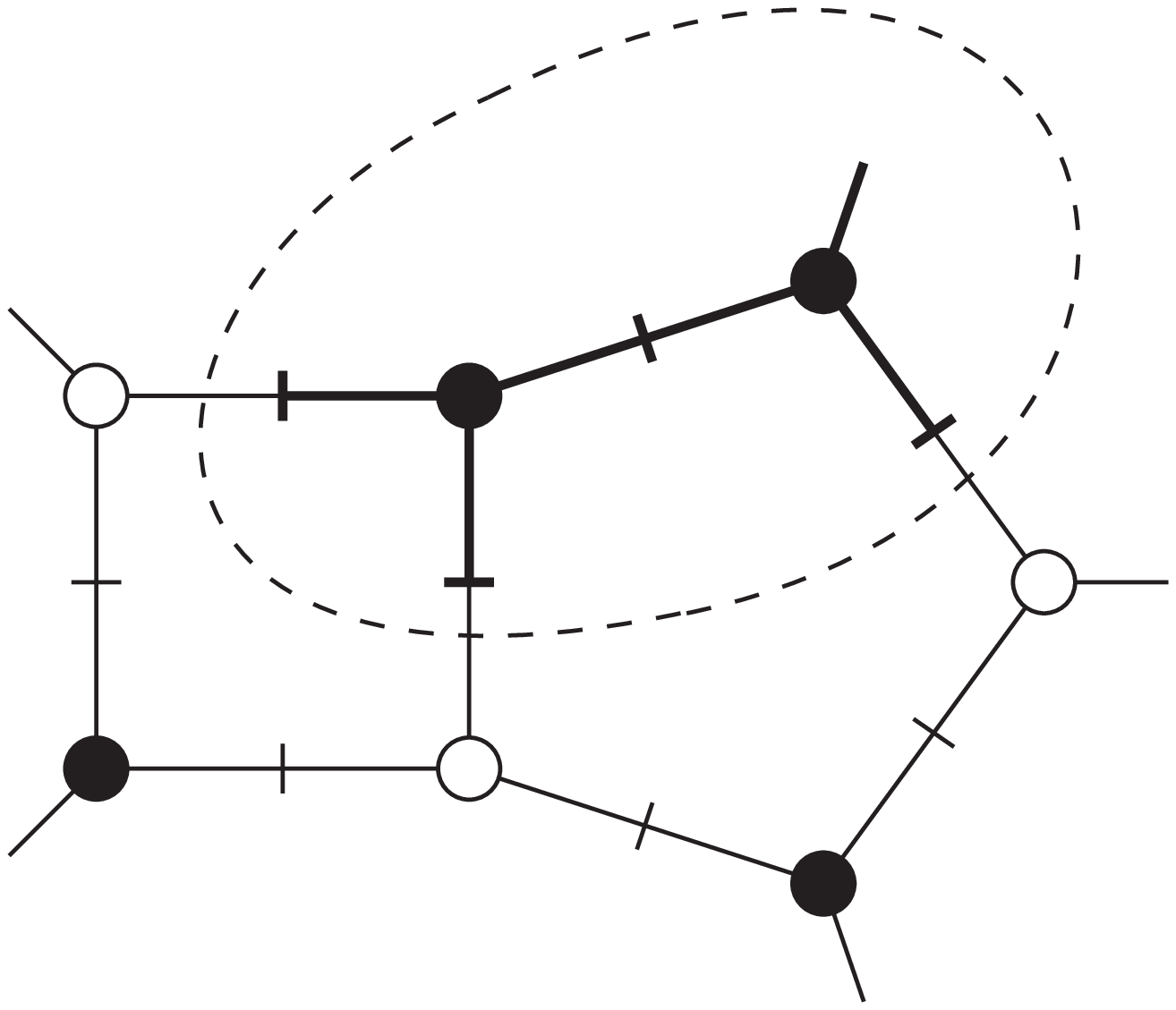}
\put(-197,15){$k_1$}
\put(-197,119){$k_2$}
\put(-54,144){$k_3$}
\put(2,68){$k_4$}
\put(-54,-9){$k_5$}
\ec
\caption{
\label{MERGESPLITSHARE}
The octacut global poles $\Global_3$ (left) and $\Global_2$ (right) of
respectively $\Pss_{3,2}(\sigma_1)$ and $\Pss_{3,2}(\sigma_3)$.  The 
encircled subdiagrams
are related by the split-and-merge operation.}
\end{figure}

This operation teaches us that octacut solutions coincide pairwise,
leaving only 10 distinct global poles. As an example, we can consider the
octacut diagrams that correspond to the global poles $\Global_3$ and $\Global_4$ of
$\Pss_{3,2}(1,2,3,4,5)$ and $\Pss_{3,2}(3,4,5,1,2)$ respectively. 
They are related by
 merging and splitting the adjacent like-chirality vertices in the four-point
 tree indicated by fat lines in \fig{MERGESPLITSHARE}.
 They therefore coincide.  We immediately
conclude that the octacut poles are identical.
The same holds for the parity-conjugate poles.

At first glance, the sharing of global poles between different pentabox integrals
would appear to preclude the use of maximal unitarity at two
loops for five-point processes.  It would seem to imply that we cannot
isolate a specific permutation by cutting all of its eight propagators
simultaneously.  As was true at four points, and as we shall explain
in following subsections, we can avoid this unhappy state of affairs
by making use of \emph{nonhomologous} contours encircling the octacut poles.
Their existence ensures that the different pentabox integrals remain distinguishable,
essentially through their behavior in the neighborhood of the global poles.

\subsection{Cancellation of Octa-Cut Residues}
\label{OCTACUTCANCELSEC}
Given that the pentabox octacut global poles coincide pairwise, it is
natural to examine the implications for the corresponding residues. 
We shall investigate whether we can construct a sum of dual
conformally invariant pentabox integrals, along with a choice of contours,
that yields a vanishing residue.  Such a combination would be a candidate
to being expressed in terms
of simpler integrals such as double boxes, factorized two-loop integrals,
and products of one-loop integrals.

\begin{figure}[!ht]
	\bc
	\includegraphics[scale=0.7]{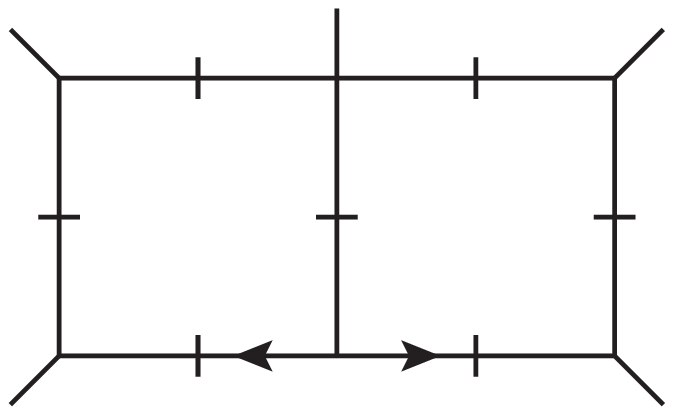}
	\put(-155,-4){$k_1$}
	\put(-157,89){$k_2$}
	\put(-85,93){$k_3$}
	\put(-14,89){$k_4$}
	\put(-14,-4){$k_5$}
	\put(-112,2){$\ell_1$}
	\put(-55,2){$\ell_2$}
	\\
	\includegraphics[scale=0.7]{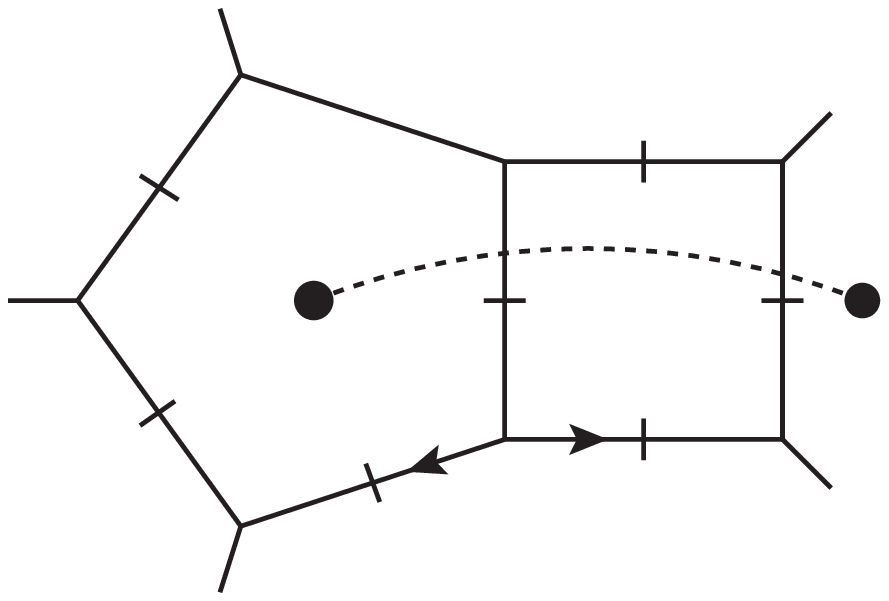}
	\put(-146,-7){$k_1$}
	\put(-194,61){$k_2$}
	\put(-146,128){$k_3$}
	\put(-14,105){$k_4$}
	\put(-14,13){$k_5$}
	\put(-107,11){$\ell_1$}
	\put(-55,19){$\ell_2$}
	\hspace*{.5in}
	\includegraphics[scale=0.7]{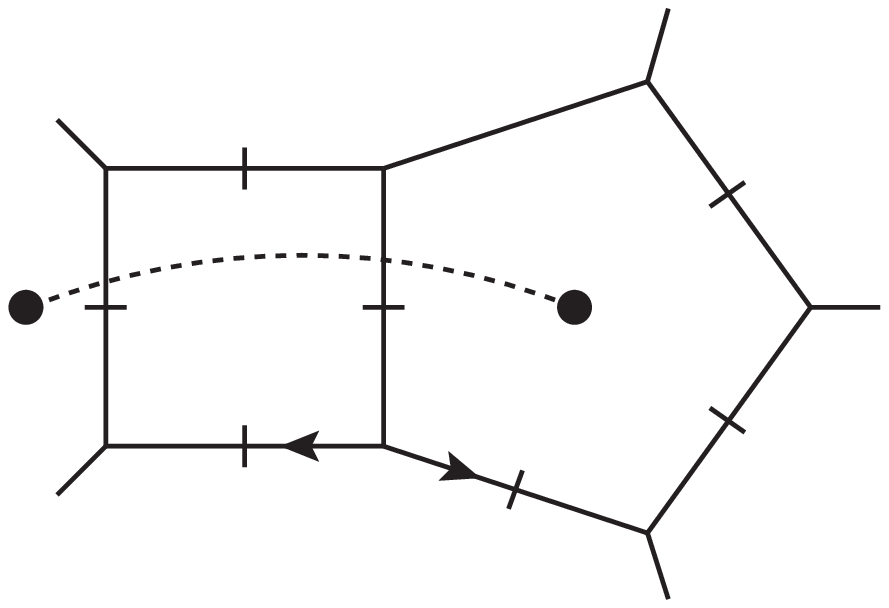}
	\put(-181,13){$k_1$}
	\put(-183,105){$k_2$}
	\put(-50,128){$k_3$}                                                   
	\put(0,61){$k_4$}
	\put(-49,-8){$k_5$}
	\put(-136,19){$\ell_1$}
	\put(-84,11){$\ell_2$}
	\ec
	\caption{
		\label{PBOXTURTLEBOXSHARE1}
		The two displayed permutations of the pentabox share a turtle-box heptacut.
		This is evident directly from the diagrams because the four-point vertex of the
		turtle box can be opened to a pair of three-point vertices to either loop.}
\end{figure}
Consider the following cyclic sum of pentabox integrals,
\begin{equation}
\begin{aligned}
&\sum_{\sigma \in Z_5}
c_{\sigma} \Pss_{3,2}[(\ell_1 + k_{\sigma(5)}^2](\sigma) =
\int\!\frac{d^D\ell_1}{(2\pi)^D}\!
\int\!\frac{d^D\ell_2}{(2\pi)^D} \\[1mm]
&\hspace*{5mm}\times\bigg[
\frac{c_{\sigma_1}(\ell_1 + k_5)^2}{
\ell_1^2(\ell_1 - k_1)^2(\ell_1 - K_{12})^2
    (\ell_1 - K_{123})^2(\ell_1 + \ell_2)^2
\ell_2^2(\ell_2-k_5)^2(\ell_2-K_{45})^2}
\\[1mm] & \hspace*{5mm}\phantom{\times\bigg[}\,
+\!\frac{c_{\sigma_2}(\ell_1+k_1)^2}{
\ell_1^2(\ell_1-k_2)^2(\ell_1-K_{23})^2(\ell_1-K_{234})^2(\ell_1+\ell_2)^2
\ell_2^2(\ell_2-k_1)^2(\ell_2-K_{51})^2}
\\[1mm] & \hspace*{5mm}\phantom{\times\bigg[}\,
+\!\!\frac{c_{\sigma_3}(\ell_1+k_2)^2}{
\ell_1^2(\ell_1-k_3)^2(\ell_1-K_{34})^2(\ell_1-K_{345})^2(\ell_1+\ell_2)^2
\ell_2^2(\ell_2-k_2)^2(\ell_2-K_{12})^2}
\\[1mm] & \hspace*{5mm}\phantom{\times\bigg[}\,
+\!\frac{c_{\sigma_4}(\ell_1+k_3)^2}{
\ell_1^2(\ell_1-k_4)^2(\ell_1-K_{45})^2(\ell_1-K_{451})^2(\ell_1+\ell_2)^2
\ell_2^2(\ell_2-k_3)^2(\ell_2-K_{23})^2}
\\[1mm] & \hspace*{5mm}\phantom{\times\bigg[}\,
+\!\frac{c_{\sigma_5}(\ell_1 + k_4)^2}{
\ell_1^2(\ell_1 - k_5)^2(\ell_1 - K_{51})^2
  (\ell_1 - K_{512})^2(\ell_1 + \ell_2)^2
\ell_2^2(\ell_2 - k_4)^2(\ell_2 - K_{34})^2}
\bigg]\,,
\end{aligned}
\label{CYCLICSUMPENTABOXES}
\end{equation}
with five free parameters $c_{\sigma}$. The sum in 
eq.~\eqref{CYCLICSUMPENTABOXES} runs over the cyclic permutations of the five
external momenta,
\begin{equation}
\begin{aligned}       
\sigma_1\equiv(1,2,3,4,5)\;, &&                                            
\sigma_2\equiv(2,3,4,5,1)\;, &&
\sigma_3\equiv(3,4,5,1,2)\;, && \\
\sigma_4\equiv(4,5,1,2,3)\;, &&
\sigma_5\equiv(5,1,2,3,4)\;.
\end{aligned}
\label{5PTPERMS}
\end{equation}
We analyze each octacut global pole individually. In order
to sketch the general features of the calculation, let us focus on the octacut
$\Global_3$ of the pentabox integral $\Pss_{3,2}(\sigma_1)$. ($\Global_4$ 
is then given by parity conjugation.) As explained above, this
particular global pole is also present in $\Pss_{3,2}(\sigma_3)$ (see
fig.~\ref{MERGESPLITSHARE}). We can reparametrize the latter integral defined in
the third line of eq. \eqref{CYCLICSUMPENTABOXES} by replacing
$\ell_2\to-\ell_1-K_{345}$ and $\ell_2\to-\ell_1+K_{12}$ and thereby align seven
out of the eight internal lines in a physically meaningful way,
\begin{equation}
\begin{aligned}
&c_{\sigma_1}\Pss_{3,2}[(\ell_1+k_5)^2](\sigma_1)+
c_{\sigma_3}\Pss_{3,2}[(\ell_2+k_1)^2](\sigma_3) = \\[1mm]
&\hspace*{15mm}
\int\frac{d^D\ell_1}{(2\pi)^D}
\int\frac{d^D\ell_2}{(2\pi)^D}                                          
\frac{1}{\ell_1^2(\ell_1-k_1)^2(\ell_1-K_{12})^2
(\ell_1+\ell_2)^2(\ell_2-k_5)^2(\ell_2-K_{45})^2} \\[1mm] &
\hspace*{5cm}\;\;
\times\bigg(
c_{\sigma_1}\frac{(\ell_1+k_5)^2}{(\ell_1-K_{123})^2}+
c_{\sigma_3}\frac{(\ell_2+k_1)^2}{(\ell_2-K_{345})^2}\bigg)\;.
\end{aligned}
\end{equation}
Indeed, the two pentaboxes now naturally share a turtle-box heptacut as
illustrated in fig.~\ref{PBOXTURTLEBOXSHARE1}. 
The heptacut has six solutions, just like the heptacut of the double-box integral.
Two of the solutions do not contain a pentabox octacut pole; each of the remaining
solutions contains one of the global poles $\Global_1,\ldots,\Global_4$.
The turtle-box heptacut solution containing $\Global_3$ has the form
\begin{equation}
\Sol_{3}:\,
\left\{
\begin{array}{ll}
	\alpha_3 = z\,, &\;\;\;
	\beta_3 = 0\,, \\[2mm]
	\alpha_4 = 0\,, &\;\;\;
	\beta_4 = \beta_4(z)\,,
\end{array}
\right.
\end{equation}
where,
\begin{align}
\beta_4(z)\equiv 
\frac{Q_2 (\Pc_3-\Pc_2)(z-\Pc_1)}{(\Pc_1-\Pc_2)(z-\Pc_3)}\,.
\label{BETA4TURTLEBOX}
\end{align}
and where $\alpha_{1,2}$ and $\beta_{1,2}$ are given by \eqn{CommonParametersPB}.
(The octacut pole of the pentaboxes is at $z=\Pc_2$, at which point $\beta_4$ becomes
$Q_2$.)
The Jacobians
associated with the changes of variables $\ell_1^\mu\to\alpha_i$ and
$\ell_2^\mu\to\beta_i$ are
\begin{align}
J_L = \det_{\mu,i}\pd{\ell_1^\mu}{\alpha_i} = -i s_{12}^2/4\,, \quad
J_R = \det_{\mu,i}\pd{\ell_2^\mu}{\beta_i} = -i s_{45}^2/4\,.
\end{align}

On this heptacut,
\begin{equation}
\begin{aligned}
&\big(
\Pss_{3,2}[(\ell_1+k_5)^2](\sigma_1)+
\Pss_{3,2}[(\ell_2+k_1)^2](\sigma_3)\big)\big|_{7\text{-cut}} =
\\[1mm] &\qquad\quad
-\frac{1}{16 s_{12}s_{45}\spaa{1}{5}\spbb{5}{2}}
\oint\frac{dz}{z(z-\Pc_1)}
\bigg(
c_{\sigma_1}
\frac{(\ell_1+k_5)^2}{(\ell_1-K_{123})^2}+
c_{\sigma_3}
\frac{(\ell_2+k_1)^2}{(\ell_2-K_{345})^2}\bigg)
\bigg|_{\Sol_3}\,.
\end{aligned}
\end{equation}
We can now evaluate the terms in parentheses; denoting the sum $\Xi(z)$ for instance,
\begin{align}
\Xi(\ell_1,\ell_2)\equiv
c_{\sigma_1}\frac{(\ell_1+k_5)^2}{(\ell_1-K_{123})^2}+
c_{\sigma_3}\frac{(\ell_2+k_1)^2}{(\ell_2-K_{345})^2}\;.
\label{XiDef}
\end{align}
By direct calculation,
\begin{align}
\Xi(z)|_{\Sol_3} = {} &
-c_{\sigma_1}\frac{\spaa{1}{5}\spbb{5}{2}(z-P_1^\bullet)}
{\spaa{1}{3}\spbb{3}{2}(z-P_2^\bullet)}-
c_{\sigma_3}\frac{\spaa{5}{1}\spbb{1}{4}(\beta_4(z)-Q_1)}
{\spaa{5}{3}\spbb{3}{4}(\beta_4(z)-Q_2)}\;.
\label{XIEXPRESSION}
\end{align}
At this stage it is not entirely clear how similar the terms in $\Xi$ are.
The first term has a pole at $z=\Pc_2$; it turns out that the second term 
also has a pole there,
\begin{align}         
\Res_{z=\Pc_2}\frac{1}{\beta_4(z)-Q_2} = {} &
\frac{(\Pc_1-\Pc_2)(\Pc_3-\Pc_2)}
{(\Pc_1-\Pc_3)Q_2}\;,
\end{align}
showing that the two pentaboxes indeed share the octacut global pole as
anticipated. We combine the two terms in eq.~\eqref{XIEXPRESSION} on a common
denominator and take the octacut residue, first imposing the turtle-box heptacut,
and then performing the contour integral in $z$. 
Remarkably, after some spinor algebra, we see
that the residues at this pentabox octacut global pole
cancel between the two integrals in question if 
$c_{\sigma_1}/c_{\sigma_3} = s_{23}/s_{34}$,
\begin{align}
\Res_{z=\Pc_2}\Xi(z)|_{\Sol_{3}} = -
\frac{s_{23}s_{51}}{\spaa{1}{3}\spbb{3}{2}}
\frac{\Pc_1-\Pc_2}{\Pc_1}
\left(\frac{c_{\sigma_1}}{s_{23}}-\frac{c_{\sigma_3}}{s_{34}}\right)\;.
\end{align}
We may fix the overall normalization so that the octacut residues are
independent of external kinematics. For the choice
$c_{\sigma_1} = s_{12}s_{23}s_{45}$ and
$c_{\sigma_3} = s_{12}s_{34}s_{45}$ it follows that $\Xi|_{\Sol_3} = s_{12} s_{45} s_{51}$,
so that,
\def\ssqz{\sqz\sqz}
\begin{equation}
\hspace*{-1mm}
\begin{aligned}
\big(
s_{12}s_{23}s_{45}\Pss_{3,2}[(\ell_1\ssqz +\ssqz k_5)^2](\sigma_1)
&\sqz +\sqz 
s_{34}s_{45}s_{12}\Pss_{3,2}[(\ell_2\ssqz +\ssqz k_1)^2](\sigma_3)\big)\big|_{7\text{-cut}} =
\frac{\Pc_1}{16}\!\oint\!\frac{dz}{z(z\ssqz -\ssqz \Pc_1)}\,.\hspace*{-1mm}
\label{OCTACUTREMOVED}
\end{aligned}
\end{equation}
The cancellation of the pole is of course equivalent to the statement that for this
choice of contour, the
octacut residues are equal in magnitude, but opposite in sign.

\begin{figure}[!h]
\bc
\includegraphics[scale=0.7]{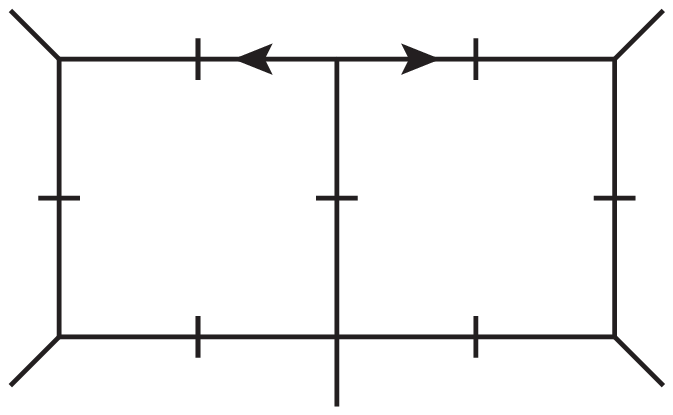}
\put(-85,-8){$k_1$}
\put(-155,-4){$k_2$}
\put(-157,89){$k_3$}
\put(-14,89){$k_4$}
\put(-14,-4){$k_5$}
\put(-112,82){$\ell_1$}
\put(-55,82){$\ell_2$}
\\
\includegraphics[scale=0.7]{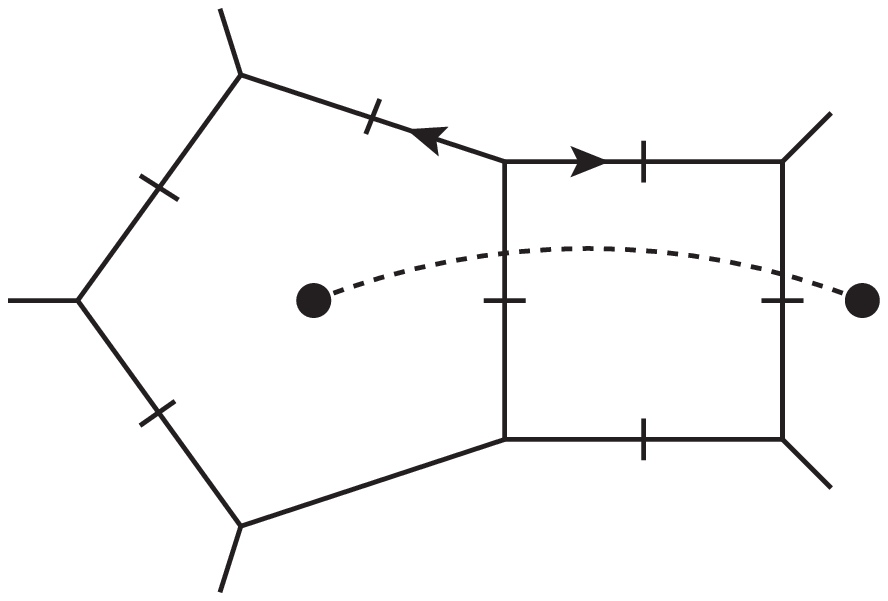}
\put(-146,-7){$k_1$}
\put(-194,61){$k_2$}
\put(-146,128){$k_3$}
\put(-14,105){$k_4$}
\put(-14,13){$k_5$}
\put(-107,108){$\ell_1$}
\put(-56,100){$\ell_2$}
\hspace*{.5in}
\includegraphics[scale=0.7]{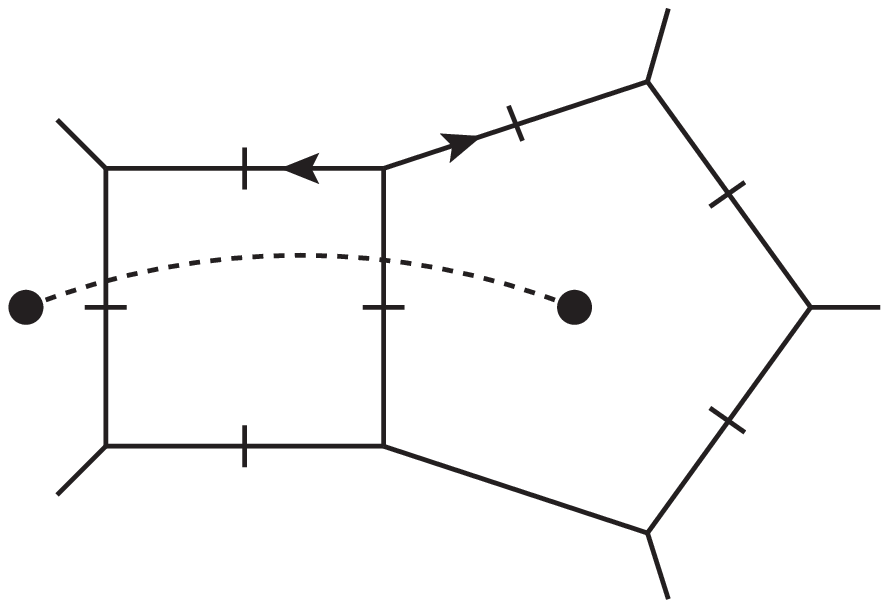}
\put(-49,-8){$k_1$}
\put(-181,13){$k_2$}
\put(-183,105){$k_3$}             
\put(-50,128){$k_4$}
\put(0,61){$k_5$}
\put(-137,100){$\ell_1$}
\put(-84,108){$\ell_2$}
\ec                                                               
\caption{
\label{PBOXTURTLEBOXSHARE2}
There is a second turtle-box heptacut which also receives contributions from
two cyclic permutations of the pentabox. The dashed lines in the left and right
diagrams indicate numerator insertions of
$(\ell_1+k_4)^2$ and $(\ell_2+k_3)^2$ respectively.}
\end{figure}
\begin{figure}[!t]
\bc
\includegraphics[scale=0.5]{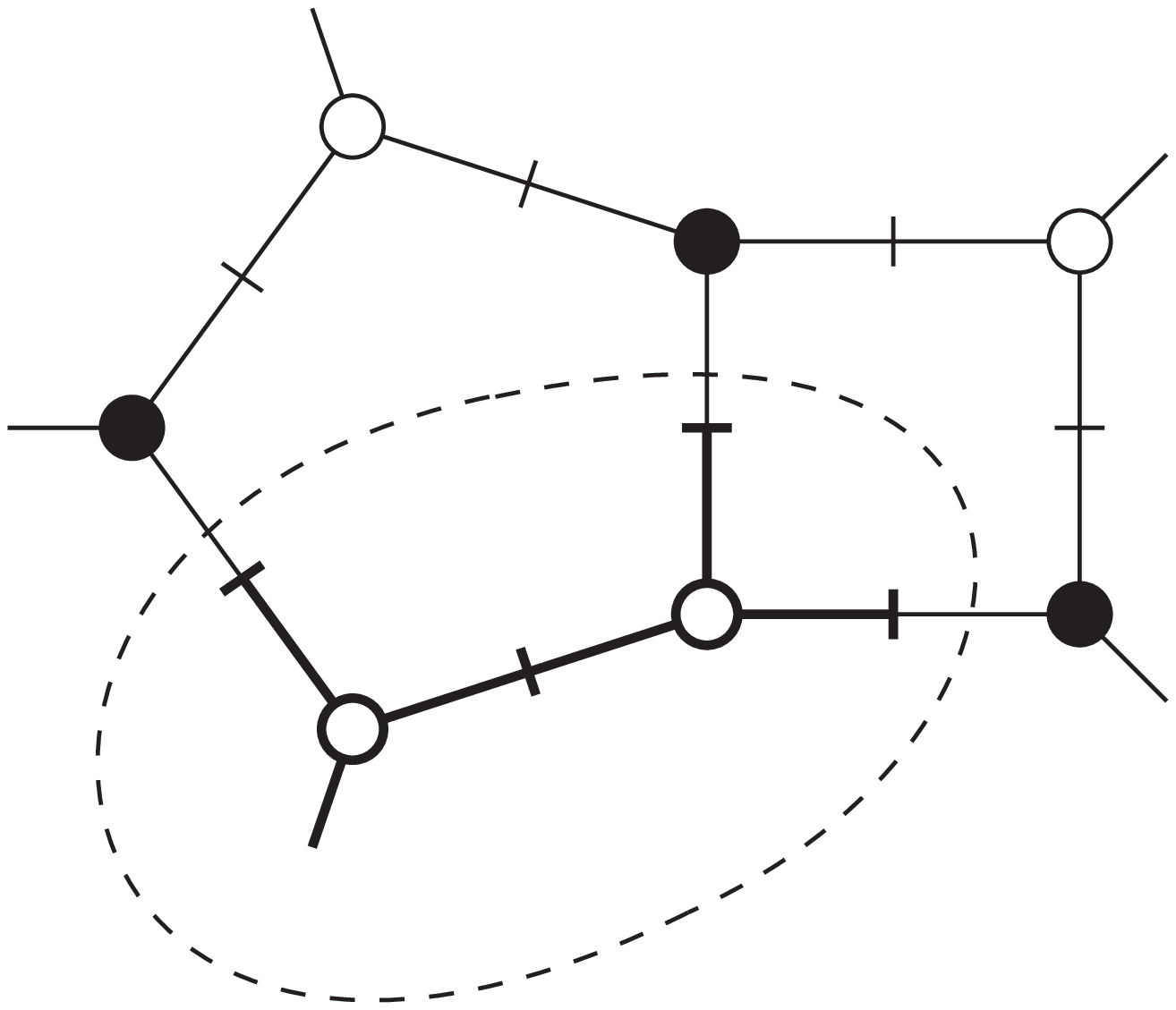}
\put(-148,16){$k_1$}
\put(-204,93){$k_2$}
\put(-148,167){$k_3$}
\put(-2,144){$k_4$}
\put(-2,41){$k_5$}
\hspace*{.8cm}
\includegraphics[scale=0.5]{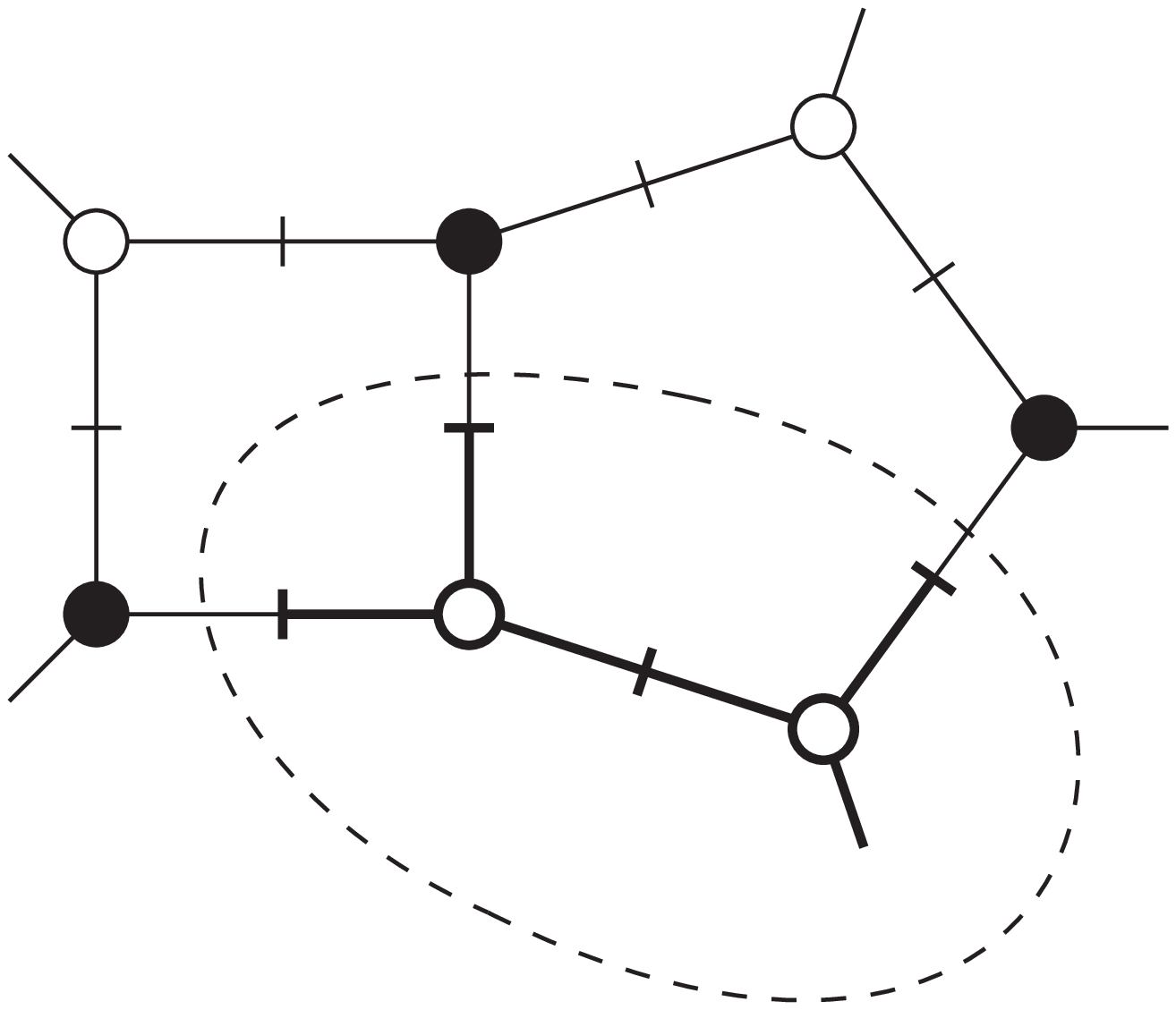}
\put(-54,16){$k_1$}
\put(-198,41){$k_2$}
\put(-198,144){$k_3$}
\put(-54,167){$k_4$}
\put(1,92){$k_5$}
\caption{
\label{OCTACUTSHARED2}
The octacut global poles $\Global_1$ and $\Global_2$ in
$\Pss_{3,2}(\sigma_1)$ are shared with $\Pss_{3,2}(\sigma_4)$. As the poles
are related by parity conjugation, only diagrams relevant for $\Global_1$ are
shown. The four-point subdiagram which is responsible for the sharing is
marked by thicker lines and surrounded by a dashed circle.}
\ec
\end{figure}

What happens with the corresponding residues evaluated at the octacut global
poles $\Global_1$ and $\Global_2$ of $\Pss_{3,2}(\sigma_1)$? Referring to
figs.~\ref{PBOXVERTICES} and \ref{OCTACUTSHARED2}, we can easily guess the answer 
from symmetry.  With an appropriate choice of contour,
these residues will cancel between $\Pss_{3,2}(\sigma_1)$ and the cyclic
permutation $\Pss_{3,2}(\sigma_4)$.  From symmetry considerations,
one possible contour corresponds to the heptacut of the turtle box shown
in \fig{PBOXTURTLEBOXSHARE2}, followed by a contour integral over the remaining 
degree of freedom.
We can show this by direct computation as
before.  First rewrite the loop momenta in the expression for $\Pss_{3,2}(\sigma_1)$
in \eqn{CYCLICSUMPENTABOXES} through the substitutions
$\ell_1\to -\ell_1+K_{123}$ and $\ell_2\to -\ell_2+K_{45}$; and the expression
for $\Pss_{3,2}(\sigma_4)$ via
$\ell_1\to -\ell_2+K_{145}$ and $\ell_2\to-\ell_1+K_{23}$.  The corresponding
terms in \eqn{CYCLICSUMPENTABOXES} then take the form,
\begin{equation}
\begin{aligned}
&c_{\sigma_1}\Pss_{3,2}[(\ell_1+k_4)^2)](\sigma_1)+
c_{\sigma_4}\Pss_{3,2}[(\ell_2+k_3)^2)](\sigma_4) = \\[1mm]
&\hspace*{15mm}
\int\frac{d^D\ell_1}{(2\pi)^D}
\int\frac{d^D\ell_2}{(2\pi)^D}
\frac{1}{\ell_1^2(\ell_1-k_3)^2(\ell_1-K_{23})^2
(\ell_1+\ell_2)^2(\ell_2-k_4)^2(\ell_2-K_{45})^2} \\[1mm] &
\hspace*{5cm}\;\;
\times\bigg(
c_{\sigma_1}\frac{(\ell_1+k_4)^2}{(\ell_1-K_{123})^2}+
c_{\sigma_4}\frac{(\ell_2+k_3)^2}{(\ell_2-K_{451})^2}\bigg)\;.
\end{aligned}
\end{equation}
\def\talpha{\tilde{\alpha}}
\def\tbeta{\tilde{\beta}}
It is convenient to introduce a modified loop momentum
parametrization, suggested by the fact that on the heptacut,
$\ell_1$ and $\ell_2$ now end up
being collinear with $k_3$ and $k_4$ respectively,
\begin{equation}
\begin{aligned}
\ell_1^\mu = {} &
\talpha_1 k_2^\mu+\talpha_2 k_3^\mu
+\frac{\talpha_3}2\spvec{2}{\sigma^\mu}{3}
+\frac{\talpha_4}2\spvec{3}{\sigma^\mu}{2} \,, \\[1mm]
\ell_2^\mu = {} &
\tbeta_1 k_4^\mu+\tbeta_2 k_5^\mu
+\frac{\tbeta_3}2\spvec{4}{\sigma^\mu}{5}
+\frac{\tbeta_4}2\spvec{5}{\sigma^\mu}{4} \,.
\end{aligned}
\end{equation}
\def\tP{\widetilde P}
\def\tPc{\tP^{\bullet}}
\def\tQ{\widetilde Q}
\def\tQc{\tQ^{\bullet}}
Correspondingly, we also define a new set of complex values,
\begin{align}
\tP_1\equiv {} & -\frac{\spaa{3}{4}}{\spaa{2}{4}}\,, &
\tP_2\equiv {} & \frac{\spbb{1}{2}}{\spbb{1}{3}}\,, &
\tP_3\equiv {} & -\frac{\spaa{3}{5}}{\spaa{2}{5}}\,, \nn \\
\tQ_1\equiv {} & -\frac{\spbb{3}{4}}{\spbb{3}{5}}\,, &
\tQ_2\equiv {} & \frac{\spaa{1}{5}}{\spaa{1}{4}}\,,
\end{align}
and their parity conjugates,
\begin{align}
\tPc_1\equiv {} & -\frac{\spbb{3}{4}}{\spbb{2}{4}}\;, &
\tPc_2\equiv {} & \frac{\spaa{1}{2}}{\spaa{1}{3}}\;, &
\tPc_3\equiv {} & -\frac{\spbb{3}{5}}{\spbb{2}{5}}\;, \nn \\
\tQc_1\equiv {} & -\frac{\spaa{3}{4}}{\spaa{3}{5}}\;, &
\tQc_2\equiv {} & \frac{\spbb{1}{5}}{\spbb{1}{4}}\;.
\end{align}

\def\tSol{\widetilde{\Sol}}
\noindent Using these definitions, the heptacut depicted in fig.~\ref{OCTACUTSHARED2}
is realized by fixing the loop-momentum parameters as follows,
\begin{align}
\tSol_{1}:\,
\left\{
\begin{array}{ll}
\talpha_3 = 0\,, &\;\;\;
\talpha_4 = z\,, \\[2mm]
\tbeta_3 = \widetilde\beta_3(z)\,, &\;\;\;
\tbeta_4 = 0\,,
\end{array}
\right. \qquad
\tbeta_3(z) = \frac{
\tQ_2(\tPc_2-\tPc_3)
(z-\tPc_1)}{
(\tPc_2-\tPc_1)
(z-\tPc_3)}\;,
\end{align}
with $\talpha_1 = \tbeta_2 = 0$ and $\talpha_2 = \tbeta_1 = 1$. The Jacobians
associated with the changes of variables $\ell_1^\mu\to\talpha_i$ and
$\ell_2^\mu\to\tbeta_i$ are
\begin{align}
\widetilde J_L = \det_{\mu,i}\pd{\ell_1^\mu}{\talpha_i} = -i s_{23}^2/4\,, \quad
\widetilde J_R = \det_{\mu,i}\pd{\ell_2^\mu}{\tbeta_i} = -i s_{45}^2/4\,.
\end{align}
We can now cut the seven shared propagators to obtain,
\begin{equation}
\begin{aligned}
&\big(
\Pss_{3,2}[(\ell_1+k_4)^2](\sigma_1)+
\Pss_{3,2}[(\ell_2+k_3)^2](\sigma_4)\big)\big|_{7\text{-cut}} =
\\[1mm] & \hspace*{15mm}
\frac{1}{16s_{23}s_{45}\spaa{3}{4}\spbb{4}{2}}\oint
\frac{dz}{z(z-\widetilde{P}_1^\bullet)}\bigg(
c_{\sigma_1}\frac{(\ell_1+k_4)^2}{(\ell_1-K_{123})^2}+
c_{\sigma_4}\frac{(\ell_2+k_3)^2}{(\ell_2-K_{451})^2}\bigg)
\bigg|_{\tSol_1}\;.
\label{eq:cyclic_sum_DCPBs_eq_1}
\end{aligned}
\end{equation}
As with the earlier combination~(\ref{XiDef}), the terms in parentheses
 combine to a constant,
$\big(\cdots\big)\big|_{\tSol_1} = s_{34}$, provided that
$c_{\sigma_1}/c_{\sigma_4} = s_{12}/s_{51}$. With a judicious choice of normalization
we obtain,               
\begin{equation}
\hspace*{-1mm}
\begin{aligned}
\big(
s_{12}s_{23}s_{45}\Pss_{3,2}[(\ell_1\ssqz +\ssqz k_4)^2](\sigma_1)\sqz +\sqz&
s_{23}s_{45}s_{51}\Pss_{3,2}[(\ell_2\ssqz +\ssqz k_3)^2](\sigma_4)\big)\big|_{7\text{-cut}} = 
\frac{\tPc_1}{16}
\!\oint\!\frac{dz}{z(z\ssqz -\ssqz \tPc_1)}\,,\hspace*{-1mm}
\end{aligned}                                                    
\end{equation}
so that the pentabox octacut pole again drops out.  In fact, because the relevant
global poles $\Global_{1,2}$ are shared by the earlier turtlebox shown 
in \fig{PBOXTURTLEBOXSHARE1}, we could also have used that heptacut, followed by the
$z$ contour integral, to demonstrate this cancellation between residues of
$\Pss_{3,2}(\sigma_1)$ and those of $\Pss_{3,2}(\sigma_4)$.

In summary, of the four
octacut poles of $\Pss_{3,2}(\sigma_1)$, the residues at the poles
$\Global_{3,4}$ cancel against two residues at global poles
of $\Pss_{3,2}(\sigma_3)$, while the residues
at the poles $\Global_{1,2}$ cancel against two residues at global poles 
of $\Pss_{3,2}(\sigma_4)$.
The pattern of cancellations extends straightforwardly
to the remaining cyclically permuted dual-conformal pentabox
integrals and their octacut residues. We conclude that in the cyclic
sum of dual-conformal pentaboxes, 
the octacut realized as
a turtle-box heptacut followed by a cut of the last propagator produces
vanishing octacut residues because of pairwise cancellations,
\begin{align}
\sum_{\rho\in\text{cyclic}}
s_{12}s_{23}s_{45}\Pss_{3,2}[(\ell_1+k_{\rho(5)})](\rho)\big|_{8\text{-cut}} = 0\,.
\end{align}
This type of contour explains how the relation~(\ref{ABDK5}) can express a sum of pentaboxes
in terms of simpler integrals, as the one-mass double boxes do not admit these particular
octacuts.   (Some, though not all, of the squared one-loop box integral terms admit these global
poles.)  The sharing of global poles which
makes this possible is highly nontrivial.

\subsection{Bowtie-Based Octa-Cuts}
In the previous subsection, we saw by direct calculation that the
linear combination of pentaboxes on the left-hand side of the
five-point ABDK relation evaluates to
zero on one combination of pentabox octacut contours.  The combination
is made up of contours realized as turtle-box heptacuts
followed by a choice of contour for the remaining degree of
freedom that puts the eighth
propagator on shell.

In this subsection, we examine a different octacut contour which
also has a transparent physical interpretation. It corresponds to
taking the hexacut of a one-mass bow-tie integral, followed by localizing
the integrand onto poles in the remaining degrees of freedom.
The latter step can be thought of as opening the two four-point vertices
to pairs of on-shell three-point vertices, or as probing the limit where both 
loop momenta become soft or collinear with an external leg.
 In contrast to the contours discussed in the previous subsection, each
 contour in the bow-tie class yields a nonvanishing residue
only for one pentabox out of the five with different cyclic orderings
of the external-momentum arguments.

\begin{figure}
	\bc
	\includegraphics[scale=0.75]{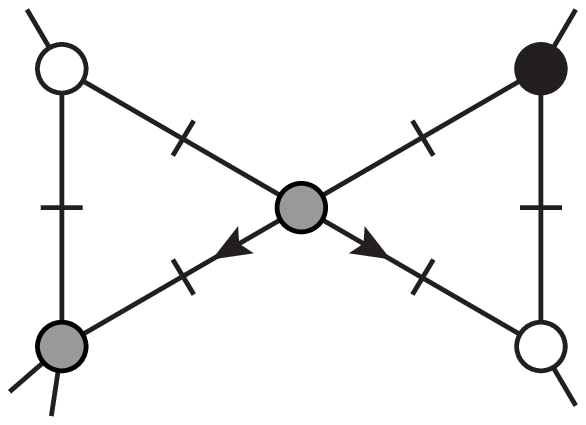}
	\put(-133,-3){$K_{12}$}
	\put(-128,96){$k_3$}
	\put(-6,96){$k_4$}
	\put(-6,-3){$k_5$}
	\put(-90,19){$\ell_1$}
	\put(-44,19){$\ell_2$}
	\hspace*{1cm}
	\includegraphics[scale=0.75]{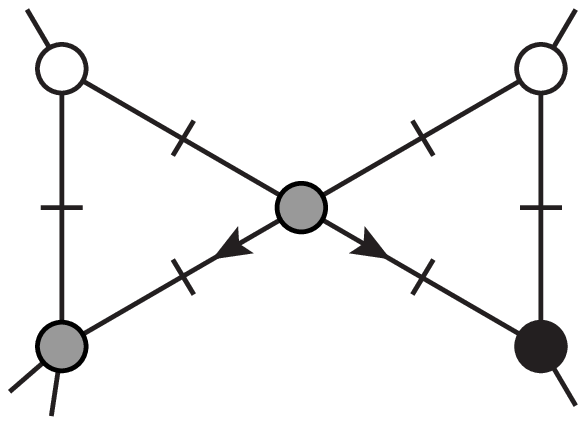}
	\put(-133,-3){$K_{12}$}
	\put(-128,96){$k_3$}
	\put(-6,96){$k_4$}
	\put(-6,-3){$k_5$}
	\put(-90,19){$\ell_1$}
	\put(-44,19){$\ell_2$}
	\caption{\label{1MABDKBOWTIE}
		The one-mass bow-tie hexacut, which has two pairs of parity conjugate branches.
		This figure only shows branches that are not related to each other by parity
		conjugation. All internal lines are on-shell.}
	\ec
\end{figure}

\begin{figure}[!ht]
	\bc
	\includegraphics[scale=0.75]{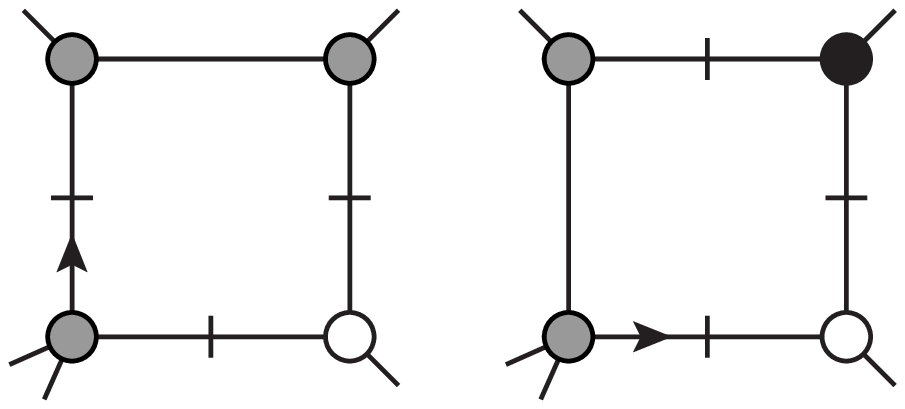}
	\put(-204,92){$k_1$}
	\put(-114,92){$k_2$}
	\put(-114,-5){$k_3$}
	\put(-204,-5){$K_{45}$}
	\put(-96,-5){$K_{12}$}
	\put(-96,92){$k_3$}
	\put(-6,92){$k_4$}
	\put(-6,-5){$k_5$}
	\put(-199,44){$\ell_1$}
	\put(-49,2){$\ell_2$}
	\put(-103,44){$\times$}
	\put(-162,-20){$I_{\square}(\sigma_4)$}
	\put(-55,-20){$I_{\square}(\sigma_1)$}
	\hspace*{1cm}
	\includegraphics[scale=0.75]{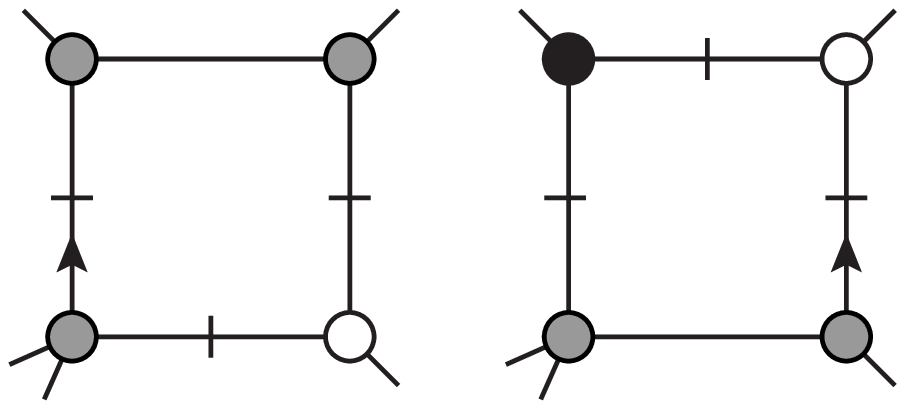}
	\put(-204,92){$k_1$}
	\put(-114,92){$k_2$}
	\put(-114,-5){$k_3$}
	\put(-204,-5){$K_{45}$}
	\put(-6,-5){$k_1$}
	\put(-96,-5){$K_{23}$}
	\put(-96,92){$k_4$}
	\put(-6,92){$k_5$}
	\put(-199,44){$\ell_1$}
	\put(-8,44){$\ell_2$}
	\put(-103,44){$\times$}
	\put(-162,-20){$I_{\square}(\sigma_4)$}
	\put(-55,-20){$I_{\square}(\sigma_2)$}
	\ec
	\caption{
		\label{ONELOOPINTSQ1}
		Products of one-loop one-mass box integrals which can support a pentabox
		octacut. A one-mass box with a massive corner $K_{12}$ is identified with the
		standard ordering $(12345)$, in agreement with the conventions for the two-loop
		integrals. The figure shows a particular branch of the hexacut.}
\end{figure}
\begin{figure}[!h]
	\bc
	\includegraphics[scale=0.75]{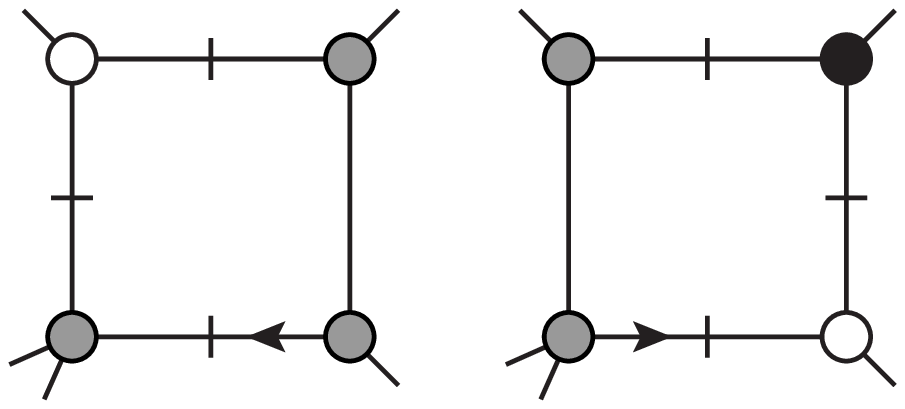}
	\put(-204,-5){$K_{12}$}
	\put(-204,92){$k_3$}
	\put(-114,92){$k_4$}
	\put(-114,-5){$k_5$}
	\put(-96,-5){$K_{12}$}
	\put(-96,92){$k_3$}
	\put(-6,92){$k_4$}
	\put(-6,-5){$k_5$}
	\put(-156,2){$\ell_1$}
	\put(-49,2){$\ell_2$}
	\put(-103,44){$\times$}
	\put(-162,-20){$I_{\square}(\sigma_1)$}
	\put(-55,-20){$I_{\square}(\sigma_1)$}
	\hspace*{1cm}
	\includegraphics[scale=0.75]{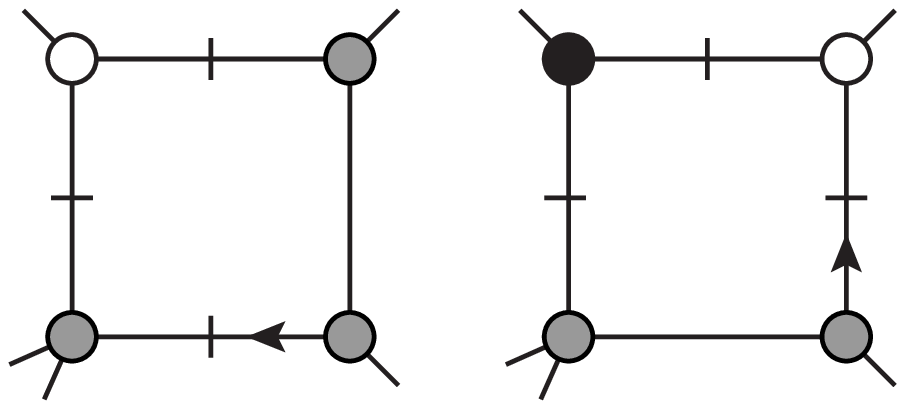}
	\put(-204,-5){$K_{12}$}
	\put(-204,92){$k_3$}
	\put(-114,92){$k_4$}
	\put(-114,-5){$k_5$}
	\put(-6,-5){$k_1$}
	\put(-96,-5){$K_{23}$}
	\put(-96,92){$k_4$}
	\put(-6,92){$k_5$}
	\put(-156,2){$\ell_1$}
	\put(-8,44){$\ell_2$}
	\put(-103,44){$\times$}
	\put(-162,-20){$I_{\square}(\sigma_1)$}
	\put(-55,-20){$I_{\square}(\sigma_2)$}
	\ec
	\caption{
		\label{ONELOOPINTSQ2}
		Products of one-loop one-mass box integrals which cannot support a pentabox
		octacut (since the propagator $(\ell_1-k_1)$ is absent), but may contribute to
		double-box octacuts, i.e. residues from Jacobian poles. The figure shows a
		particular configuration of chiral, antichiral and nonchiral vertices.}
\end{figure}

Consider the one-mass bow-tie hexacut with the standard ordering of the
external legs and standard labeling of loop momenta, shown in fig.~\ref{1MABDKBOWTIE}.
One pentabox integral along with one one-mass double-box integral share
this hexacut.  What other integrals can share it?  
As we are interested in analyzing a cross-order integral
relation, we are led to consider products of one-loop integrals.
There is an obvious candidate to share this cut: 
a product of one-mass box integrals. 
If we examine the right-side loop in either of the diagrams in~\fig{1MABDKBOWTIE},
we see that we can complete the three propagators to a one-loop box in one of
two ways, with the massive leg carrying either $K_{12}$ or $K_{23}$.  We can
complete the three propagators on the left-side loop in two ways as well,
with the massive leg carrying either $K_{45}$ or $K_{12}$.  Overall, this
leaves us with four possible combinations of one-mass boxes,
 which are shown in
figs.~\ref{ONELOOPINTSQ1} and~\ref{ONELOOPINTSQ2}.  We have chosen the loop-momentum
labelings in order to align six of the eight internal lines between the pentabox and
the products of one-loop integrals in a physical way.
This gives us the following expressions, 
\begin{align}
I_{\square}(\sigma_4)\times I_{\square}(\sigma_1)\equiv {} &
\int\frac{d^D\ell_1}{(2\pi)^D}
\frac{1}{\ell_1^2(\ell_1-k_1)^2(\ell_1-K_{12})^2(\ell_1-K_{123})^2}
\nn \\ & \qquad\quad\times
\int\frac{d^D\ell_2}{(2\pi)^D}
\frac{1}{\ell_2^2(\ell_2-k_5)^2(\ell_2-K_{45})^2(\ell_2-K_{345})^2}\,, \\[1mm]
I_{\square}(\sigma_4)\times I_{\square}(\sigma_2)\equiv {} &
\int\frac{d^D\ell_1}{(2\pi)^D}
\frac{1}{\ell_1^2(\ell_1-k_1)^2(\ell_1-K_{12})^2(\ell_1-K_{123})^2}
\nn \\ & \qquad\quad\times
\int\frac{d^D\ell_2}{(2\pi)^D}
\frac{1}{\ell_2^2(\ell_2-k_5)^2(\ell_2-K_{45})^2(\ell_2+k_1)^2}\,, \\[1mm]
I_{\square}(\sigma_1)\times I_{\square}(\sigma_1)\equiv {} &
\int\frac{d^D\ell_1}{(2\pi)^D}
\frac{1}{\ell_1^2(\ell_1-K_{12})^2(\ell_1-K_{123})^2(\ell_1+k_5)^2}
\nn \\ & \qquad\quad\times
\int\frac{d^D\ell_2}{(2\pi)^d}
\frac{1}{\ell_2^2(\ell_2-k_5)^2(\ell_2-K_{45})^2(\ell_2-K_{345})^2}\,, \\[1mm]
I_{\square}(\sigma_1)\times I_{\square}(\sigma_2)\equiv {} &
\int\frac{d^D\ell_1}{(2\pi)^D}
\frac{1}{\ell_1^2(\ell_1-K_{12})^2(\ell_1-K_{123})^2(\ell_1+k_5)^2}
\nn \\ & \qquad\quad\times
\int\frac{d^D\ell_2}{(2\pi)^D}
\frac{1}{\ell_2^2(\ell_2-k_5)^2(\ell_2-K_{45})^2(\ell_2+k_1)^2}\,.
\end{align}
In these equations, the $\sigma_i$ refer to the orderings given
in \eqn{5PTPERMS}, 
with the massive leg made up of the first two arguments
as in \eqn{BoxPermutationDefinition}.

\begin{figure}[!h]
\bc
\includegraphics[scale=0.75]{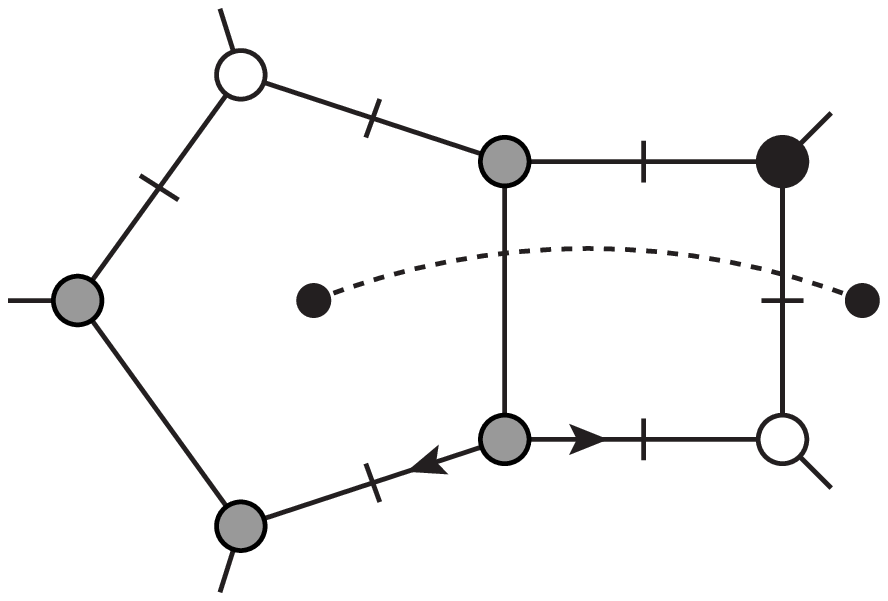}
\put(-154,-5){$k_1$}
\put(-206,64){$k_2$}
\put(-154,134){$k_3$}
\put(-14,110){$k_4$}
\put(-14,18){$k_5$}
\put(-116,13){$\ell_1$}
\put(-59,22){$\ell_2$}
\hspace*{7mm}
\includegraphics[scale=0.75]{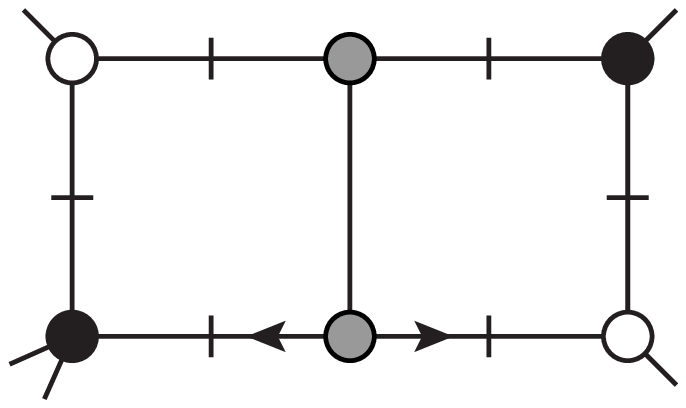}
\put(-157,18){$K_{12}$}
\put(-157,111){$k_3$}
\put(-3,111){$k_4$}
\put(-3,18){$k_5$}
\put(-110,23){$\ell_1$}
\put(-50,23){$\ell_2$}
\caption{
\label{1MABDKBOWTIETWOLOOP}
The one-mass bow-tie hexacut embeds several genuine two-loop integrals. Our
basis includes the dual conformal pentabox and the one-mass double box. The
figure shows a particular kinematical configuration.}
\ec
\end{figure}

The hexacut shown in \fig{1MABDKBOWTIE} defines the 
two-dimensional algebraic variety,
\begin{equation}
\begin{aligned}
\Sol\equiv
\big\{\,(\ell_1,\ell_2)\in \CP^4\times\CP^4\;\big|\; &
\ell_1^2 = 0\,,\;
(\ell_1-K_{12})^2 = 0\,,\;
(\ell_1-K_{123})^2 = 0\,,\; \\ &
\ell_2^2 = 0\,,\;
(\ell_2-k_5)^2 = 0\,,\;
(\ell_2-K_{45})^2 = 0\,\big\}\;.
\label{1MBOWTIECUTEQ}
\end{aligned}
\end{equation}
As in the four-point case there are four classes of hexacut solutions,
each parametrized by two complex variables $(z_1,z_2)\in\C^2$. By parity, it is
sufficient to consider the cuts depicted in fig.~\ref{1MABDKBOWTIE}. Here we
focus mainly on the kinematics of the first of the two cuts, see
fig.~\ref{1MABDKBOWTIETWOLOOP}.

\def\Kfmu{K_{12}^{\flat,\mu}}
\def\Kf{K_{12}^\flat}
As in the four-point case with the massless bow-tie integral~(\ref{BOWTIECUTEQ}),
the one-mass bow-tie hexacut is just a product of independent cuts of one-loop triangle
integrals. This makes it straightforward to write down solutions.
The natural loop-momentum parametrization for this problem is
\def\halpha{\hat\alpha}
\begin{equation}
\begin{aligned}
\label{LOOPPARAMFLAT}
\ell_1^\mu = {} &
\halpha_1 \Kfmu+\halpha_2 k_3^{\mu}
-\frac{\halpha_3\spaa{1}{\Kf}}{2\,\spaa23}\spvec{k_3}{\sigma^\mu}{\Kf}
-\frac{\halpha_4 \spbb{1}{\Kf}}{2\,\spbb23} \spvec{\Kf}{\sigma^\mu}{k_3}\;, \\[1mm]
\ell_2^\mu = {} &
\beta_1 k_4^\mu+\beta_2 k_5^\mu+
\frac{\beta_3}2\spvec{k_4}{\sigma^\mu}{k_5}+
\frac{\beta_4}2\spvec{k_5}{\sigma^\mu}{k_4}\;,
\end{aligned}
\end{equation}
where the flattened vector $\Kf$ (projected along the direction of
momentum $k_3$) is defined by,
\begin{align}
\label{MUTUALLYPROJECTING}
\Kfmu \equiv {} &
K_{12}^\mu-\frac{K_{12}^2}{\gamma_1}k_3^\mu\;.
\end{align}
In this equation, $\gamma_1\equiv 2K_{12}\cdot k_3$. We can then
write down the solutions to the hexacut equations.  All solutions share
the following parameter values,
\begin{equation}
\halpha_1 = 1\,,\;\;
\halpha_2 = 0\,,\;\;
\beta_1 = 0\,,\;\;
\beta_2 = 1\,.
\end{equation}
The solution
 $\Sol_1$ corresponding to the left diagram in fig.~\ref{1MABDKBOWTIE} is
 given by the following values of the remaining parameters,
\begin{align}
\halpha_3 = 0\,,\;\;
\halpha_4 = z_1\,,\;\;
\beta_3 = z_2\,,\;\;
\beta_4 = 0\,.
\end{align}
The parity-conjugate solution $\Sol_2$ is given by,
\begin{align}
\halpha_3 = z_1\,,\;\;
\halpha_4 = 0\,,\;\;
\beta_3 = 0\,,\;\;
\beta_4 = z_2\,.
\end{align}
The solution $\Sol_3$ corresponding to the right diagram 
in fig.~\ref{1MABDKBOWTIE} is
given by,
\begin{align}
\halpha_3 = 0\,,\;\;
\halpha_4 = z_1\,,\;\;
\beta_3 = 0\,,\;\;
\beta_4 = z_2\,.
\end{align}
The last solution $\Sol_4$ is the parity-conjugate to $\Sol_3$,
and is given by,
\begin{align}
\halpha_3 = z_1\,,\;\;
\halpha_4 = 0\,,\;\;
\beta_3 = z_2\,,\;\;
\beta_4 = 0\,.
\end{align}

\def\invbowJ{J_{\bowtie}^{-1}}
\def\tGamma{\widetilde{\Gamma}}
The bow-tie hexacut one-mass double-box and pentabox integrals are,
\begin{equation}
\begin{aligned}
\Pss_{2,2}[1](\sigma_1)\big|_{6\text{-cut}} = {} &
\oint_{\tGamma}
d^2z\frac{\invbowJ(z_1,z_2)}{(\ell_1+\ell_2)^2}
\bigg|_{\Sol_i}\;, \\[1mm]
\Pss_{3,2}[(\ell_1+k_5)^2](\sigma_1)\big|_{6\text{-cut}} = {} &
\oint_{\tGamma}
d^2z\frac{\invbowJ(z_1,z_2)(\ell_1+k_5)^2}{(\ell_1-k_1)^2(\ell_1+\ell_2)^2}
\bigg|_{\Sol_i}\,,
\end{aligned}
\end{equation}
and similarly for the products of one-mass boxes. In these equations,
$\invbowJ(z_1,z_2)$ is the net inverse Jacobian from performing
the hexacut
(including the Jacobians from the change of variables from the loop momenta
to the parameters $\halpha_i$ and $\beta_i$).  It has the same value for
all solutions,
\begin{align}
\invbowJ(z_1,z_2)\equiv -\frac{1}{16\gamma_1 s_{45}}\frac{1}{z_1z_2}\;.
\label{1MBOWTIEJACOBIAN}
\end{align}
The contour $\tGamma$ is in general a weighted sum of contours surrounding
global poles within the four solutions.

For the most general treatment, parity-odd terms such as the pentagon on the
one-loop side and Levi-Civita numerators that integrate to zero must be included as well. 
 Here we instead
construct linear combinations of residues in order to project out parity-odd
terms from the integrands on both sides of the ABDK relation~(\ref{ABDK5}).  
On the two-loop side, 
it suffices to take
parity-even contours that encircle two parity-conjugate global poles with the
same weight. Both factors in a product of one-loop one-mass boxes with a
Levi-Civita insertion in either loop also integrate to zero, although the
product of Levi-Civita contractions is really a Gram determinant. Accordingly,
in general we have to encircle at least four global poles to produce a consistent contour.
It is easy to show that the poles must be two parity-conjugate pairs which are
in turn related by parity-conjugation of either the left or right loop.  This
leads us to consider sums over the four solutions,
\begin{equation}
\begin{aligned}
\Pss_{2,2}[1](\sigma_1)\big|_{6\text{-cut}} = {} &
\sum_{i=1}^4 \oint_{\tGamma_i}
c_i^{(\Gamma)} d^2z\frac{\invbowJ(z_1,z_2)}{(\ell_1+\ell_2)^2}
\bigg|_{\Sol_i}\;, \\[1mm]
\Pss_{3,2}[(\ell_1+k_5)^2](\sigma_1)\big|_{6\text{-cut}} = {} &
\sum_{i=1}^4 \oint_{\tGamma_i}
c_i^{(\Gamma)} d^2z\frac{\invbowJ(z_1,z_2)(\ell_1+k_5)^2}{(\ell_1-k_1)^2(\ell_1+\ell_2)^2}
\bigg|_{\Sol_i}\,,
\end{aligned}
\label{SumOverSolutions}
\end{equation}
where $\tGamma_i$ is the image in $\Sol_i$
of a given contour in $\Sol_1$ under the parity conjugation,
under the parity operation on the right-side loop, or under the combined
operation.  We will choose different
sets of global poles, and corresponding contours, to isolate different terms.  
(In this analysis, there will be a unique contour enclosing each global pole.)
The coefficients $c_i^{(\Gamma)}$
will be chosen in order to enforce the absence of parity-odd terms 
on the right-hand side.  These coefficients could in principle be different for
different global poles.  Some of the global poles will be shared between
different solutions $\Sol_i$; we must be careful
to take only one copy of such global poles in the sum.

\newcommand{\eqinsum}{\overset{\Sigma}{=}}
We determined the coefficients of the dual-conformal pentabox integrals 
in the previous subsection.  
We are left with the task of determining the one-mass double-box and squared
one-loop one-mass box coefficients in the relation.  We start with the hexacut,
\begin{equation}
\begin{alignedat}{10}
&\sum_{i=1}^4 c_i^{(\Gamma)} \oint_{\widetilde\Gamma_i}d^2z\,\invbowJ(z_1,z_2)\bigg[
\frac{c_{\sigma_1}(\ell_1+k_5)^2}{(\ell_1-k_1)^2(\ell_1+\ell_2)^2}+
\frac{c_{1\text{m},\sigma_1}}{(\ell_1+\ell_2)^2}
\bigg]\bigg|_{\Sol_i} = \hspace*{-6cm} \\
& \hspace*{13mm}
\sum_{i=1}^4 c_i^{(\Gamma)} \oint_{\tGamma_i}d^2z\,\invbowJ(z_1,z_2)\bigg[
&&\frac{c_{\sigma_4\times\sigma_1}}{(\ell_1-k_1)^2(\ell_2-K_{345})^2}+
\frac{c_{\sigma_4\times\sigma_2}}{(\ell_1-k_1)^2(\ell_2+k_1)^2} \\
& &&
+\frac{c_{\sigma_1\times\sigma_1}}{(\ell_1+k_5)^2(\ell_2-K_{345})^2}+
\frac{c_{\sigma_1\times\sigma_2}}{(\ell_1+k_5)^2(\ell_2+k_1)^2}
\bigg]\bigg|_{\Sol_i}\,.
\label{BOWTIE_CONTOUR}
\end{alignedat}
\end{equation}
and re-express it in terms of the unfixed variables on the different solutions.
The solution $\Sol_1$ gives the following contribution to the equation,
\def\tm{\!-\!}
\def\tp{\!+\!}
\begin{equation}
\hspace*{-15mm}
\begin{alignedat}{10}
c_i^{(\Gamma)} &\oint_{\tGamma_1}
d^2z\,\frac1{16 s_{45} z_1 z_2}
\biggl[ && \hspace*{-15mm}
  \!\frac{c_{\sigma_1} P_2\Qc_2}{s_{12} s_{23} (z_1\tm P_2)(z_2\tm\Qc_2)}
\!-\!\frac{c_{1\text{m},\sigma_1}\spbb23}
          {\spaa23\spaa45\spbb12\spbb35^2 (z_1\tm P_4) (z_2\tm\Qc_2)}\biggr] 
          \eqinsum
\hspace*{-8cm}
	\\
&\hspace{10mm} c_i^{(\Gamma)} \oint_{\tGamma_1}
d^2z\;\frac1{16 s_{45} z_1 z_2}
\biggl[ &&\frac{c_{\sigma_4\times\sigma_1}}
               {s_{12} \spaa23\spbb13 \spaa34 \spbb35 (z_1\tm P_2)(z_2\tm\Qc_2)}\\
& &&-\frac{c_{\sigma_4\times\sigma_2}}
      {s_{12}\spbb13 \spaa14 \spaa23 \spbb15 (z_1\tm P_2)(z_2\tm\Qc_1) }\\
& &&-\frac{c_{\sigma_1\times\sigma_1}\spbb23}
          {s_{34} \spaa23\spaa45 \spbb12 \spbb35^2 (z_1\tm P_4)(z_2\tm\Qc_2)}\\
& &&-\frac{c_{\sigma_1\times\sigma_2}\spbb23}
    {\spaa14\spaa23\spaa45  \spbb12\spbb15\spbb34\spbb35 (z_1\tm P_4)(z_2\tm\Qc_1)}
  \biggr]\,.
\end{alignedat}\hspace*{-10mm}
\label{OneMassHexacutI}
\end{equation}
where the notation `$\eqinsum$' means that the equality holds only after summing
over all four solutions.
In this expression,
we have introduced labels for two additional complex values and their
parity conjugates,
\begin{equation}
P_4 \equiv \frac{\spaa34 \spbb23 \spbb45}{\spaa23 \spbb12 \spbb35} \,,
\qquad
\Pc_4 \equiv \frac{\spbb34 \spaa23 \spaa45}{\spbb23 \spaa12 \spaa35} \,.
\end{equation}
From solution $\Sol_2$, we obtain the spinor conjugate of \eqn{OneMassHexacutI}.
From solution $\Sol_3$, we obtain,
\begin{equation}
\hspace*{-10mm}\begin{alignedat}{10}
&-c_i^{(\Gamma)} \oint_{\tGamma_3}
\frac{d^2z}{16 s_{45} z_1 z_2}
\\&\hspace*{5mm}\times
\biggl[  -\frac{c_{\sigma_1}\spbb35 (z_1\tm P_4)}
       {\spaa12 (z_1\tm P_2) \big(\spaa23 \spbb12 \spbb13 \spbb34 
       	                        (z_1\tm P_2) (z_2\tp {\Qc_2}{}^{-1})
                             \tp \gamma_1 \spbb14 \spbb23 (z_2\tm Q_1)\big)}
\hspace*{-11cm}\\& \hspace*{5mm}\hphantom{\times\biggl[}\;
+ \frac{c_{1\text{m},\sigma_1} \spbb23 \spbb13}
         {\spaa45 \spbb34 \big(\spaa23 \spbb12 \spbb13 \spbb34
         	     (z_1\tm P_2) (z_2\tp\Qc_2{}^{-1})
                               \tp \gamma_1 \spbb14 \spbb23 (z_2\tm Q_1)\big)}\biggr] \eqinsum
\hspace*{-11cm}\\
&\hspace{10mm}-c_i^{(\Gamma)} \oint_{\tGamma_3}
\frac{d^2z}{16 s_{45} z_1 z_2}
\biggl[ &&
\frac{c_{\sigma_4\times\sigma_1}}
{s_{12}\spaa23 \spaa35\spbb13 \spbb34  (z_1\tm P_2)(z_2\tm Q_2)}
\\
& &&-\frac{c_{\sigma_4\times\sigma_2}}
       {s_{12}\spaa23 \spaa15\spbb13 \spbb14  (z_1\tm P_2)(z_2\tm Q_1) }\\
& &&-\frac{c_{\sigma_1\times\sigma_1} \spbb23}
    {s_{35}\spaa23 \spaa45 \spbb12 \spbb34^2 (z_1\tm P_4)(z_2\tm Q_2)}\\
& &&-\frac{c_{\sigma_1\times\sigma_2} \spbb23}
     {\spaa23 \spaa45 \spbb12 \spbb34 \spbb14 \spbb35 \spaa15 (z_1\tm P_4)(z_2\tm Q_1)}
 \biggr]\,.
\end{alignedat}
\label{OneMassHexacutIII}
\end{equation}
From solution $\Sol_4$, we obtain the spinor conjugate of this equation.
The minus signs on both sides of \eqn{OneMassHexacutIII} arise from the relative ordering 
of the six variables we integrate in order to obtain this form, compared to the
canonical order,
\begin{equation}
d\halpha_1 \wedge d\halpha_2 \wedge d\halpha_3 \wedge d\halpha_4
\wedge d\beta_1 \wedge d\beta_2 \wedge d\beta_3 \wedge d\beta_4\,;
\end{equation} 
in \eqn{OneMassHexacutI}, we must permute one variable ($\beta_4$) twice to the left, whereas
in \eqn{OneMassHexacutIII}, we must permute one variable ($\beta_3$) once to the left.
One might be tempted to cancel the minus signs on both sides of \eqn{OneMassHexacutIII},
but that would alter the relative signs between different solutions.

Most of the singularities in \eqns{OneMassHexacutI}{OneMassHexacutIII} are manifest;
in order to see what singularities may arise from the more intricate denominators
on the left-hand side of \eqn{OneMassHexacutIII}, consider two limits,
\begin{equation}
\begin{aligned}
\big(\spaa23 \spbb12 \spbb13 \spbb34 &(z_1\tm P_2) (z_2\tp {\Qc_2}{}^{-1})
\tp \gamma_1 \spbb14 \spbb23 (z_2\tm Q_1)\big)\big|_{z_1=0} =\\
&\spaa35 \spbb13 \spbb23 \spbb45 (z_2-Q_2)\,,\\
\big(\spaa23 \spbb12 \spbb13 \spbb34 &(z_1\tm P_2) (z_2\tp {\Qc_2}{}^{-1})
\tp \gamma_1 \spbb14 \spbb23 (z_2\tm Q_1)\big)\big|_{z_2=0} =\\
&\spaa23 \spbb12 \spbb13 \spbb35 (z_1-P_4)\,.
\end{aligned}
\end{equation}
Thus, taking the residue at $z_1=0$ will reveal a pole at $z_2=Q_2$; and taking
the residue at $z_2=0$ will reveal a pole at $z_1=P_4$.

We can now enumerate the global poles.  Sixteen poles are located within the bulk
of a single solution; we can group these into four sets, where the poles in each
set are related by parity and the right-loop parity operation, with $(z_1,z_2)$ values,
\begin{equation}
\begin{aligned}
\textrm{I}: &\quad (P_2,\Qc_2);\quad (\Pc_2,Q_2);\quad (P_2,Q_2);\quad (\Pc_2,\Qc_2);\\
\textrm{II}: &\quad (P_2,\Qc_1);\quad (\Pc_2,Q_1);\quad (P_2,Q_1);\quad (\Pc_2,\Qc_1);\\
\textrm{III}: &\quad (P_4,\Qc_2);\quad (\Pc_4,Q_2);\quad (P_4,Q_2);\quad (\Pc_4,\Qc_2);\\
\textrm{IV}: &\quad (P_4,\Qc_1);\quad (\Pc_4,Q_1);\quad (P_4,Q_1);\quad (\Pc_4,\Qc_1).
\end{aligned}
\label{PoleQuartets}
\end{equation}
The sum in \eqn{SumOverSolutions} is over the global poles within each of these sets.  
We will determine the appropriate coefficients below.  
The last set has residues only for the one-loop box squared terms; as in the four-point
case in \sect{FourPointABDKSection}, we do not consider them.

Six poles are each shared between two 
solutions; we can group these into three pairs
\begin{equation}
\begin{aligned}
\textrm{V}: &\quad (P_2,0);\quad (\Pc_2,0);\\
\textrm{VI}: &\quad (P_4,0);\quad (\Pc_4,0);\\
\textrm{VII}: &\quad (0,\Qc_2);\quad (0,Q_2).
\end{aligned}
\label{PolePairs}
\end{equation}
The first pole in the first two pairs is shared between solutions $\Sol_1$ and $\Sol_3$,
and the second is shared between solutions $\Sol_2$ and $\Sol_4$.  In the last pair, the
first pole is shared between $\Sol_1$ and $\Sol_4$, and the second between $\Sol_2$ and
$\Sol_3$.  We can avoid double counting by picking the poles out of $\Sol_{1,2}$ in all
three cases, setting $c^{(\Gamma)}_{3,4}=0$.

Finally, one global pole, at $(z_1,z_2)=(0,0)$, is shared between all four solutions.
We can avoid overcounting the pole by setting $c^{(\Gamma)}_{2,3,4} = 0$.

In the one-loop box squared terms, there are three distinct numerators that give
rise to vanishing integrals: an insertion of a parity-odd numerator in either integral,
or a simultaneous insertion in both.  These are the integrals which the sum in
\eqn{SumOverSolutions} is intended to eliminate.  For the first set in \eqn{PoleQuartets},
for example, we have the following constraints,
\begin{equation}
\begin{aligned}
&\sum_{i=1}^4 c_i^{(\Gamma)} \oint_{\tGamma_i}d^2z\,\invbowJ(z_1,z_2)
\frac{\varepsilon(\ell_1,k_1,k_2,k_3)}{(\ell_1-k_1)^2(\ell_2-K_{345})^2} = 0\,,\\
&\sum_{i=1}^4 c_i^{(\Gamma)} \oint_{\tGamma_i}d^2z\,\invbowJ(z_1,z_2)
\frac{\varepsilon(\ell_2,k_3,k_4,k_5)}{(\ell_1-k_1)^2(\ell_2-K_{345})^2} = 0\,,\\
&\sum_{i=1}^4 c_i^{(\Gamma)} \oint_{\tGamma_i}d^2z\,\invbowJ(z_1,z_2)
\frac{\varepsilon(\ell_1,k_1,k_2,k_3)\varepsilon(\ell_2,k_3,k_4,k_5)}
     {(\ell_1-k_1)^2(\ell_2-K_{345})^2} = 0\,.\\
\end{aligned}
\end{equation}
Evaluating the integrands on the various solutions (and omitting 
overall $z_i$-independent factors), 
these equations become,
\begin{equation}
\begin{aligned}
0&= -c_1^{(\Gamma)}\oint_{\tGamma_1}\!d^2z\,
  \frac{1}{\spaa34\spbb35 z_2 (z_1\tm P_2)(z_2\tm \Qc_2)}
+c_2^{(\Gamma)}\oint_{\tGamma_2}\!d^2z\,
  \frac{1}{\spaa35\spbb34 z_2 (z_1\tm\Pc_2)(z_2\tm Q_2)}\\
&+ c_3^{(\Gamma)}\oint_{\tGamma_3}\!d^2z\,
  \frac{1}{\spaa35\spbb34 z_2 (z_1\tm P_2)(z_2\tm Q_2)}
-c_4^{(\Gamma)}\oint_{\tGamma_4}\!d^2z\,
  \frac{1}{\spaa34\spbb35 z_2 (z_1\tm\Pc_2)(z_2\tm\Qc_2)}
\,,\\
0&= c_1^{(\Gamma)}\!\oint_{\tGamma_1}\!d^2z\,
   \frac{1}{\spaa23\spbb13 z_1 (z_1\tm P_2)(z_2\tm\Qc_2)}
\\ &\hspace*{5mm}
-c_2^{(\Gamma)}\!\oint_{\tGamma_2}\!d^2z\,
   \frac{1}{\spaa13\spbb23 z_1 (z_1\tm\Pc_2)(z_2\tm Q_2)}\\
&\hspace*{5mm} + c_3^{(\Gamma)}\!\oint_{\tGamma_3}\!d^2z\,
   \frac{1}{\spaa23\spbb13 z_1 (z_1\tm P_2)(z_2\tm Q_2)}
\\&\hspace*{5mm}
- c_4^{(\Gamma)}\!\oint_{\tGamma_4}d^2z\,\
   \frac{1}{\spaa13\spbb23 z_1 (z_1\tm\Pc_2)(z_2\tm\Qc_2)}
\,,\\
0&= -c_1^{(\Gamma)}\oint_{\tGamma_1}\!d^2z\,
\frac{1}{ (z_1\tm P_2)(z_2\tm\Qc_2)}
-c_2^{(\Gamma)}\oint_{\tGamma_2}\!d^2z\,
\frac{1}{(z_1\tm\Pc_2)(z_2\tm Q_2)}\\
&-c_3^{(\Gamma)}\oint_{\tGamma_3}\!d^2z\,
\frac{1}{(z_1\tm P_2)(z_2\tm Q_2)}
-c_4^{(\Gamma)}\oint_{\tGamma_4}d^2z\,\
\frac{1}{(z_1-\Pc_2)(z_2-\Qc_2)}
\,.
\end{aligned}
\hspace*{-10mm}\end{equation}
Evaluating the residues on set~I of the poles in \eqn{PoleQuartets}, we find the equations,
\begin{equation}
\begin{aligned}
0 &= \frac{1}{s_{34}} 
  \bigl(c_1^{(\Gamma)} - c_2^{(\Gamma)}- c_3^{(\Gamma)} + c_4^{(\Gamma)} \bigr)\,,\\
0 &= -\frac{1}{s_{23}} 
  \bigl(c_1^{(\Gamma)}- c_2^{(\Gamma)} + c_3^{(\Gamma)} - c_4^{(\Gamma)}\bigr)\,,\\
0 &= -
  \bigl(c_1^{(\Gamma)}+ c_2^{(\Gamma)} + c_3^{(\Gamma)} + c_4^{(\Gamma)}\bigr)\,.
\end{aligned}
\end{equation}
Solving these equations, we find,
\begin{equation}
c_2^{(\Gamma)} = c_1^{(\Gamma)},\quad
c_3^{(\Gamma)} = -c_1^{(\Gamma)},\quad
c_4^{(\Gamma)} = -c_1^{(\Gamma)}.
\end{equation}
The solution is the same for the other pole sets in \eqn{PoleQuartets}.  For the pairs in
\eqn{PolePairs}, we find,
\begin{equation}
c_2^{(\Gamma)} = c_1^{(\Gamma)}.
\end{equation}

\def\onehalf{{\textstyle\frac12}}
The pole sets in \eqns{PoleQuartets}{PolePairs} are shared between the left- and
right-hand sides of \eqn{SumOverSolutions}.  With the coefficients $c_i^{(\Gamma)}$ determined,
we can obtain equations for the coefficients of the various integrals by matching
sums of residues at the different pole sets.  From set~I, we obtain,
\begin{equation}
c_{\sigma_4\times \sigma_1} = \onehalf s_{34} c_{\sigma_1}\,;
\label{c41sol}
\end{equation}
using the value chosen for $c_{\sigma_1}$ in the previous subsection, we find,
\begin{equation}
c_{\sigma_4\times\sigma_1} = \onehalf s_{12} s_{23} s_{34} s_{45}\,.
\end{equation}
From set~II, we obtain,
\begin{equation}
c_{\sigma_4\times \sigma_2} = \onehalf s_{51} c_{\sigma_1}
=\onehalf s_{12} s_{23} s_{45} s_{51}\,.
\label{c42sol}
\end{equation}
From set~III, we obtain a relation between $c_{\sigma_1\times\sigma_1}$ and 
$c_{1\text{m},\sigma_1}$,
\begin{equation}
c_{\sigma_1\times\sigma_1} = \onehalf s_{34}c_{1\text{m},\sigma_1}\,.
\label{c11sol}
\end{equation}
As noted above, the set~IV consists of poles that appear only in the one-loop box
squared terms, and as in the four-point case, we set these aside.  
The pair~V gives no new information beyond \eqns{c41sol}{c42sol}.
From the pair~VI, after substituting \eqn{c11sol}, we obtain a relation for
$c_{\sigma_1\times\sigma_2}$,
\begin{equation}
c_{\sigma_1\times\sigma_2} = \onehalf s_{51}c_{1\text{m},\sigma_1}\,.
\label{c12sol}
\end{equation}
From the pair~VII, we obtain a solution for $c_{1\text{m},\sigma_1}$
after substituting \eqns{c41sol}{c11sol},
\begin{equation}
c_{1\text{m},\sigma_1} = \frac{s_{34}s_{45}}{s_{12}s_{23}} c_{\sigma_1} =
s_{34} s_{45}^2\,.
\end{equation}
With this value, we also find,
\begin{equation}
\begin{aligned}
c_{\sigma_1\times\sigma_1} &= \onehalf s_{34}^2 s_{45}^2\,,\\
c_{\sigma_1\times\sigma_2} &= \onehalf s_{34} s_{45}^2 s_{51}\,.
\end{aligned}
\label{c11n12sol}
\end{equation}
The remaining pole, at $(z_1,z_2)=(0,0)$, gives no additional equations.

As in the four-point case, the integrands of the squared one-loop box terms are
symmetric under the interchange $\ell_1\leftrightarrow\ell_2$, whereas the integrands
of the pentaboxes and one-mass double boxes are not.  This means that in symmetrizing
the integrands following the discussion at the end of \sect{FourPointABDKSection},
the left-hand side's residues will acquire an extra factor of $1/2$.  The same will 
also be true of the product of non-identical one-loop boxes, because the residue extraction
above will find non-vanishing residues only for one of the two terms.  Accordingly,
the factors of $1/8$ noted at the beginning of this section for the pentaboxes and
one-mass double boxes will become $1/16$, whereas the one-loop box squared terms
and the product of different one-loop boxes will have factors of $1/32$ in front
of their residues.  The relative factor of $1/2$ is precisely what is seen above
in eqs.~(\ref{c41sol}--\ref{c11n12sol}).

The matching of residues thus leads precisely to the
integral coefficients present in the five-point ABDK relation~(\ref{ABDK5}).
 In other words, we have reconstructed the
parity-even leading-localization part of the ABDK relation
in the maximally supersymmetric Yang--Mills theory.
We will not discuss the details, but we have also checked explicitly that an analysis 
of the hexacut, including all
parity-even and -odd contributions, yields a residue-by-residue match of 
the leading-localization terms in the relation, without the need for summing over sets
of residues as in the discussion above.

\section{Discussion and Conclusions}
\label{Conclusions}

In this paper, we have studied the two-loop ABDK/BDS relation from the
viewpoint of maximal generalized unitarity. The coefficients of integrals
in an amplitude in this approach are given by multivariate contour integrals
of products of trees. The multivariate contour integrals are taken around
global poles, and unlike the single-variable case, there can be several
non-homologous contours surrounding a given global pole. We gave a simple example
of this in \sect{DEGENERATERESIDUES}. Different integrals can share a global pole, but in
some instances have different residues with respect to different contours surrounding
the pole.

It turns out that the left- and right-hand sides of the ABDK relation (\ref{ABDK4})
and~(\ref{ABDK5})
do indeed share global poles, and have residues with respect to a common contour.
This allows us to match contributions on both sides of the relation for the
global poles of the planar two-loop integrals. The matches allow us to determine
the coefficients of the one-loop box squared terms on the right-hand
side of eqs.~(\ref{ABDK4}) and (\ref{ABDK5}), and of the pentabox on the left-hand
side of \eqn{ABDK5}.   We leave to future work puzzles associated with residues appearing only
on the right-hand side, not shared by planar two-loop integrals.
In addition, the right-hand sides also have terms not detectable in the maximal
localizations of the integrals that we perform,
such the one-loop amplitude in $D=4-4\eps$ dimensions.
It would be interesting to see
if these are also accessible to generalized-unitarity techniques. The analysis
in this paper suggests that maximal generalized-unitarity techniques can be
used to search for new integral or amplitude identities beyond the ones dictated
by dual conformal invariance.

\acknowledgments
We have benefited from discussions with Simon Caron-Huot, David Skinner, 
Jaroslav Trnka, and Yang Zhang. KJL is
grateful for the hospitality of the Institute for Advanced Study in Princeton
and the Institut de Physique Th\'eorique, CEA Saclay, where part of this work
was carried out. MS thanks SLAC National Accelerator Laboratory and the Institut
de Physique Th\'eorique, CEA--Saclay for hospitality during several phases of
this project.  HJ's work is supported in part by the Swedish Research Council under 
grant 621--2014--5722, the Knut and Alice Wallenberg Foundation under 
grant KAW~2013.0235 (Wallenberg Academy Fellowship), and the CERN-COFUND Fellowship 
program (co-funded via Marie Curie Actions grant 
PCOFUND--GA--2010--267194 under 
the European Union’s Seventh Framework Programme).
DAK's work has been supported by the European Research Council
under Advanced Investigator Grant
ERC--AdG--228301.  DAK also thanks the Moore Visiting Scholars program,
 the Moore Center for Theoretical Cosmology and Physics (under grant \#776), and
 the Walter Burke Institute for Theoretical Physics for their support.  
The research leading to these results has received funding from the European Union Seventh Framework Programme (FP7/2007--2013) under grant agreement no.~627521.


\end{document}